\documentclass[prb,superscriptaddress,twocolumn,showpacs]{revtex4}
\usepackage{graphicx}
\usepackage{times}
\usepackage{color}
\usepackage{amsmath,amssymb}
\usepackage{multirow}
\usepackage{dcolumn}

\newcommand{\idop}{\mathbb{I}}
\newcommand{\id}{\mathbf{1}}

\newcommand{\ie}{{\it i.~e.}}

\newcommand{\fs}[4]{%
F^{#1#2#3}_{#4}
}
\newcommand{\qnum}[1]{\lfloor #1 \rfloor}

\newcommand{\bra}[1]{\left\langle{#1}\right|}
\newcommand{\ket}[1]{\left|{#1}\right\rangle}

\allowdisplaybreaks[4]

\begin{document}

\title{Anyonic quantum spin chains: Spin-1 generalizations and topological stability}

\author{C. Gils}
\affiliation{Theoretische Physik, Eidgen\"ossische Technische Hochschule Z\"urich, 8093 Z\"urich, Switzerland}
\affiliation{Department of Mathematics and Statistics, University of Saskatchewan, Saskatoon S7N 5E6, Canada}
\author{E. Ardonne}
\affiliation{Nordita, Royal Institute of Technology and Stockholm University,
Roslagstullsbacken 23, SE-106 91 Stockholm, Sweden}
\affiliation{Department of Physics, Stockholm University, AlbaNova University Center, SE-106 91 Stockholm, Sweden}
\author{S. Trebst}
\affiliation{Institute for Theoretical Physics, University of Cologne, 50937 Cologne, Germany}
\affiliation{Microsoft Research, Station Q, University of California, Santa Barbara, CA 93106, USA}
\author{D.A. Huse}
\affiliation{Physics Department, Princeton University, Princeton, NJ 08544, USA}
\author{A.W.W. Ludwig}
\affiliation{Physics Department, University of California,  Santa Barbara, CA 93106, USA}
\author{M. Troyer}
\affiliation{Theoretische Physik, Eidgen\"ossische Technische Hochschule Z\"urich, 8093 Z\"urich, Switzerland}
\author{Z. Wang}
\affiliation{Microsoft Research, Station Q, University of California, Santa Barbara, CA 93106, USA}
\date{\today}

\begin{abstract}
There are many interesting parallels between systems of interacting non-Abelian anyons and
quantum magnetism, occuring in  ordinary SU(2) quantum magnets.
Here we consider theories of so-called  su(2)$_k$ anyons, well-known
deformations of SU(2), in which only  the first $k+1$ angular momenta of SU(2) occur.
In this manuscript, we discuss in particular
anyonic generalizations of ordinary SU(2) spin chains with 
an emphasis on anyonic spin $S=1$ chains. We find that the overall phase diagrams for these anyonic spin-1 chains closely mirror the phase diagram of the ordinary bilinear-biquadratic spin-1 chain including anyonic generalizations of the Haldane phase, the AKLT construction, and supersymmetric quantum critical points. A novel feature of the anyonic spin-1 chains is an additional topological symmetry that protects the gapless phases.
Distinctions further arise 
in the form of an even/odd effect in the deformation parameter $k$ 
when considering su(2)$_k$ anyonic theories with $k\ge5$, as well as
for the special case of the  su(2)$_4$ theory for which the spin-1 representation plays a special role.
We also address anyonic generalizations of spin-1/2 chains with a focus 
on the topological protection provided for their gapless ground states. 
Finally, we put our results into context of earlier generalizations of SU(2) quantum spin chains, in particular  so-called (fused) Temperley-Lieb chains. 
\end{abstract}


\maketitle

\tableofcontents

\section{Introduction}

Ever since the early days of condensed matter physics, quantum magnets have played an integral role in shaping our understanding of interacting quantum many-body systems. Following the experimental discovery of the high-temperature superconductors whose undoped parent compounds typically are antiferromagnets,
the study of quantum magnets has further intensified yielding a plethora of deeper insights. 
Early on, quantum spin chains -- typically one-dimensional arrangements of SU(2) spins -- have become prototypical systems that proved to be fruitful ground for analytical descriptions and quasi-exact numerical analysis \cite{Affleck}. One seminal result was the exact solution of the antiferromagnetic spin-1/2 Heisenberg chain via the Bethe ansatz and its description in terms of conformal field theory. Another crucial contribution was Haldane's realization \cite{Haldane} that the antiferromagnetic spin-1 Heisenberg chain forms a gapped state with characteristic zero-energy edge states for open boundary conditions -- a principle observation that holds true for all half-integer and integer spin chains.
 More recently, it has been found that  the gapped Haldane phase of the spin-1 chain is 
an example of a symmetry protected topological phase \cite{Pollmann,Chen11} making it a
one-dimensional cousin of topological insulator states in two and three dimensions \cite{TopoInsulator}, which have attracted much recent interest.

Over the years, a plethora of physical systems that connect to
the elementary physics of quantum spin chains have been identified, including transition metal oxides \cite{Hase93}, 
Au quantum wires on semiconducting surfaces \cite{Claessen11}, or ultra-cold atoms in optical lattices \cite{Schwager12}.
Recently, it has been realized that
certain `deformations' of quantum spins
can be used to describe some of the more peculiar topological properties of exotic quasiparticles, so-called non-Abelian anyons, that arise in certain topologically ordered systems, including certain fractional quantum Hall states\cite{MooreRead}, $p_x + ip_y$ superconductors \cite{ReadGreen}, heterostructures of topological insulators and superconductors \cite{FuKanePplusIP}, heterostructures of spin-orbit coupled semiconductors and superconductors \cite{SauSOSemi} and possibly certain Iridates \cite{Iridates} which may effectively realize the Kitaev honeycomb model \cite{Kitaev06}.
To be more specific, the
deformations of quantum spins
are representations of the anyon theories 
called
su(2)$_k$, which can be described as
theory of ordinary SU(2) quantum spins
that is deformed
in such a way that only the first $k+1$ (generalized)  angular momenta 
\[
	j = 0, \frac{1}{2}, 1, \ldots, \frac{k}{2} \,
\]
can occur.
These generalized angular momenta capture the non-Abelian properties of the anyonic quasiparticles present in the su(2)$_k$ theory. For instance, 
the non-Abelian nature of the so-called Majorana fermion is captured by the generalized angular momentum 1/2 of the su(2)$_2$ theory.
The same holds for so-called Ising anyons, while Fibonacci anyons can be  represented by the generalized angular momentum 1 of the su(2)$_3$ theory.
Similar to the coupling of two ordinary spins,  a pair of generalized angular momenta can be combined (or `fused') into a new set of joint quantum numbers. For instance, for $k\ge2$, two generalized angular momenta $1/2$ can be combined to form either a state with generalized angular momentum $0$ or  a state with generalized angular momentum $1$, which is written as 
\begin{equation} 
   1/2 \times 1/2 = 0 + 1 \,,
\end{equation}
reminiscent of two ordinary spin 1/2's forming either a singlet or triplet state.
Similarly, two generalized angular momenta $1$ can be combined into
\begin{equation}
   1 \times 1 = 0 + 1 + 2 \,
   \label{eq:fusion-spin1}
\end{equation}
for deformation parameters $k\ge4$. For lower values of $k$, the rules differ because the number of representations is limited by $k$. In particular, for $k=3$,
one finds $1 \times 1 = 0 + 1$, while for $k=2$, one has $1 \times 1 = 0$. Finally, for 
$k=1$, the general momentum $1$ is not allowed.
For the anyonic theories,  the above equations are often referred to as fusion rules.

The many-body physics of a set of 
interacting
non-Abelian anyons can 
be captured by a Hamiltonian that is formed by pairwise interactions which assign
energies to the different outcomes in the above fusion rules. Such an approach is a straightforward generalization of the conventional Heisenberg model, whose pairwise interaction term $J \vec{S}_i \cdot \vec{S}_j$ is simply a projector onto the singlet state, which is energetically favored for antiferromagnetic couplings ($J<0$) or penalized for ferromagnetic couplings ($J>0$). 

The first step in this direction was taken by some of us for anyonic spin-1/2 chains in Ref.~(\onlinecite{Feiguin_07}) and later generalized to spin-1 chains in Ref.~(\onlinecite{su2k_short}) by the current group of authors. The careful analysis of the ground states of these one-dimensional systems has resulted in a number of insights. First, anyonic spin-1/2 chains typically form gapless ground states which can be described in terms of conformal field theory \cite{Feiguin_07}. These gapless states turn out to be protected by a topological symmetry inherent to the anyon chains
that renders them stable against local perturbations\cite{Feiguin_07,Trebst_08}. Moreover,
 these gapless states can in fact be interpreted as edge states
that reveal the true ground state of a two-dimensional set of anyons -- a novel topological liquid that is separated by the original topological liquids (of which the anyons are excitations) by an edge \cite{su2k_short}. This picture has been verified
by a careful analysis of ladder systems, in which multiple chains are coupled\cite{Liquids}.

Going beyond spin-1/2 chains, we began to study
the physics of anyonic spin-1 chains with first results being reported in a preceding (much more condensed) paper \cite{su2k_short}. In the manuscript at hand, we provide an in-depth discussion of these anyonic spin-1 chains. We find that many of the  distinctive features
of ordinary SU(2) spin-1/2 and spin-1 chains also hold for their anyonic cousins. For instance, anyonic spin-1 chains exhibit a gapped topological phase for antiferromagnetic couplings -- the anyonic generalization of the Haldane phase. Exploring  the phase diagram of chains of pairwise interacting spin-1 anyons,
we find a striking resemblance of the anyonic phase diagram to the one of the ordinary bilinear-biquadratic spin-1 chain.
 In particular, we find multiple  gapless phases (and phase transitions) in addition to the gapped Haldane phase. For the former, 
a similar topological protection mechanism and edge state interpretation holds as for the gapless phases of the anyonic spin-1/2 chains \cite{su2k_short}.

The focus of this manuscript  is to provide an exhaustive description of 
the 
phase diagram(s) of the anyonic spin-1 chains. Our exploration of these systems has led to a 
large amount
of results as the phase diagrams turned out to be much richer than initially anticipated. In particular, we find two families of phase diagrams depending on whether the deformation parameter $k$ of the su(2)$_k$ anyonic theories is odd or even. Moreover, we obtain a distinct phase diagram for $k=4$, a result that can be explained by the special role played by the generalized angular momentum $1$ in the su(2)$_4$ theory.

In order to guide
the reader through these various results we have taken some care to structure the manuscript as follows:
We will start with an introduction to the anyonic su(2)$_k$ theories and a description of the anyonic generalization of the Heisenberg model in Section II. The following sections will then give a detailed  expos\'e of our results, devoting Sec. III to the discussion of anyonic spin-1 chains with odd  deformation parameters $k\ge5$, followed by a discussion of the case of even deformation parameters $k\ge6$ in Sec. IV. In Sec. V we will turn to the case of $k=4$ for which the spin-1 representation plays a special role and a 
rich phase diagram is obtained. We will then turn to anyonic spin-1/2 chains and discuss their physics, in particular their topological stability in Sec. VI. We will end with a broader discussion of our results, in particular in light of other deformations of conventional spin chains such as continuous su(2)$_q$ deformations or so-called (fused) Temperley-Lieb spin chains. The main  part of the  manuscript is followed by an appendix that provides the  technical details of our calculations.

\section{The anyonic quantum spin chain Hamiltonians}

In light of the recent interest in topological phases of matter, it is of great importance
to gain an understanding topological models in their simplest incarnation, and we will
thus study one-dimensional chains of interacting non-Abelian anyons. In this section,
we will briefly explain the models by drawing parallels with ordinary
one-dimensional spin chains. Moreover, we will explain why the `topological' nature of these models goes beyond the fact that they are constructed from `topological' particles, namely non-Abelian anyons.

One of the prototypical one-dimensional spin chain models  is the Heisenberg model, in
which SU(2) spins interact via a `spin-spin' interaction of the type $\vec{S}_i \cdot \vec{S}_j$,
where the labels $i$ and $j$ denote the locations of the interacting spins. Often,
one restricts the interaction to nearest-neighbor, or next-nearest-neighbor
pairs of
spins. For the description of the anyonic quantum spin chains, it will be beneficial to
think of this interaction in terms of the 
total
spin of the two interacting spins.
In this paper, we will only consider nearest neighbor interactions.

As a first example, we look at conventional SU(2) spin-$1/2$, and consider the total
spin
$\vec{S}_T=$ $(\vec{S}_i+\vec{S}_{i+1})$
of two interacting spins $\vec{S}_i$ and $\vec{S}_{i+1}$,
whose magnitude is characterized by the eigenvalue of
$(\vec{S}_T)^2=$
$(\vec{S}_i+\vec{S}_{i+1})^2$. Because the total spin $\vec{S}_T$ can be either $0$ or $1$,
with $\vec{S}_T^2$ eigenvalues $0$ and $2$, we can write
\begin{equation}
(\vec{S}_i+\vec{S}_{i+1})^2 = 0 P^{(0)}_{i} + 2 P^{(1)}_{i} \ ,
\end{equation}
where the projection operator $P^{(s)}_{i}$ projects onto the total spin $s$ channel of the
two spins $\vec{S}_{i}$ and $\vec{S}_{i+1}$.
 Evaluating the left hand side, one obtains 
\begin{equation}
\vec{S}_i \cdot \vec{S}_{i+1} = P^{(1)}_{i} - \frac{3}{4} \idop_{i} =  -P^{(0)}_{i} + \frac{1}{4} \idop_{i} \ ,
\end{equation}
where in the last step we used that we can rewrite the
identity operator as $\idop_{i} = P^{(0)}_{i} + P^{(1)}_{i}$,
which holds in the case of spin-$1/2$.
We conclude that the Heisenberg interaction assigns energy to two
interacting spins, depending on their combined spin, and the Heisenberg
Hamiltonian can be written in terms of projectors as
\begin{equation}
H = J \sum_{i} P^{(0)}_i \ ,
\end{equation}
where $J=1$ corresponds to an antiferromagnetic coupling,
and $J=-1$ to the ferromagnetic version. 

For spin-$1$, one can similarly write the bilinear and bi-quadratic terms
$\vec{S}_i \cdot \vec{S}_{i+1}$ and $(\vec{S}_i \cdot \vec{S}_{i+1})^2$ respectively,
in terms of the projection operators $P^{(1)}_{i}$ and $P^{(2)}_{i}$.
 In particular, the relations
\begin{align}
(\vec{S}_i + \vec{S}_{i+1})^2 &= 2 P_{i}^{(1)} + 6 P_{i}^{(2)} \nonumber \\
(\vec{S}_i + \vec{S}_{i+1})^4 &= 4 P_{i}^{(1)} + 36 P_{i}^{(2)} \ ,
\end{align}
can be rewritten as
\begin{align}
(\vec{S}_{i}\cdot \vec{S}_{i+1}) &= P_{i}^{(1)} + 3 P_{i}^{(2)} -2 \idop_{i} \nonumber \\
(\vec{S}_{i}\cdot \vec{S}_{i+1})^2 &= -3P_{i}^{(1)} - 3P_{i}^{(2)} + 4 \idop_{i} \ . 
\end{align}
 Consequently,
the bilinear-biquadratic spin-$1$ Hamiltonian
\begin{equation}
\label{eq:bilbiq}
H_{\rm{bb}} = \sum_{i} \cos(\theta_{\rm bb}) (\vec{S}_{i}\cdot \vec{S}_{i+1})
+ \sin(\theta_{\rm bb}) (\vec{S}_{i}\cdot \vec{S}_{i+1})^2
\end{equation}
 can be expressed in terms of the projectors $P^{(1)}_{i}$ and $P^{(2)}_{i}$ as follows, 
\begin{eqnarray}
	H_{\rm bb} & = & \sum_{i} J_2 P^{(2)}_{i} + J_1 P^{(1)}_{i} \nonumber \\ 
	                  & = & \sum_{i} \cos\theta_{2,1} P^{(2)}_{i} - \sin\theta_{2,1} P^{(1)}_{i} \ .
\label{eq:spin1-hamiltonian}
\end{eqnarray}
Here, the relation between the two angles $\theta_{2,1}$ and $\theta_{\rm bb}$
is given by
\begin{align}
\tan\theta_{2,1} &= \frac{\tan\theta_{\rm bb}-1/3}{1-\tan\theta_{\rm bb}} &
\tan\theta_{\rm bb} &= \frac{\tan\theta_{2,1}+1/3}{1+\tan\theta_{2,1}} \ .
\end{align}

We will now shift
our attention to anyonic degrees of freedom. Details about anyon models, in particular those of type
su(2)$_k$,
can be found in
appendix~\ref{app:su2k-anyons}.
A general introduction can be found, e.g.,  in references
Ref. (\onlinecite{Kitaev06,ZHWang,bonderson-thesis}).
Here, we will only introduce those concepts that are necessary for defining
the   chain Hamiltonians.
The Hamiltonians for the anyon chains that we will consider in this paper are of the form
of Eq. (\ref{eq:spin1-hamiltonian}).  The projectors  $P^{(j)}_{i}$
in that equation have however a different meaning for anyons (as compared to ordinary spins)
which will be defined in Eq. (\ref{eq:generalprojector}) below.

Anyons are labeled by generalized angular momenta, or - in the language of anyons models - `topological charges'. These generalized angular momenta correspond to quantum numbers, just as in the case of ordinary spin degrees of freedom.
 The notion of combined spin, or
tensor product of spins,
corresponds to the notion of `fusion' in the language of anyons,  
and can in general result in more than one type of
anyon. The possible outcomes are called `fusion channels'. The generalization of the
Heisenberg interaction for spins to the anyonic case 
is to assign an energy to two
interacting anyons  based on
their fusion channel. How 
this  is done  in practice,
will be described in more detail below
and in appendix~\ref{app:hamiltonians}.

The  class of anyons considered in  this paper
is derived from SU(2) where spin-$S$ ranges from 
$S=0,1/2,1,3/2,\ldots$. In contrast, su(2)$_k$ anyons contain only a subset of generalized angular momenta, namely 
\begin{displaymath}
j=0,\frac{1}{2},1,\ldots,\frac{k}{2}.
\end{displaymath}
The truncation, characterized by the `level' $k$, has two important
consequences which we will describe  in the following.

The first consequence concerns the fusion rules of the anyons. 
 The tensor product
of two SU(2)  spins $S_1$ and $S_2$
decomposes as
$$S_1 \otimes S_2 = |S_1-S_2| \oplus \cdots \oplus (S_1+S_2) \ .$$
The process of taking
tensor products is associative, and the same is true for the fusion rules.
Because of  the
truncation in the su(2)$_k$ theory,
 the SU(2) tensor product rule  has to be modified. 
It turns
out that there is only one way of doing this, consistent with the requirement that
the fusion rules are associative. In particular, the fusion rules of su(2)$_k$ anyons
read
\begin{equation}
j_1 \times j_2 = |j_1 - j_2| + (|j_1 - j_2|+1) + \cdots + \min (j_1+j_2,k-j_1-j_2).
\label{eq:su2k-fusion}
\end{equation}

The second important consequence of the truncation follows from the fusion rules.
The dimension of the Hilbert space of a number  $N$ of ordinary SU(2) spin-$1/2$'s 
is equal to $2^N$, and the spins can add up to a maximum spin of $N/2$.
In contrast, the
dimension of the
Hilbert space of a number  $N$ of  $j=1/2$ anyons in the 
su(2)$_k$ theory is smaller than $2^N$.  
In appendix~\ref{app:su2k-anyons}, it is shown that
 the dimension of the
Hilbert space for $N$ $j=1/2$ anyons
grows  as
$d_{1/2}^N$, 
asymptotically for large $N$,
where
$d_{1/2} = 2 \cos\left( \frac{\pi}{k+2}\right)$
is the so-called {\it quantum dimension} of the $j=1/2$ anyon. 
For $1<k<\infty$, this implies that
the effective number of degrees of freedom for each anyon is irrational.
This is less mysterious than it sounds: all this is saying is that one
can not think of the Hilbert space of $N$ anyons as a tensor product of
$N$ one-anyon Hilbert spaces. 

Because the Hilbert space does not have a tensor product structure, an alternative 
description of the state space and the Hamiltonian acting on it is needed.
We will describe here how this can be done,
but leave the details for the appendices
where we also give an
explicit description of the Hamiltonians studied
in this paper.

\begin{figure}[b]
  \begin{center}
\includegraphics[width=.5\linewidth]{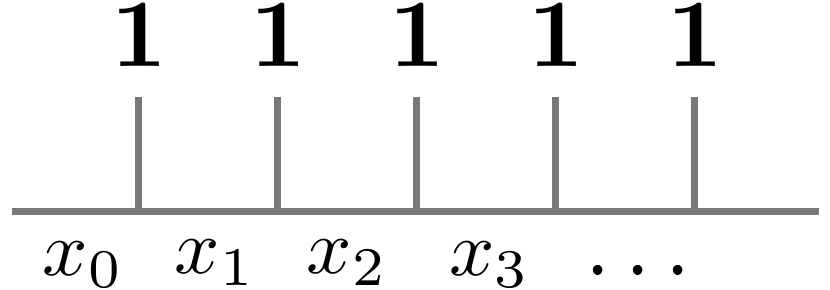}
\caption{The anyonic spin-1 chain.}
  \label{Fig:Spin1-Chain}
\end{center}
\end{figure}

The Hilbert space of a chain of anyons can be described in terms of a so-called
`fusion tree'.
In Figure~\ref{Fig:Spin1-Chain}, the fusion tree for a chain of `spin-1' anyons is
displayed. The lines in the fusion tree carry a label  indicating 
the type of anyon the line corresponds to.
The lines coming from above correspond to the spin-1 anyons which
constitute the chain. The horizontal lines, labeled by $x_i$, are the actual degrees
of freedom. The possible `values' of the $x_i$ are the same as those of the anyons present in the
anyon model, namely $x_i = 0, 1/2, \ldots ,k/2$, in case of su(2)$_k$ anyons. 
The $x_i$ cannot be chosen arbitrarily,
but may only take values such that the fusion rules are obeyed at the
trivalent points. For example,
the anyon type $x_1$ has to appear in the fusion product of $x_0 \times 1$, the anyon type
$x_2$ appears in the fusion product $x_1 \times 1$, and so on. 
Each labeling of the fusion tree that is consistent with the fusion rules 
corresponds to a (orthonormal) state in the Hilbert space, and these states
span this space.   

Typically, we will use periodic boundary conditions $x_{L} = x_{0}$, which implies
that $x_0$ has to appear in the fusion product $x_{L-1} \times 1$, where $L$
denotes the number of sites of the chain. States in the Hilbert space will be written
as $\ket{x_{0},x_{1},\ldots,x_{L-1}}$.

The Hamiltonian assigns an  energy based on 
the fusion channel of two neighboring anyons  in the chain.
However, in the  above discussed representation
of the Hilbert space (see Figure~\ref{Fig:Spin1-Chain}), the fusion channel of two
neighboring anyons is not explicit. To remedy this problem, we 
employ
a local basis transformation  which changes 
the order in which
the anyons  are fused. This is permissible
because of the associativity of the fusion rules.
For ordinary  SU(2) spins, 
 this basis transformation is described in terms
of the Wigner $6j$-symbols. In the case of anyons, this basis transformation
is described by what are known as the $F$-symbols. A detailed discussion of the  
$F$ symbols, as well as explicit representations for 
su(2)$_k$ anyons  can be found in
appendix~\ref{app:f-symbols}.

\begin{figure}[t]
  \begin{center}
\includegraphics[width=\linewidth]{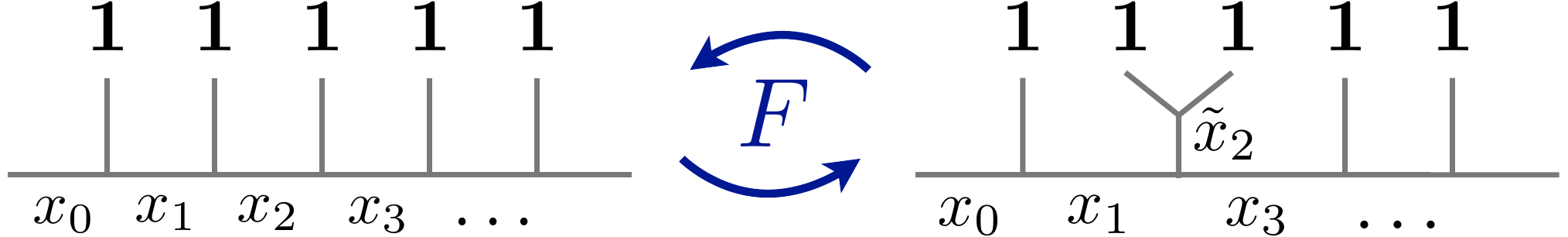}
\caption{The basis transformation for the anyonic spin-1 chain.}
  \label{Fig:Spin1-BasisTransformation}
\end{center}
\end{figure}

The basis transformation is depicted in Figure~\ref{Fig:Spin1-BasisTransformation}.
On the left hand side, $x_1$ is fused with a spin-$1$ anyon, resulting in $x_2$,
which is subsequently fused with the next spin-$1$ anyon, resulting in anyon
type $x_3$. After the basis transformation, one first fuses the two spin-$1$ anyons,
resulting in $\tilde{x}_2$, which is fused with $x_1$, resulting 
in the anyon type
$x_3$.  
Both bases are
equivalent; however,
in the second
basis, the fusion channel of the two spin-$1$ anyons is 
explicit, namely $\tilde{x}_2$. Thus, after performing this basis transformation, one can
assign the appropriate energy based on the value of $\tilde{x}_2$. Subsequently, one
transforms back to the original basis. 
The operator projecting onto the anyon$-j$ channel of two neighboring anyons $i$ and $i+1$
is thus given by
\begin{equation}
P^{(j)}_{i}   = F_{i}^{-1} \Pi_i^{(j)} F_{i} \ ,
\label{eq:generalprojector}
\end{equation}
where $F_{i}$ is shorthand for the local basis transformation
depicted in Figure~\ref{Fig:Spin1-BasisTransformation}. The operator
$\Pi_{i}^{(j)}$ projects onto the fusion channel $\tilde{x}_i = j$, i.e., 
 the fusion of two anyons into an anyon of type $j$ is penalized with energy $E=1$,
while the other possible fusion channels are assigned $E=0$. For explicit matrix representations of $P_i^{(j)}$ we refer to
appendix~\ref{app:hamiltonians}.

It is important to realize that the form of the projector \eqref{eq:generalprojector}
is universal and applicable to anyonic chains composed of arbitrary types
of anyons. Changing to a different anyon model will merely result in a
different structure of the Hilbert space and different $F$-symbols.

\subsection{Topological symmetry}
\label{sec:top-sym}

In this section, we present a detailed discussion of  the `topological symmetry operator'.
The Hamiltonians considered in this paper commute with the topological symmetry
operator, and the associated symmetry plays a crucial role in the analysis
of the anyonic chain models. 
In  `equation'
\eqref{topsym1}, a chain of type-$j$ anyons with periodic boundary conditions
 is displayed (in this particular case, $L=3$).
\begin{equation}
\label{topsym1}
\includegraphics[width=0.5\columnwidth]{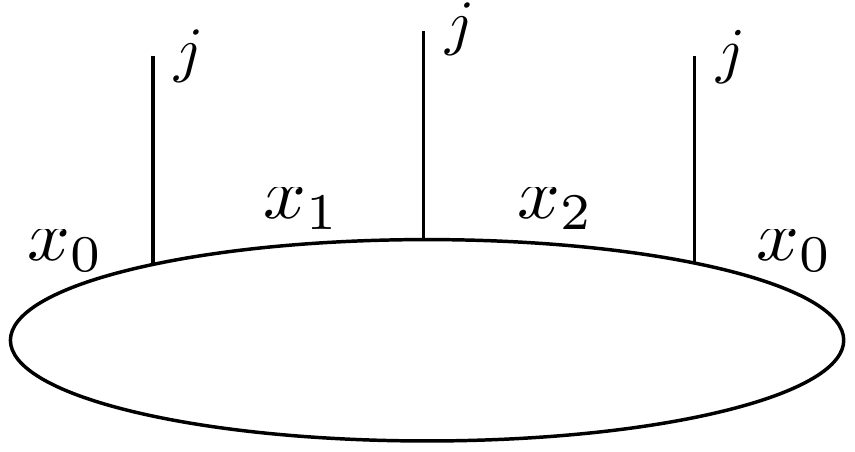}
\end{equation}
For each type of anyon $l$, there exists a topological operator
$Y_l$.
The action of this operator $Y_l$ on the state
$\ket{x_0,x_1,\ldots,x_{L-1}}$, displayed in \eqref{topsym1} for $L=3$,
can be described as follows. First, an additional anyon of type $l$
is created inside the spine of the fusion tree, as displayed in \eqref{topsym2}.
\begin{equation}
\label{topsym2}
\includegraphics[width=0.5\columnwidth]{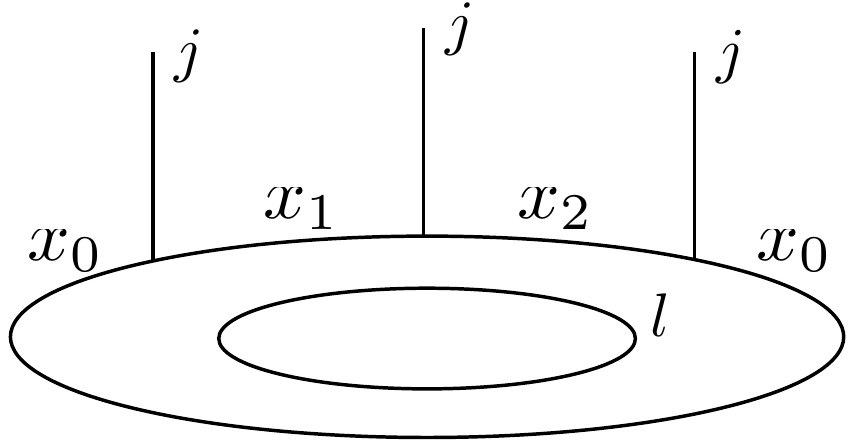}
\end{equation}
This additional spin-$l$ anyon is `merged' with the fusion diagram
by applying an $F$-matrix, namely,
$\left(F^{x_0,x_0,l}_{l}\right)_{0}^{x'_0}$, resulting in the state
$$
\sum_{x'_0} \left(F^{x_0,x_0,l}_{l}\right)_{0}^{x'_0} \ket{x_0,x_1,\ldots,x_{L-1}}
$$
as depicted in \eqref{topsym3}. 
\begin{equation}
\label{topsym3}
\includegraphics[width=0.5\columnwidth]{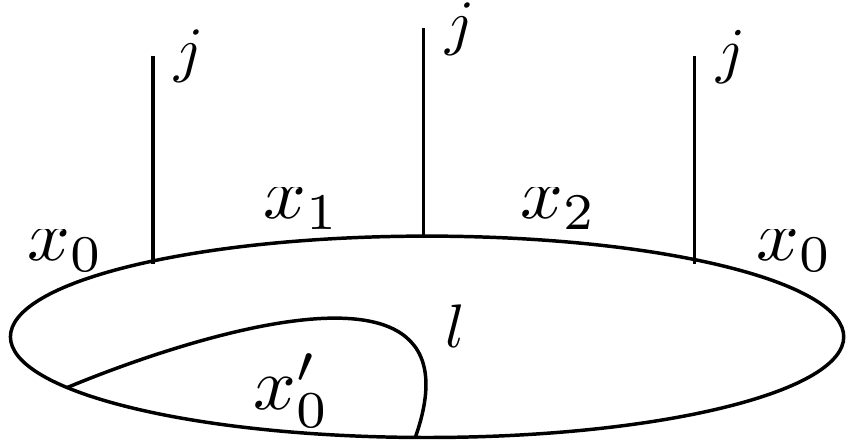}
\end{equation}
Next, one `moves' the additional spin-$l$ anyon around the ring,
by applying additional $F$-matrices. After the first step, one obtains the state
$$
\sum_{x'_0,x'_1}
\left(F^{x_0,x_0,l}_{l}\right)_{0}^{x'_0}
\left(F^{j,x_1,l}_{x'_0}\right)_{x_0}^{x'_1}
\ket{x_0,x_1,\ldots,x_{L-1}}
$$
as illustrated in \eqref{topsym4}.
\begin{equation}
\label{topsym4}
\includegraphics[width=0.5\columnwidth]{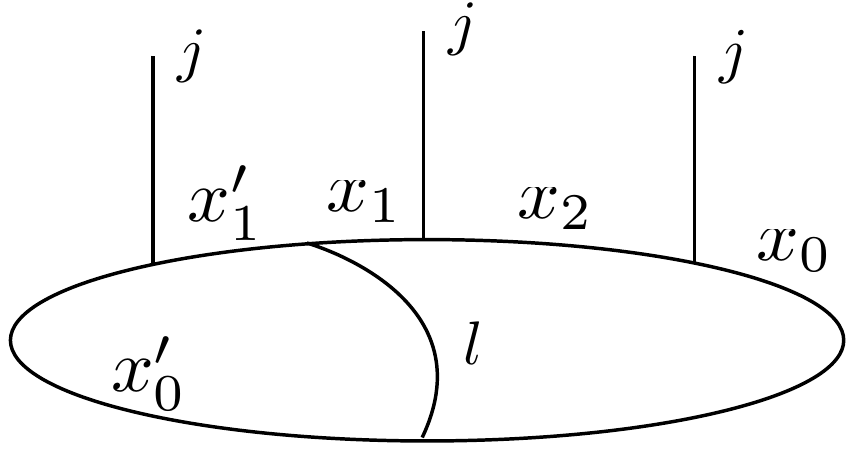}
\end{equation}
Another move of this sort gives
$$
\sum_{x'_0,x'_1,x'_2}\hspace{-2mm}
\left(F^{x_0,x_0,l}_{l}\right)_{0}^{x'_0}
\left(F^{j,x_1,l}_{x'_0}\right)_{x_0}^{x'_1}
\left(F^{j,x_2,l}_{x'_1}\right)_{x_1}^{x'_2}
\ket{x_0,x_1,\ldots,x_{L-1}}
$$
as shown in \eqref{topsym5}.
\begin{equation}
\label{topsym5}
\includegraphics[width=0.5\columnwidth]{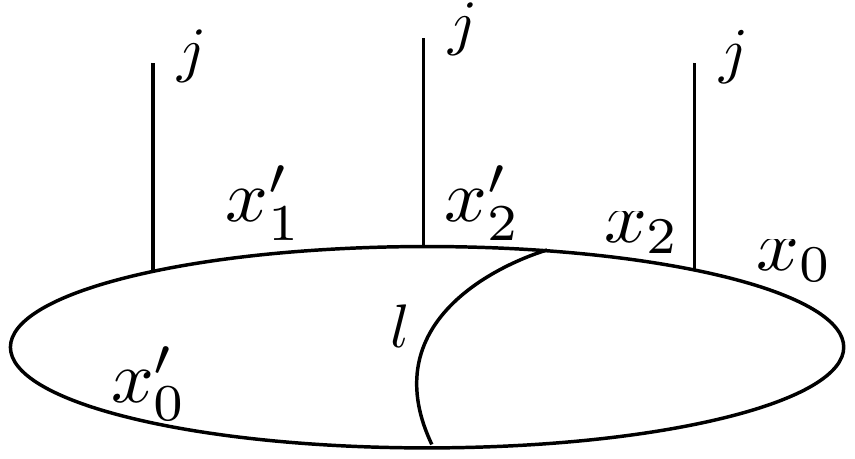} 
\end{equation}
Finally, after $L$ steps, one has come full circle, giving rise to the states
$$
\sum_{x'_0,x'_1,\ldots,x'_L}\hspace{-4mm}
\left(F^{x_0,x_0,l}_{l}\right)_{0}^{x'_0}
\prod_{i=0}^{L-1}
\left(F^{j,x_{i+1},l}_{x'_i}\right)_{x_i}^{x'_{i+1}}
\ket{x_0,x_1,\ldots,x_{L-1}}
$$
as depicted in \eqref{topsym6}, for $L=3$.
\begin{equation}
\label{topsym6}
\includegraphics[width=0.5\columnwidth]{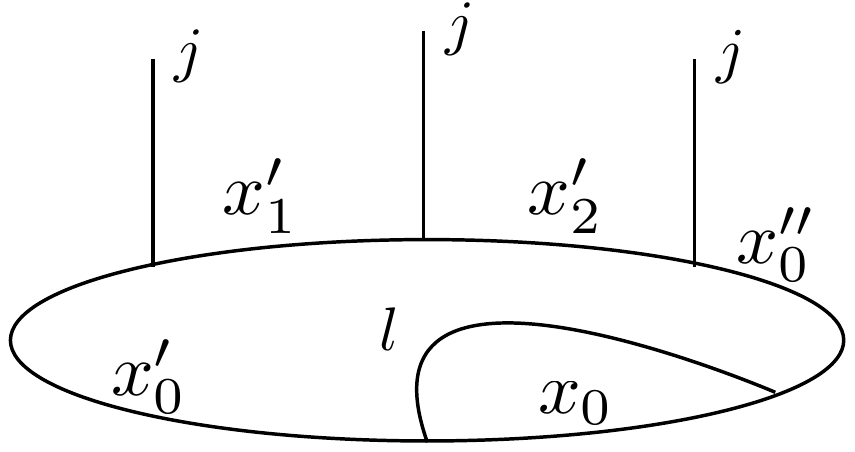}
\end{equation}
From the general properties of anyon models (see e.g. Ref. ( \onlinecite{ZHWang,Kitaev06})), we find that
$x''_0 = x'_0$ (the overall topological quantum number of an isolated
set of anyons can not change).
We can now remove the additional spin-$l$ anyon in the same way as we added it,
thereby finishing the
operation of acting with $Y_l$ on the state $\ket{x_0,x_1,\ldots,x_{L-1}}$.
Thus, we obtain the expression
\begin{align}
&Y_l \ket{x_0,x_1,\ldots,x_{L-1}} = \nonumber \\
& \sum_{x'_0,x'_1,\ldots,x'_{L-1}}\prod_{i=0}^{L-1}
\left(F^{j,x_{i+1},l}_{x'_i}\right)_{x_i}^{x'_{i+1}}
\ket{x_0,x_1,\ldots,x_{L-1}} \ .
\end{align}
We can now state the matrix elements of the topological operator $Y_l$
in the fusion tree basis
\begin{equation}
\bra{x'_0,x'_1,\ldots,x'_{L-1}} Y_l \ket{x_0,x_1,\ldots,x_{L-1}} =
\prod_{i=0}^{L-1} \left( F^{j, x_i,l}_{x'_{i+1}}\right)^{x'_i}_{x_{i+1}} \ .
\end{equation}

The above definition of the topological operator does not depend on whether the additional spin-$l$ anyon is encircled by the anyon chain (as in Figures \eqref{topsym2}-\eqref{topsym6}) or whether the additional spin-$l$ anyon encircles the entire anyon chain.
 When using the latter description of the topological operator, 
one can think of the additional
spin-$l$ anyon as going around the `fusion product' of all the spin-$j$ anyons constituting the
anyonic chain, or better, encircling the flux through the chain.
 This flux through the chain is related to the additional spin-$l$ anyon as follows,
\begin{equation}
\includegraphics[height=1.5cm]{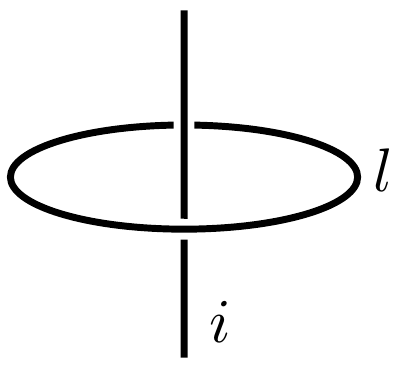}
\hspace{.25cm}
\raisebox{.65cm}{{\Large $= \frac{S_{l,i}}{S_{0,i}}$}}
\hspace{.25cm}
\includegraphics[height=1.5cm]{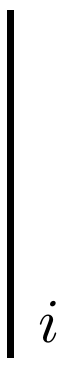} \ ,
\label{eq:topsymev}
\end{equation}
where $i$ denotes the flux going through the chain, and the matrix $S$ is the
modular $S$-matrix of the anyon model. For a derivation of  Eq.~\eqref{eq:topsymev},
see e.g. Ref. (\onlinecite{Kitaev06,ZHWang}), and the explicit form of $S$ in the case of su(2)$_k$ anyons
is given in appendix~\ref{app:su2k-anyons}.

The definition of the topological operator contains    
elements of the $F$ matrices only. This is also true for the anyonic spin Hamiltonians
we consider in this paper. It follows that the operators $Y_l$ commute with the
Hamiltonian and that  a topological quantum number can be assigned to all the
eigenstates. This has far reaching consequences for the stability of the
critical phases. Excited states which are relevant in the renormalization group
sense (i.e., have energy smaller than 2) may  lie in a different topological sector
than the ground state and thus do not drive the system into a different phase.
In addition, we will see that the operators $Y_l$ play an important role in
the zero-energy ground states at the AKLT point in the Haldane-gapped phase of the
spin-$1$ models.

\section{Anyonic su(2)$_k$  spin-$\bf1$ chains: odd $k\ge 5$}
\label{bil_biq_chain}

\subsection{Introduction}

We will start our discussion of anyonic quantum spin chains with the anyonic version 
 of the ordinary SU(2) spin-1 Heisenberg chain, which
has long been appreciated as one of the paradigmatic spin chain models. 
For antiferromagnetic couplings the spin-1 chain is well known  to form
a gapped phase, in distinction from the gapless spin-1/2 Heisenberg chain\cite{Haldane}. 

In the following sections, we discuss in detail the anyonic su(2)$_k$ 
deformations of the ordinary SU(2) spin-1 chain. We will see that much of the seminal 
features of the SU(2) spin-1 chain carry over to these anyonic deformations with
a number of new subtleties arising. One is a dependence of the observed phases
and phase diagrams on the deformation parameter $k$. In particular, we find an 
even/odd effect in $k$ for $k\ge5$ and a distinctive  behavior for $k=4$. 
We have therefore split our discussion of the anyonic spin-1 chains into three
different sections. We will  address anyonic spin-1 chains with {\em odd} 
$k\ge5$ in the remainder of this section, in which we will also give a brief recount
of the phase diagram of the ordinary SU(2) spin-1 chain. The subsequent section will
 be devoted to the case of $k\ge6$ with $k$ being {\em even}. Finally,  an entire section is devoted to a detailed account of the physics for the special
case of $k=4$.

\subsection{The ordinary SU(2) Heisenberg spin-$\bf 1$ chain}

\begin{figure}[t]
\begin{center}
\includegraphics[width=.8\columnwidth]{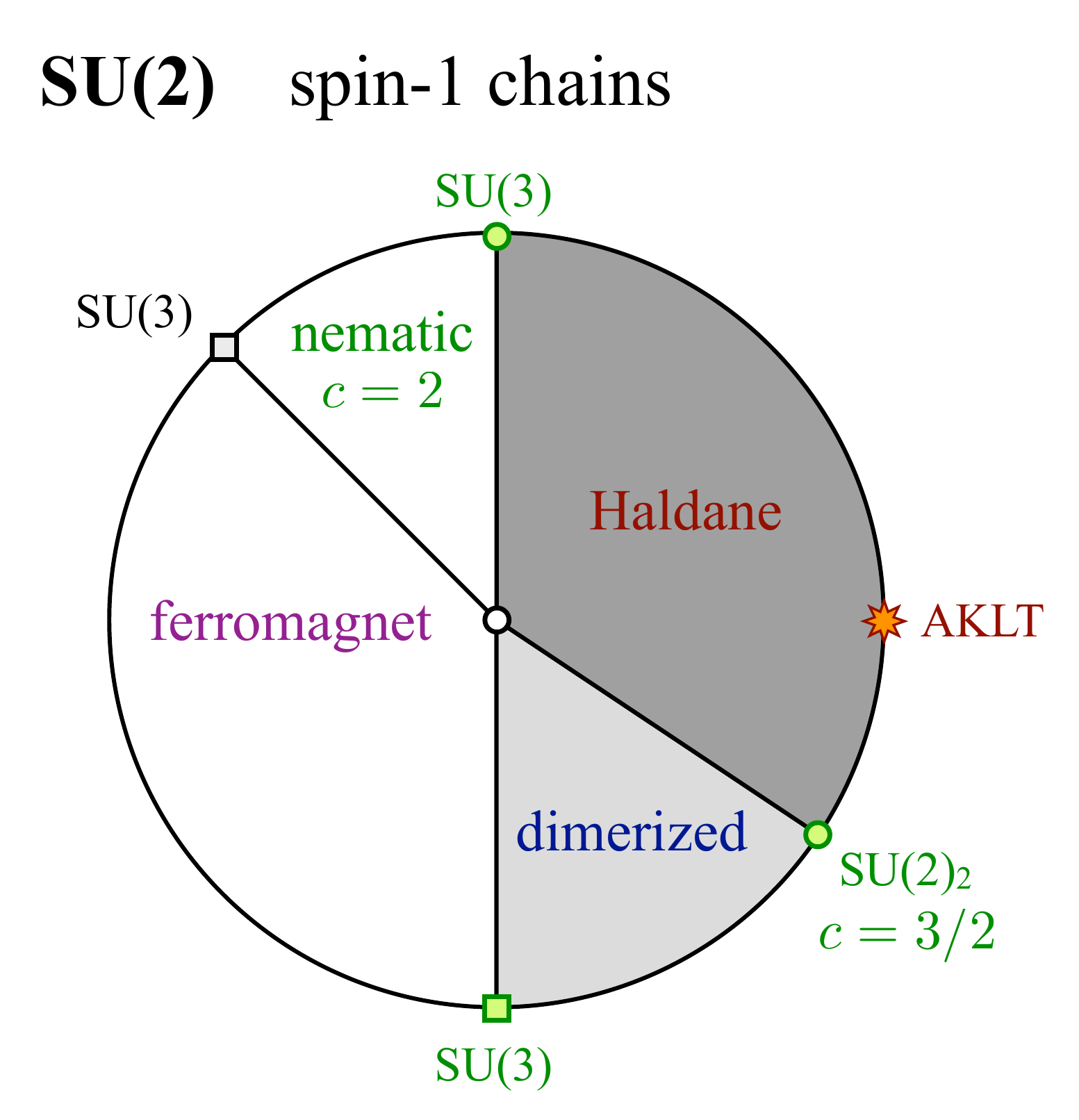}
\caption{(color online)
	      Phase diagrams of the ordinary SU(2) spin-1 chain in a projector representation \eqref{eq:spin1-hamiltonian} 
	      with $J_1=-\sin(\theta_{2,1})$ and $J_2=\cos(\theta_{2,1})$. }
  \label{fig:su2_phase_diagram}
\end{center}
\end{figure}

Before addressing the physics of the anyonic spin chains we briefly recapitulate 
the phase diagram of the ordinary SU(2) spin-1 Heisenberg chain. While the 
latter is typically discussed as a circle phase diagram in terms of bilinear and 
biquadratic spin exchange, we will recast the phase diagram in terms of the
projector representation in Eq.~\eqref{eq:spin1-hamiltonian} - the  generic representation of 
anyonic spin chains. 
Fig.~\ref{fig:su2_phase_diagram} shows the phase diagram in the 
projector representation of Eq.~\eqref{eq:spin1-hamiltonian}. It contains four
different phases, of which two are gapped phases and two are gapless phases.
The well known Haldane phase \cite{Haldane} extends in the parameter regime 
		$-\arctan (2/3) < \theta_{2,1} < \pi/2$
and includes the so-called Affleck-Kennedy-Lieb-Tasaki (AKLT) point
\cite{AKLT} at $\theta_{2,1} = 0$ 
(in which only the projector $P^{(2)}$ is present in the Hamiltonian),
at which the exact
form of the ground-state wave function in terms of a valence bond solid state
can be obtained. 
The conventional  (gapped)  Heisenberg chain (bilinear in spin-1 operators) with antiferromagnetic coupling 
corresponds to $\theta_{2,1} = -\arctan(1/3)$.
The second gapped phase is a (spontaneously)  dimerized phase \cite{DimerizedPhase}
that occurs in the parameter regime 
		$-\pi/2 < \theta_{2,1} < -\arctan(2/3)$.
The phase transition at $\theta_{2,1} = -\arctan(2/3)$ between the two gapped  phases 
is described by the  su(2)$_2$ conformal field theory with central charge $c=3/2$,
which happens to possess $N=1$ supersymmetry 
-- a result that can be obtained 
by means of a (nested) Bethe Ansatz \cite{Date}. 

At the other end of the Haldane gapped phase, $\theta_{2,1} = \pi/2$, there is a phase
transition to  gapless  phase that extends over the range
		$\pi/2 < \theta_{2,1} < 3\pi/4$.
This critical phase can be described by a conformal field theory with central charge $c=2$. 
There are characteristic quadrupolar (nematic) spin correlations \cite{Laeuchli}
in this phase, as well as  a three sublattice structure \cite{FathTriple} resulting 
in soft modes at momenta $K= 0, 2\pi/3, 4\pi/3$.
At the transition from the gapped Haldane phase to this critical nematic phase at
$\theta_{2,1} = \pi/2$, 
the system has enhanced SU(3) symmetry. This point in the phase diagram of the
spin-1 SU(2) chain represents actually  the SU(3) chain with
a fundamental representation at each site, which is known to be 
described by the $SU(3)_1$ conformal field theory.
(This chain is again exactly
solvable by a Bethe Ansatz \cite{LaiSutherland,Uimin}.)

Finally, there is a gapless ferromagnetic phase, extending over the parameter range $3\pi/4<\theta_{2,1}< 3\pi/2$.
The phase transitions from this phase to both the adjacent dimerized phase as well as the
nematic phase are first order. In the vicinity of the transition between the dimerized 
and ferromagnetic phase, early analytical work \cite{ChubukovPRB} suggested the possibility
of an intermediate nematic phase, which, however has later been found to not materialize 
\cite{FathNematic,Laeuchli,Grover}.

\subsection{Phase diagram of the anyonic spin-1 chains -- overview}

\begin{figure}[t]
\begin{center}
\includegraphics[width=.8\columnwidth]{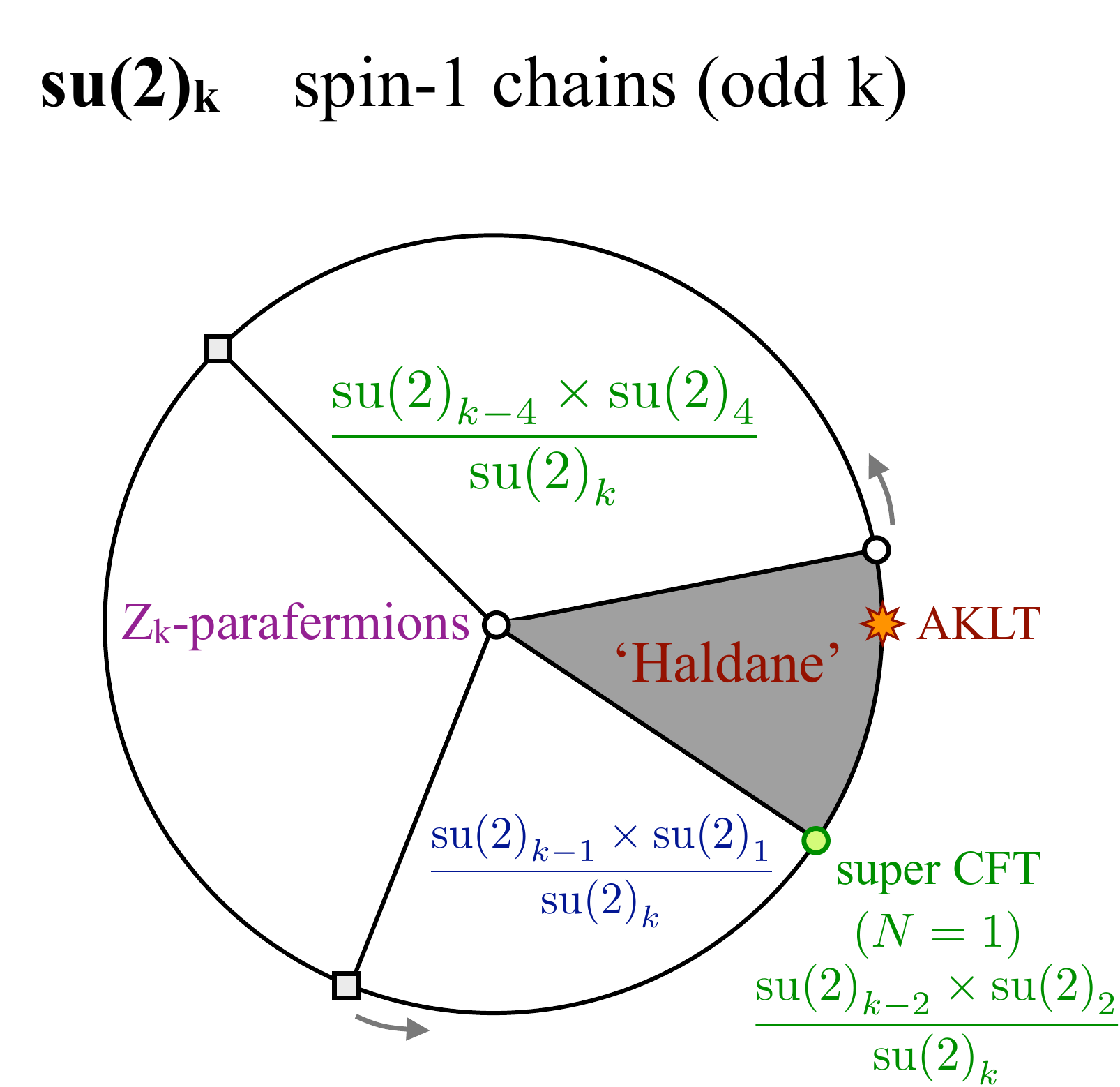}
\caption{(color online)
	      Phase diagrams of the anyonic su(2)$_k$ spin-$1$ chain with odd $k$ 
	      in a projector representation \eqref{eq:spin1-hamiltonian} where  $J_1=-\sin(\theta_{2,1})$ 
	      and $J_2=\cos(\theta_{2,1})$. 
	      With increasing (odd) index $k \geq 5$ the phase boundaries move as indicated
	      by the arrows.}
  \label{fig:oddk_phase_diagram}
\end{center}
\end{figure}

In this section, we provide an overview of the phase diagram of the
anyonic spin-1 chains for odd $k\geq 5$. This phase diagram bears great
resemblance to the corresponding phase diagram of the SU(2)
spin-1 Heisenberg chain
(Fig.~\ref{fig:su2_phase_diagram}).
The generic phase diagram for the su(2)$_k$
spin-1 chain is given in Figure~\ref{fig:oddk_phase_diagram}.
In Figures~\ref{Fig:su(2)_5} and \ref{Fig:su(2)_7}, we display the phase diagrams for
$k=5$ and $k=7$, as well as the characteristic spectra of the four
different phases and the ($N=1$) 
super-symmetric critical point which separates
the Haldane gapped phase and the phase which will be  called
``$Z_2$ sublattice phase'' (this is the phase intervening between the Haldane phase
and the $Z_k$-parafermion phase, and it encompasses the angles
 $\theta_{2,1} \lesssim -0.19\pi \approx -\arctan(2/3)$).

The spin-1 anyonic spin chain is gapped in a finite region around $\theta_{2,1}=0$.
This gapped phase is the anyonic analogue of the Haldane gapped phase,
and the point $\theta_{2,1}=0$ is equivalent to the AKLT point. At this point, the
Hamiltonian penalizes the fusion of two neighboring anyons in the
spin-2 channel. The ground states with periodic boundary conditions
can be found exactly at this point, for all $k$, and the ground state degeneracy is $(k+1)/2$.

For $\theta_{2,1} < 0$, there is a phase transition at $\theta_{2,1} \approx -0.19 \pi$
into an extended critical region. The position of this phase transition did not show
any appreciable dependence on the value of $k$
(remember that $k\geq 5$ throughout this section).
This gapless
region occurs where the ordinary SU(2) spin-1 chain is in
the gapped dimerized phase. This difference in behavior is the most remarkable
distinction between the ordinary SU(2)  spin-1 chain, and the anyonic spin-1 chains. 

The critical point at $\theta_{2,1} \approx -0.19 \pi \approx -\arctan(2/3)$,
separating the Haldane phase and the extended critical region, is described
in terms of 
an $N=1$ 
super-symmetric minimal conformal model.

For angles $\theta_{2,1} > 0$, there is a phase transition from the Haldane phase into another
extended critical region which bears some resemblance to the extended nematic
region in case of the ordinary spin-1 chain. In particular, this phase has a $Z_3$
sublattice structure. The location of the phase transition
does depend on $k$, and moves towards $\theta_{2,1} = \pi/2$ with increasing $k$.

Finally, there is an extended critical region in the vicinitiy of $\theta_{2,1} = \pi$, the
point where the fusion of two neighboring anyons into the spin-2 channel is favored.
This critical phase is the anyonic analogue of the ferromagnetic phase of the ordinary spin-1 chain,
and the critical behavior is described by the $Z_k$ parafermion conformal field theory. 

The phase transitions from the ferromagnetic phase to the neighboring extended
critical regions are first order. The phase transition into the
anyonic version of  the nematic phase
occurs at $\theta_{2,1} = 3\pi/4$, independent of the value of $k$. The location of the other phase transition
depends on $k$, and moves towards $\theta_{2,1} = 3\pi/2$ for increasing $k$.

Below, we will discuss in detail each of the phases mentioned above. We will focus
on the topological properties and the similarities to the ordinary SU(2) spin-1 chain.

\subsection{Critical phases}

We investigate the phase diagram of our model numerically using exact diagonalization.
In our
analysis, we follow a standard procedure to determine the conformal
field theory describing the behavior of the extended critical regions and the
critical points: the numerically obtained spectrum is first shifted (by some constant offset) such that the ground
state has zero energy zero. The spectrum is then rescaled such that the energy of the lowest lying excitation
matches the  energy of the lowest lying excitation of the conformal field theory describing
the phase. The so obtained energy spectrum is finally compared to the energy spectra of  candidate CFTs.
The CFT (if any)  which matches the numerically obtained energy levels is the one 
describing the system at the angle $\theta$.
We note that the list of candidate CFTs is limited:
If the chain is critical, each energy level in the spectrum corresponds to a
field in the applicable conformal field theory. These fields satisfy fusion rules
which have to be compatible with the fusion rules of the underlying su(2)$_k$
theory. This constraint restricts the candidate conformal field theories that
could describe the 
criticality of anyonic quantum chains.

The eigenenergies in a system of finite size described by a conformal
field theory take the form
\cite{Cardy}
\begin{equation}
E = E_1L +\frac{2\pi v}{L} \left (-\frac{c}{12}+ h+\bar{h} \right ),
\label{CFT_energy_levels}
\end{equation}
where the velocity $v$ is an overall scale factor, and $c$ is the central 
charge of the CFT.
The scaling dimensions $h+\bar{h}$ take the form $h=h^0+n$, $\bar{h}=\bar{h}^0+\bar{n}$,
with $n$ and $\bar{n}$ non-negative integers, and $h^0$ and $\bar{h}^0$ are the
holomorphic and antiholomorphic conformal weights
of the primary fields in the given CFT. 
The momenta $K$ (in units $2\pi/L$) are such that
$K=h-\bar{h}+ K_0$ or $K=h-\bar{h}+ K_0+L/2$,
where $K_0$ is a constant shift of the momentum that
 determines at which momentum the primary field occurs.
This shift can be determined from the numerics, and is not fixed by
conformal symmetry. Thus, different microscopic
realizations of the same conformal field theories can give rise to different
values for $K_0$.

As  explained in section~\ref{sec:top-sym}, the anyonic spin chains have
a topological symmetry; all the states in the spectrum can therefore be
assigned a topological quantum number. 
The possible eigenvalues of the topological symmetry operator, 
also denoted as topological quantum numbers,
are in one-to-one correspondence with
the types of anyons which appear in the particular anyon theory considered.

\subsubsection{$Z_k$-parafermion phase}

We begin the discussion of the phase diagram given in Figure~\ref{fig:oddk_phase_diagram}
with the $Z_k$-parafermion phase which corresponds to the gapless ferromagnetic phase
in the  SU(2) spin-$1$ chain. In the anyonic spin-$1$ chains, this phase
contains the point $\theta_{2,1} = \pi$ where  it is favorable for two
neighboring anyons to be in the spin-$2$ channel. One of the phase boundaries of this phase
is located at $\theta_{2,1} = 3\pi/4$. The location of the other phase boundary depends on $k$:
with increasing $k$, it moves towards the 
location of the phase boundary in the SU(2) spin-$1$ chain (at angle
 $\theta_{2,1} = 3\pi/2$).

\begin{figure*}[ht]
  \begin{center}
  \includegraphics[width=.45\linewidth]{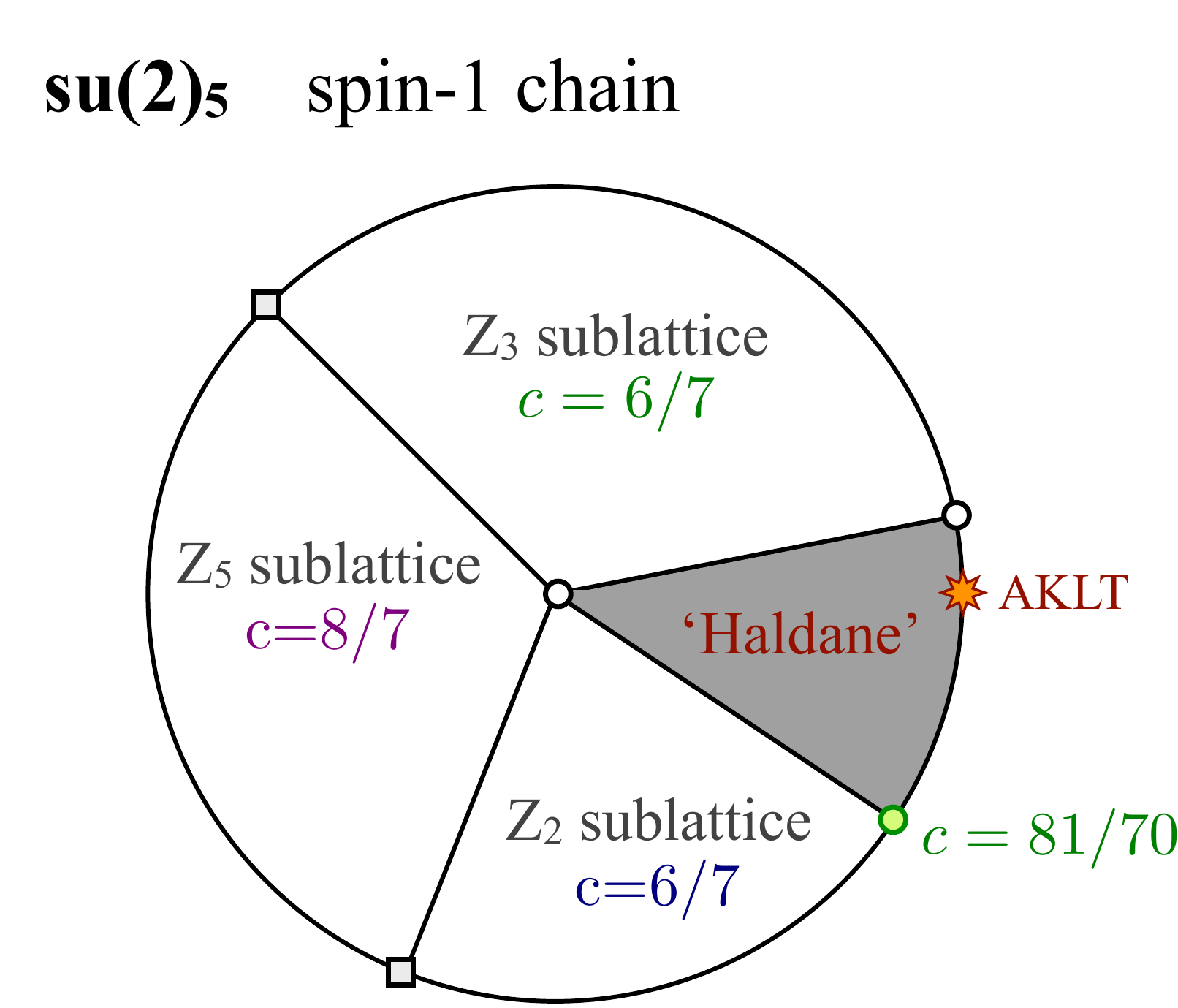}
  \hskip 0.06 \linewidth
  \includegraphics[width=.48 \linewidth]{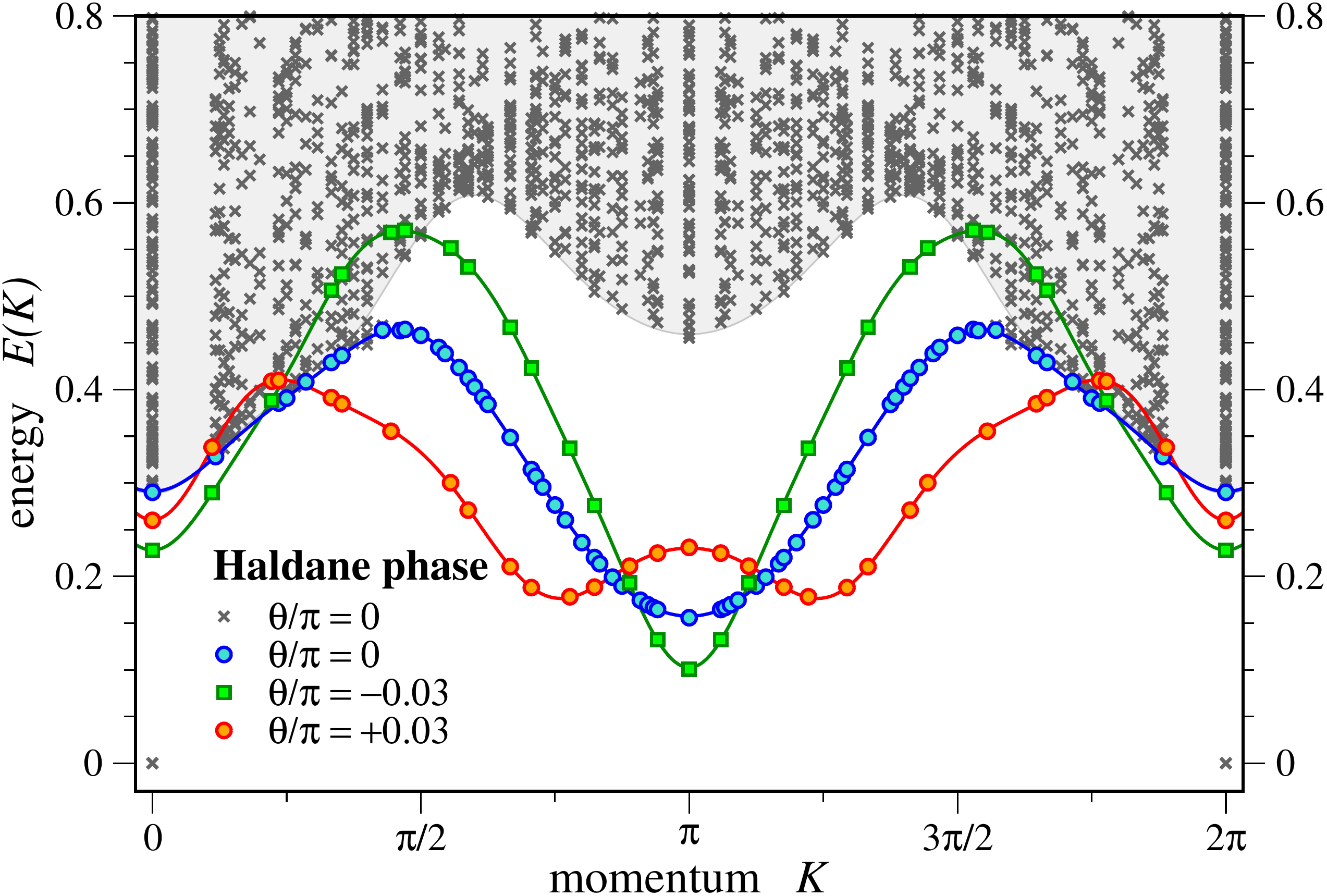} 
  \vskip 3mm
  \includegraphics[width=.48\linewidth]{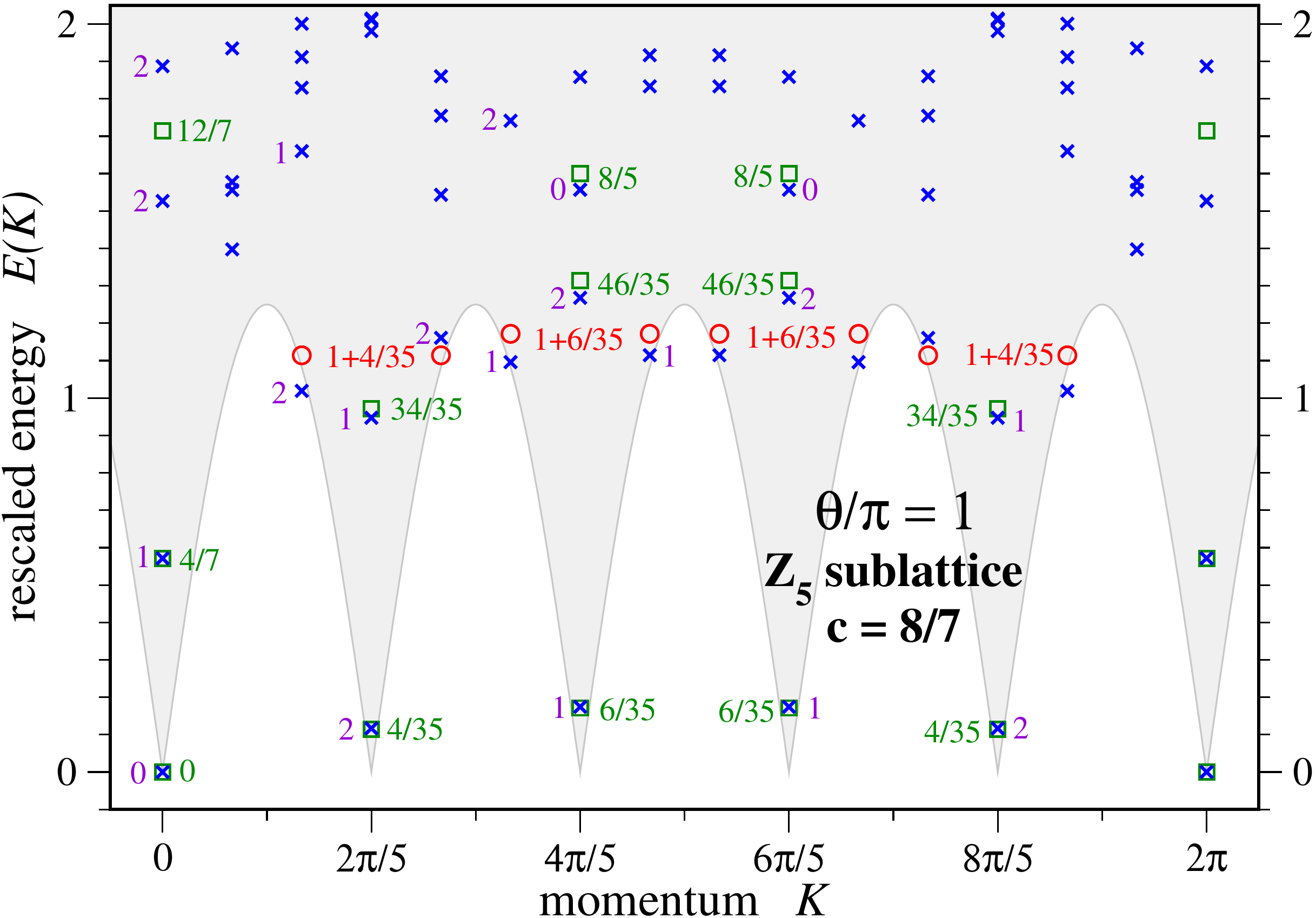} 
  \hskip 0.03 \linewidth
  \includegraphics[width=.48 \linewidth]{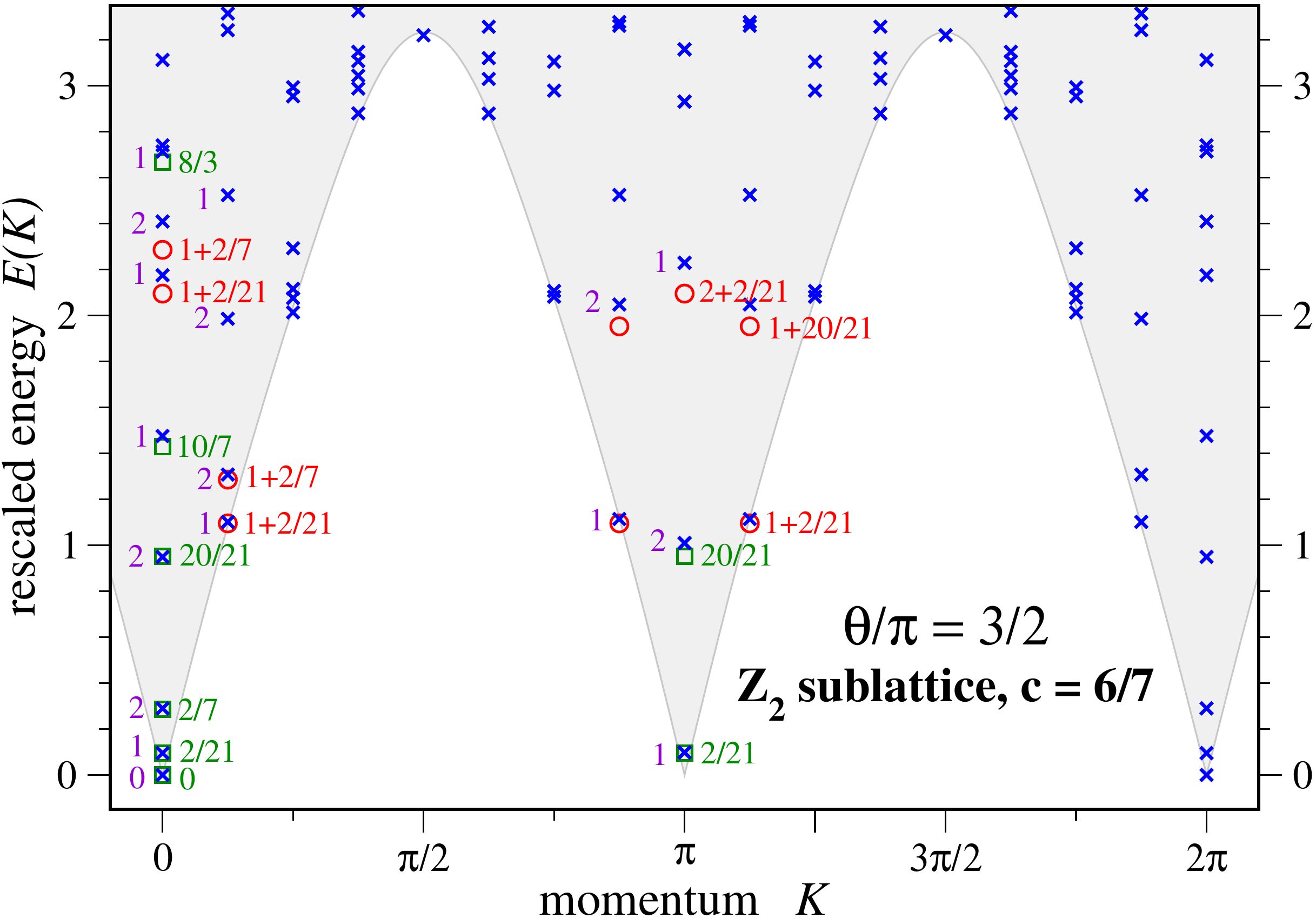} 
  \vskip 1mm
  \includegraphics[width=.48 \linewidth]{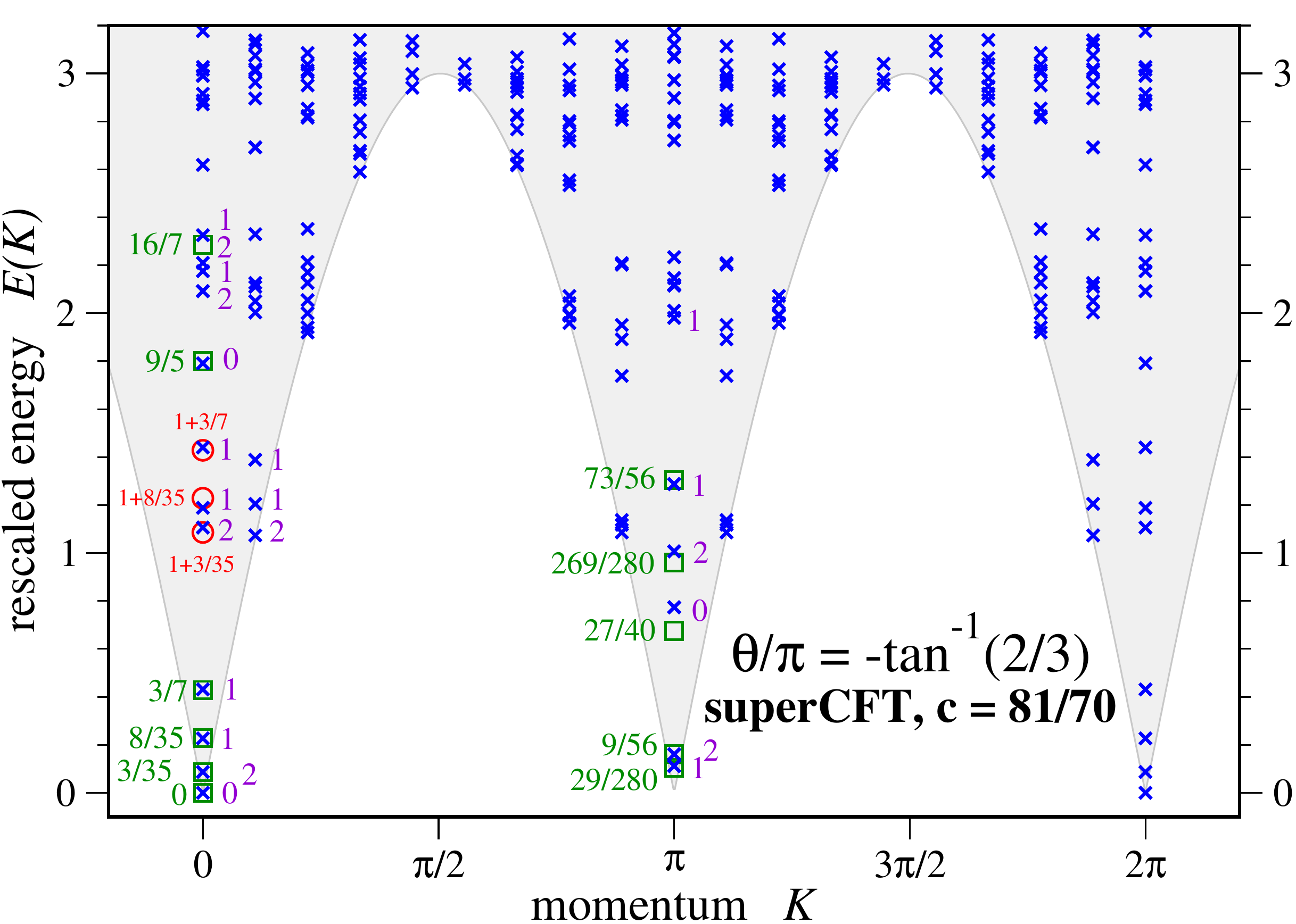} 
  \hskip 0.03 \linewidth
  \includegraphics[width=.48 \linewidth]{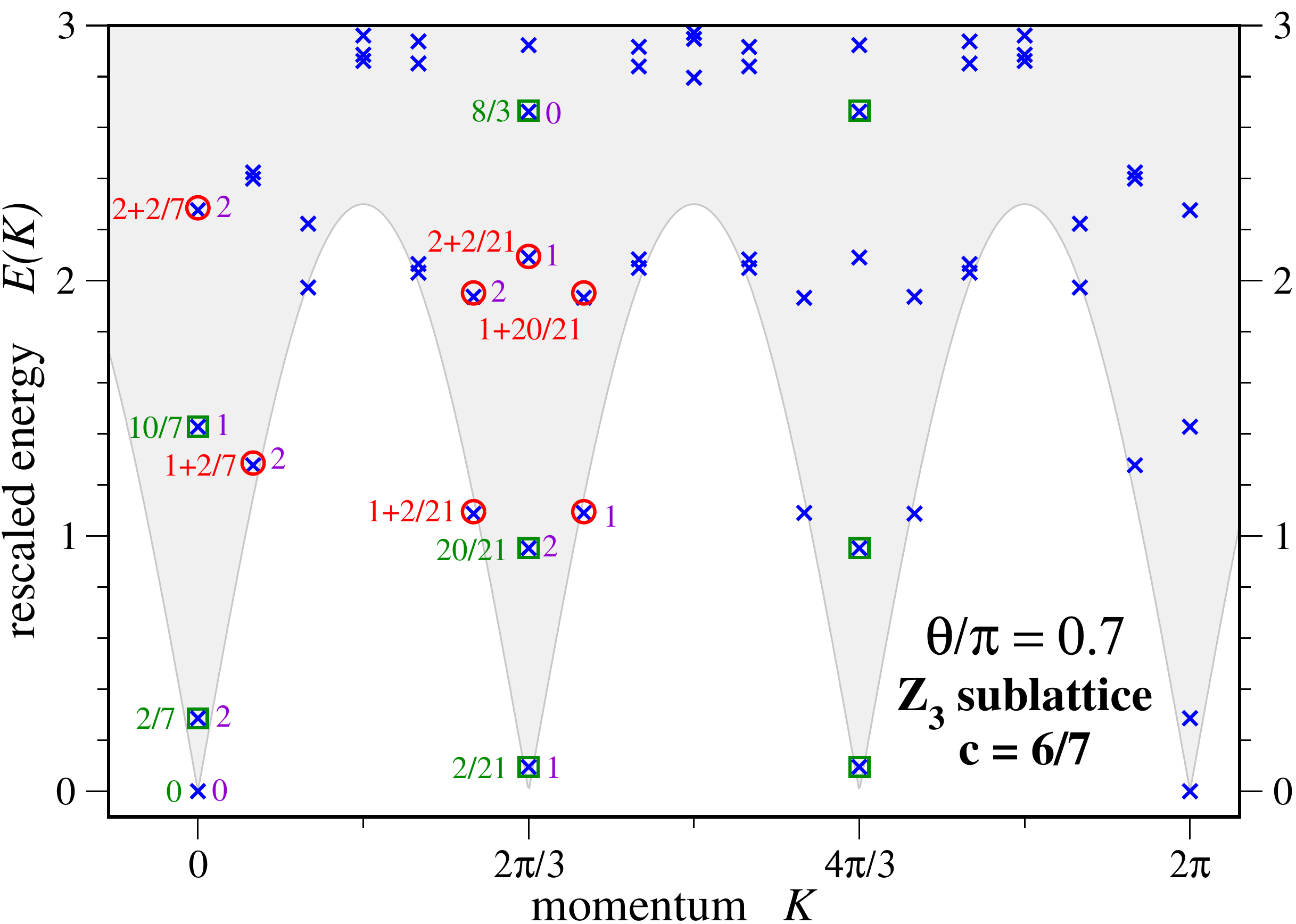} 
     \caption{(color online) 
                    {\bf The su(2)$_5$ spin-1 chain:} The energy spectra for the various phases of the phase diagram
                  are  shown in the upper left panel. For the critical phases/point the energy spectra have been rescaled
                    to match the conformal field theory prediction given in Eq.~\eqref{CFT_energy_levels}. 
                    Green squares indicate the location of the primary fields, red circles the descendant fields.
                    The energies predicted by conformal field theory are given in green (red) for primary (descendant) fields.
                    The topological symmetry sector is indicated by the violet index. 
                    Data shown are for system sizes $L=18$ and $L=15$, respectively.}
     \label{Fig:su(2)_5}
  \end{center}
\end{figure*}

\begin{figure*}[ht]
  \begin{center}
  \includegraphics[width=.45\linewidth]{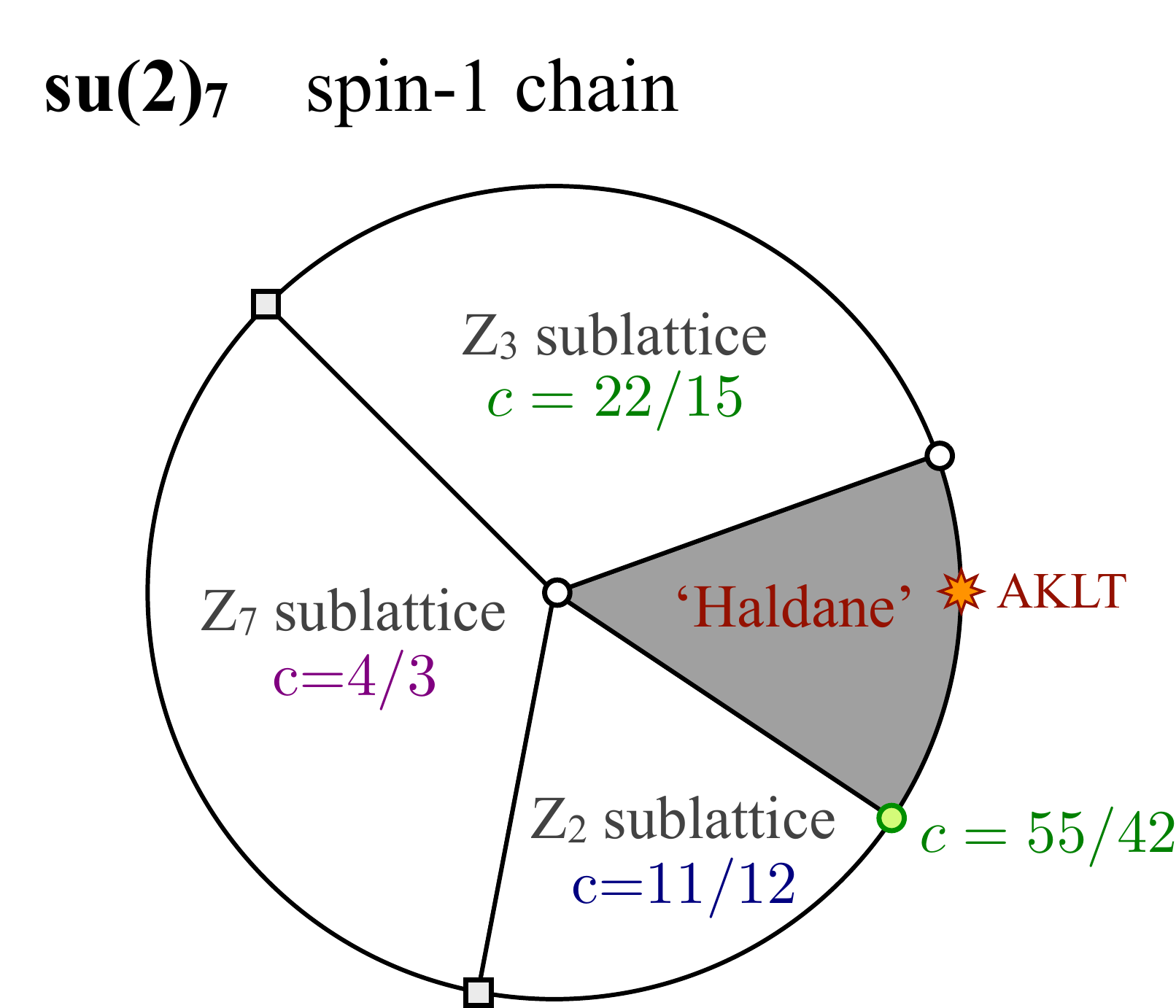}
  \hskip 0.06 \linewidth
  \includegraphics[width=.48 \linewidth]{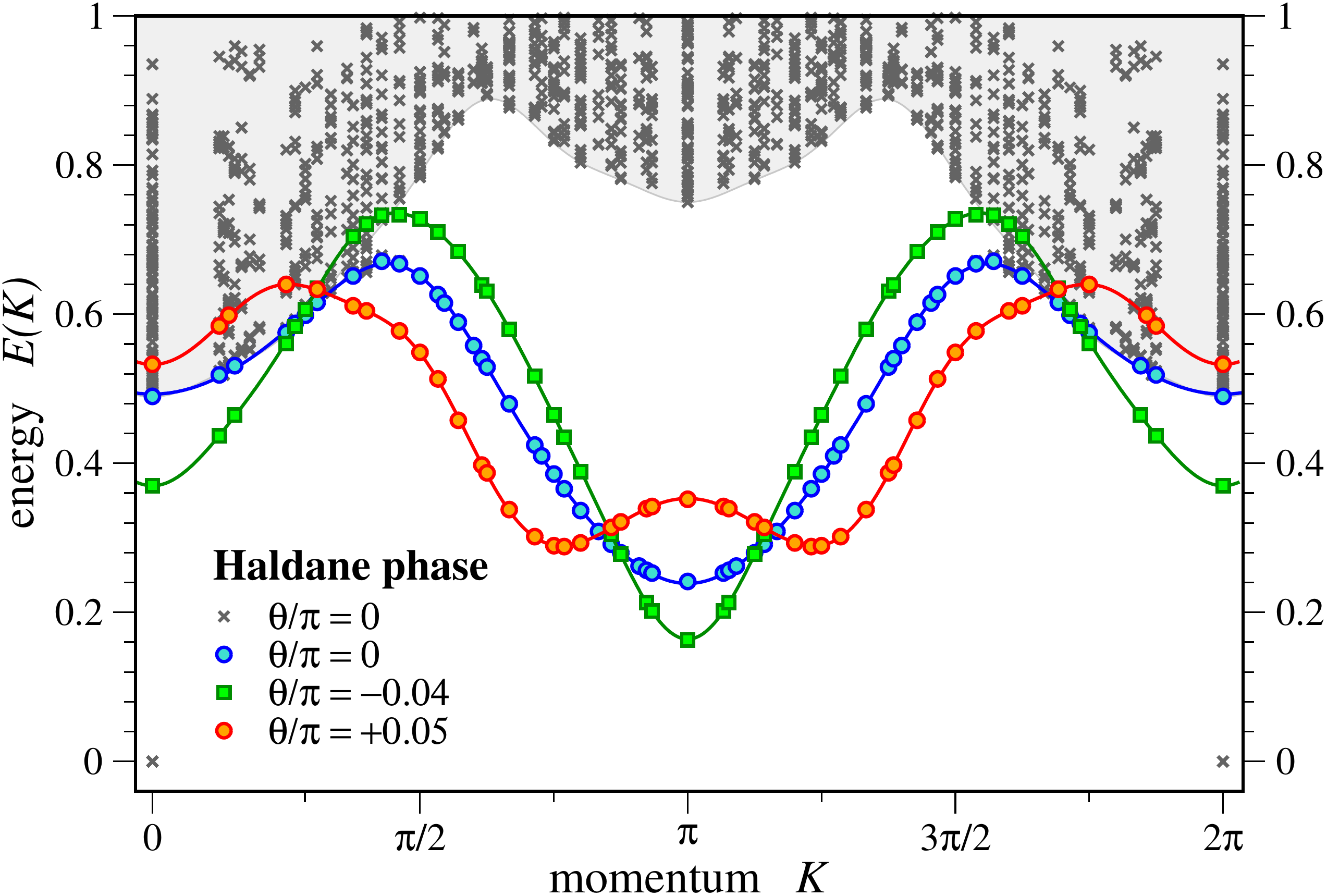} 
  \vskip 3mm
  \includegraphics[width=.48 \linewidth]{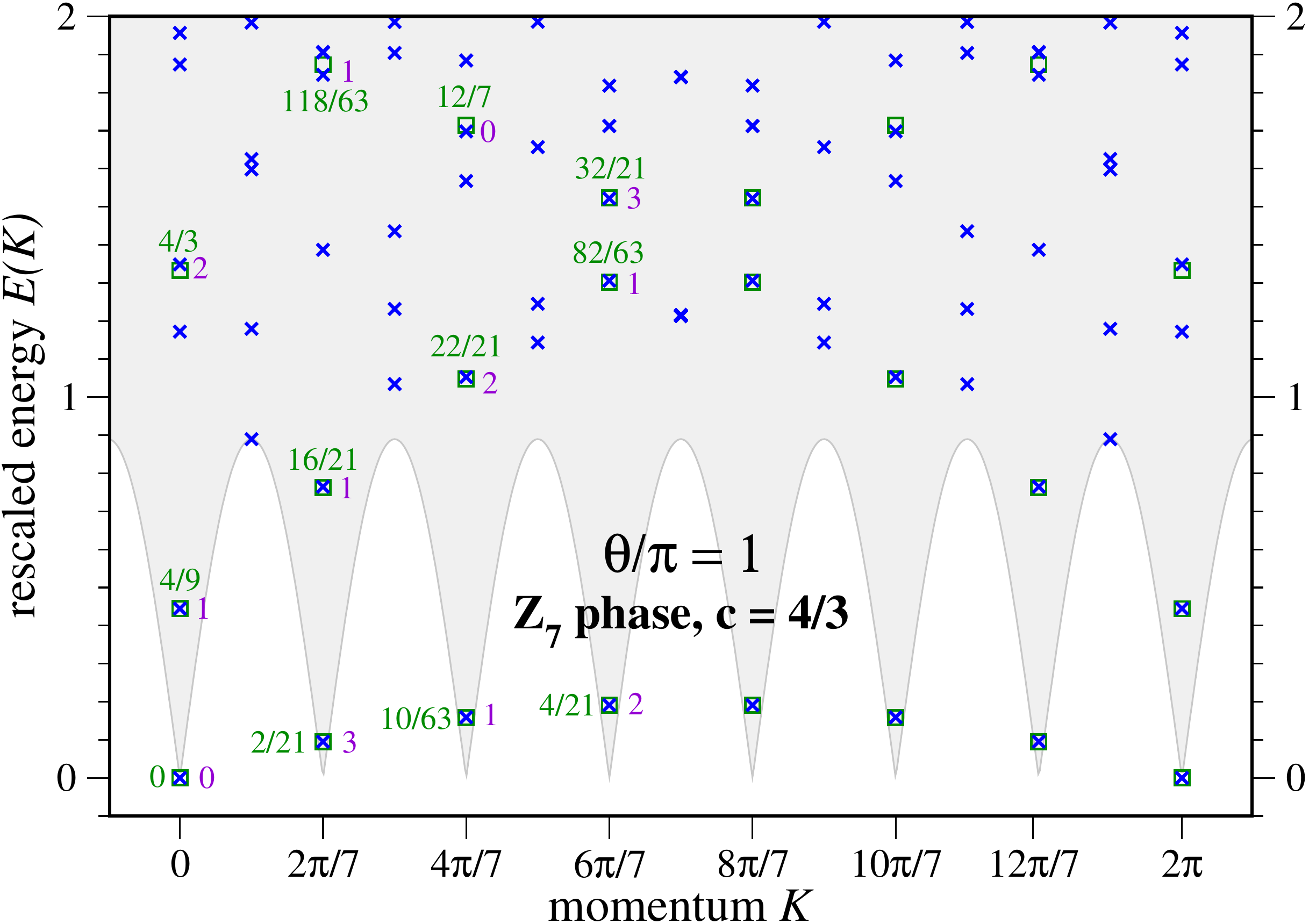} 
  \hskip 0.03 \linewidth
  \includegraphics[width=.48 \linewidth]{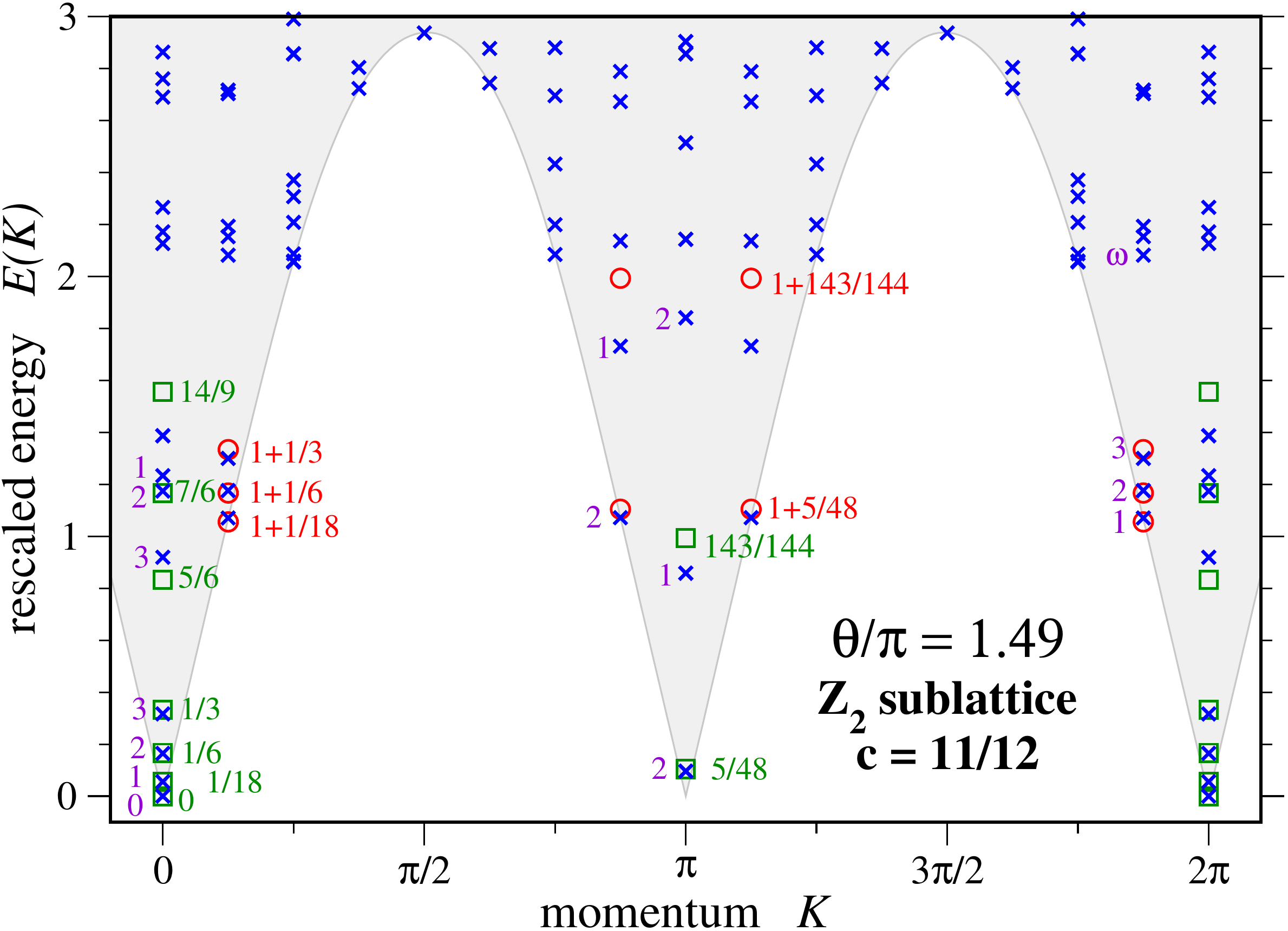} 
  \vskip 1mm
  \includegraphics[width=.48 \linewidth]{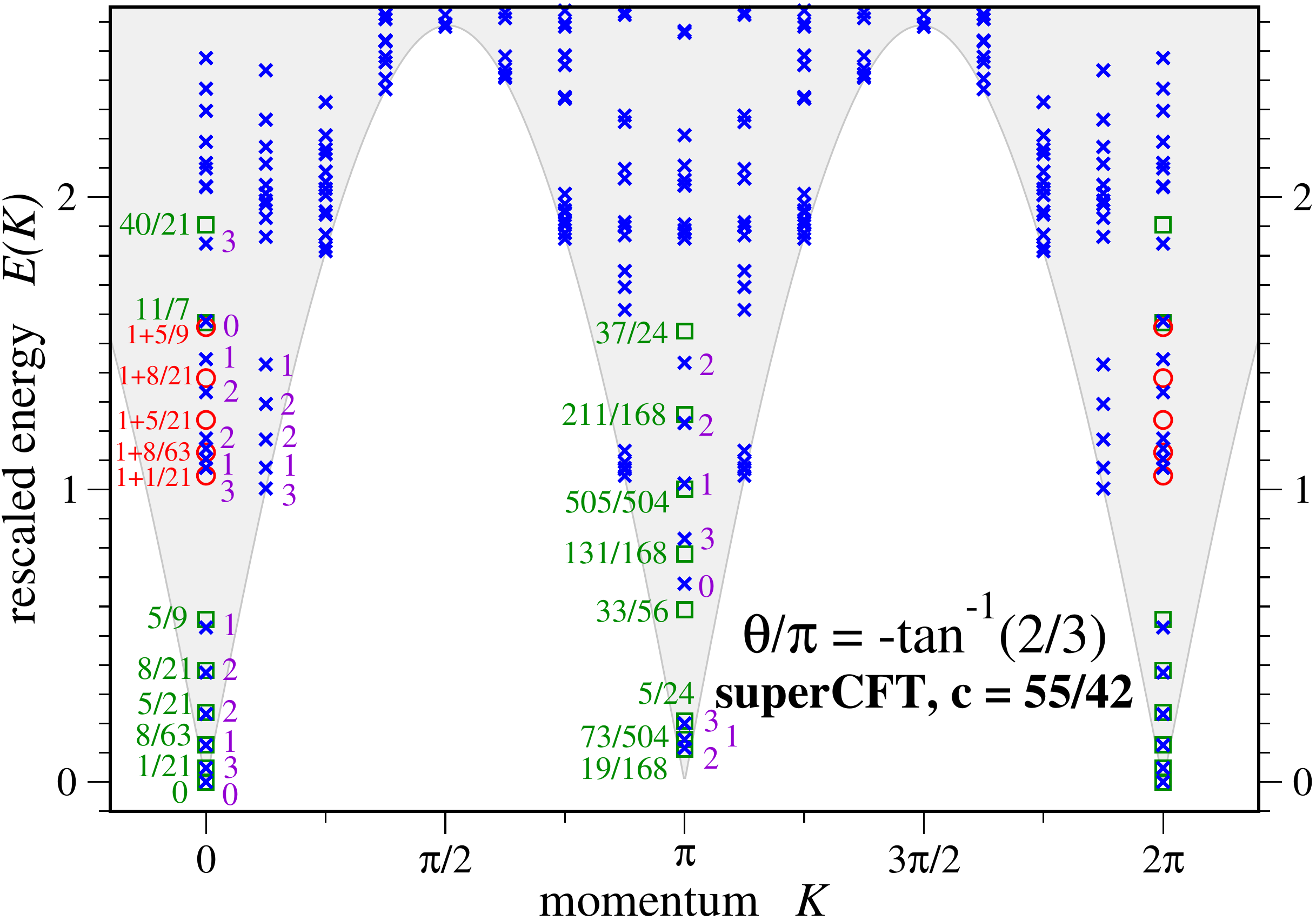} 
  \hskip 0.03 \linewidth
  \includegraphics[width=.48 \linewidth]{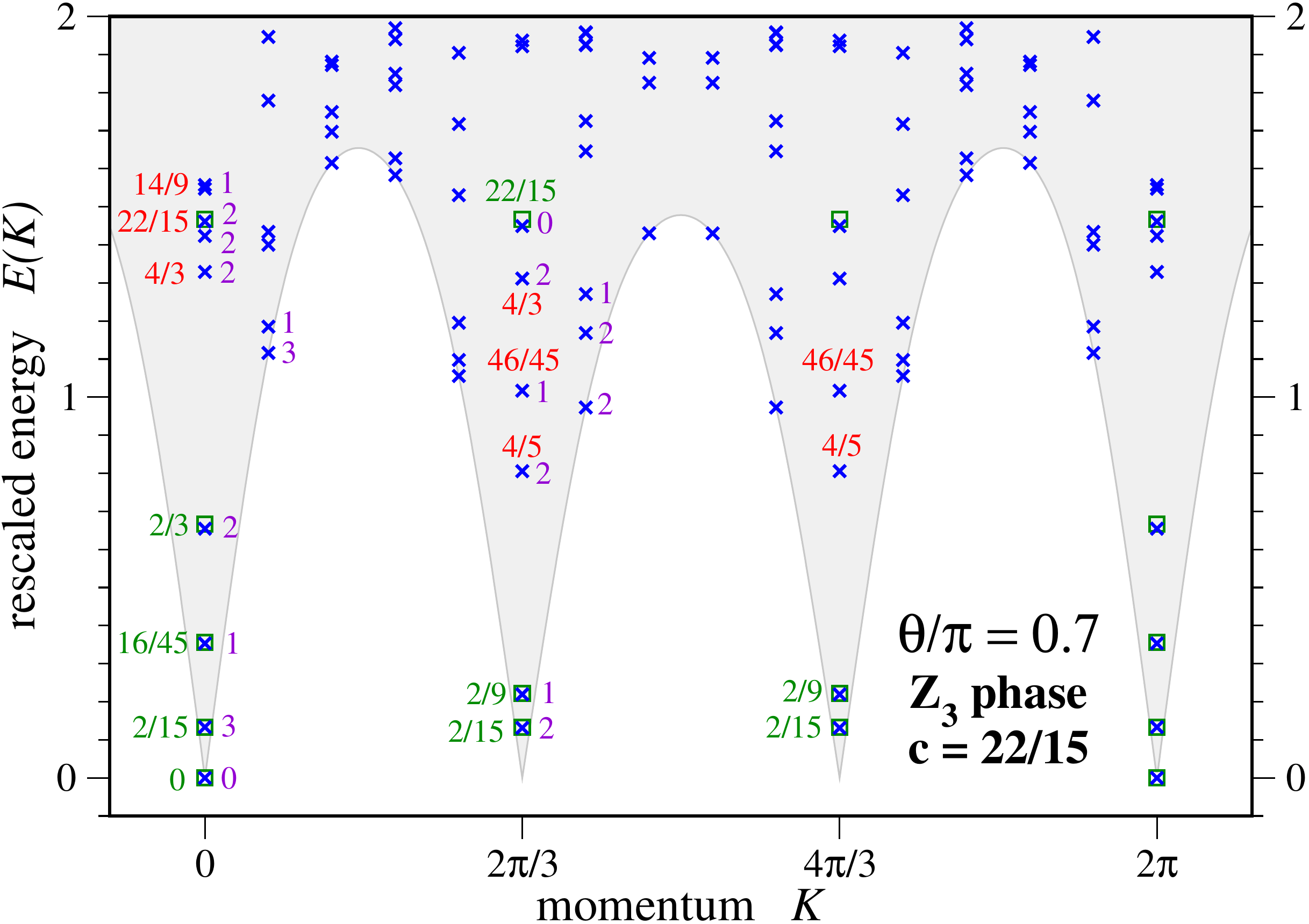} 
     \caption{(color online) 
                    {\bf The su(2)$_7$ spin-1 chain:} The energy spectra for the various phases of the phase diagram
                 are shown in the upper left panel. For the critical phases/point the energy spectra have been rescaled
                    to match the conformal field theory prediction given in Eq.~\eqref{CFT_energy_levels}. 
                    Green squares indicate the location of the primary fields, red circles the descendant fields.
                    The energies predicted by conformal field theory are given in green (red) for primary (descendant) fields.
                    The topological symmetry sector is indicated by the violet index. 
                    Data shown are for system sizes $L=16$ and $L=14$, respectively.}
     \label{Fig:su(2)_7}
  \end{center}
\end{figure*}

The spectra at angle $\theta_{2,1} = \pi$ for $k=5$ and $k=7$ are displayed in the middle panel of 
Figures~\ref{Fig:su(2)_5} and \ref{Fig:su(2)_7}, respectively.
The energy spectra were rescaled such that  the energy of the
lowest excitation matches the energy predicted by the $Z_k$ parafermion conformal
field theory\cite{zf85}.
Some details of this CFT
are reviewed in appendix~\ref{app:zk-pf}.
In the Figures, we indicate the locations of
the energies of the states corresponding to the primary fields by green squares, while 
blue crosses correspond the numerically obtained energy levels. We find good agreement between numerically obtained energy spectra and the $Z_k$ parafermion CFTs 
for both the su(2)$_5$ and su(2)$_7$ anyon models. For su(2)$_5$,
we also indicate the location of a few descendant fields that  match the numerical prediction.
 Generally, the identification of descendant fields is more difficult  due to finite size effects.

The fields of the $Z_k$ parafermion theory carry two labels, $(l,m)$ that take the values $l=0,1,\ldots (k-1)$, and $m=0,2,\ldots 2(k-1)$.
The momentum and topological quantum number of the fields is determined by the
labels $m$ and $l$, respectively. The topological quantum number simply is given by
$l$. For the momentum, the following relation holds: $K = \frac{2m \pi}{k}$.

\clearpage

\begin{table*}[t]
\begin{tabular}{|l|c|c|c|c|c|}
\hline \hline
\multirow{2}{*}{\bf gapless theory} & \multirow{2}{*}{\bf coset description} &
\multicolumn{4}{c|}{\bf central charge} \\
& & $k$ & $k=5$ & $k=7$ & SU(2) $(k \to \infty)$ \\
\hline \hline
$Z_k$ phase & su(2)$_k/u(1)_{2k}$ & $c=2\frac{k-1}{k+2} $ & $c=8/7$ & $c=4/3$ & $c=2$ \\
$Z_2$ phase & su(2)$_{k-1}\times su(2)_1/su(2)_k$ & $c=1-\frac{6}{(k+1)(k+2)}$ & $c=6/7$
& $c=11/12$ & $c=1$ \\
$Z_3$ phase & su(2)$_{k-4}\times su(2)_4/su(2)_k$ & $c=2-\frac{24}{(k-2)(k+2)}$ & $c=6/7$
& $c=22/15$ & $c=2$ \\
superconformal point  & su(2)$_{k-2}\times su(2)_2/su(2)_k$  & $c=\frac{3}{2}-\frac{12}{k(k+2)}$
& $c=81/70$ & $c=55/42$ & $c=3/2$ \\
\hline \hline
\end{tabular}
\caption{Critical theories in the su(2)$_k$ spin-1 chains for $k \geq 5$.}
\end{table*}

We find that there are no {\it relevant}  primary fields which have the same set of quantum numbers 
as the identity field. This implies that there are no relevant operators that can 
be added to the Hamiltonian to drive a phase
transition if  both translational and topological symmetry are left unbroken. This phase is an example
of a critical phase whose criticality is protected by the topological symmetry.

\subsubsection{$Z_2$-phase: $(A,D)$ modular invariant of coset
 su(2)$_{k-1}\times su(2)_1/su(2)_k$}

Upon increasing $\theta_{2,1}$, one encounters a first order transition from the
$Z_k$ parafermion phase into a different extended critical phase that has
a $Z_2$ sublattice symmetry. We  identified the CFTs describing these critical
phases for $k=5$ and $k=7$ as Virasoro  conformal minimal models, with central charge
$c= 1-\frac{6}{(k+1)(k+2)}$. However, the field content describing the criticality is
not the `usual' minimal model -- the diagonal $(A,A)$ modular invariant -- but the so-called $(A,D)$ modular invariant which
contains a different number of fields. Details of these different modular invariants
can be found in \cite{cardyNPB1986,ciz87a,ciz87b}. For our purposes, it suffices to notice that 
some of the primary fields in the $(A,A)$ invariant do not appear
in the $(A,D)$ invariant while others  appear twice. The details of this CFT are
summarized in table~\ref{table_modular} in appendix~\ref{app:minmod}.
The scaling dimensions of the fields are given in equation~\eqref{h_minmodel}.

Again, it is possible to  identify the topological sectors and the momenta at which
the various fields occur from the labels of the fields. As discussed in appendix~\ref{app:minmod},
the fields can be labeled by $(r,s)$, where $s$ takes the values $s=1,3,\ldots,k$.
The topological sector is determined by $(s-1)/2$, while the momenta are fixed by the
$r$ label. In particular, for $k=5$, the fields with labels $r=1,5$ occur at $K=0$, while 
the fields with label $r=3$ occur both at $K=0,\pi$. For $k=7$, the fields with $r=1,3,5,7$
occur at $K=0$, while the fields at $r=4$ are doubly degenerate and occur $K=\pi$. 
    
\subsubsection{$Z_3$-phase: coset su(2)$_{k-4}\times su(2)_4/su(2)_k$} 

At $\theta_{2,1} = 3\pi/4$, there is a first order transition between the $Z_k$ `ferromagnetic' phase
and a critical region of criticality that exhibits a $Z_3$ sublattice symmetry. We determined that the CFT describing the $Z_3$ critical region is  a series of coset models
with $S_3$ symmetry, namely $\frac{su(2)_4 \times su(2)_{k-4}}{su(2)_k}$.  In appendix~\ref{app:s3_models}, we list some details of
these coset models, in particular, the scaling dimensions of the primary fields (a detailed analysis  can be found in~\cite{fz87,zf87}).
 The primary fields are
labeled by two integers $(r,s)$. As was the case for the $Z_2$ critical phase, only
a subset of the fields appear in the spectrum, namely those with $r+s$ even. In addition,
the label $s$ has to be odd, and it determines the topological quantum number via $(s-1)/2$. 
The location of the second endpoint of this $Z_3$ critical region (i.e., the transition to the Haldane gapped phase) 
is found to vary with $k$.

In Figures~\ref{Fig:su(2)_5} and \ref{Fig:su(2)_7}, we display representative energy
spectra for this phase (angle $\theta_{2,1} = 0.7 \pi$). In these spectra, we indicate the topological
sectors of some of the low lying fields and give the scaling dimensions of the primary fields.

\subsubsection{Superconformal critical point}

The transition between the $Z_2$ phase and the Haldane gapped phase
occurs at the angle $\theta_{2,1} \approx -0.19 \pi$, which shows little dependence
on the level $k$.
The critical point itself is described by a $N=1$ superconformal minimal model\cite{fqs84},
$\frac{su(2)_2\times su(2)_{k-2}}{su(2)_k}$. Details on this theory can be
found in appendix~\ref{app:supercft}. In the limit of $k\rightarrow\infty$, this
theory approaches the su(2)$_2$ theory, which describes the critical point in the
SU(2) spin-$1$ bilinear-biquadratic spin chain. 

In the spectra for $k=5$ and $k=7$ of the anyonic spin-$1$ chain at this critical point, we indicate
the scaling dimensions and topological sectors of the primary fields which are labeled by $(r,s)$. Like in the other coset models (excluding
the $Z_k$ parafermion theory), the label $s$ is associated with the su(2)$_k$ denominator
of the coset and hence labels the topological sector. The momentum at which the
primary fields appear is determined by $K = (r+s\bmod 2) \pi$. 

The superconformal critical point separates the Haldane gapped phase from the
$Z_2$ sublattice critical region. Therefore, we expect that there will be a relevant perturbation
which drives the phase transition between these two different phases,
and that this perturbation does not break any symmetries. 
A relevant perturbation is a field which has the same quantum numbers as the 
ground state and whose scaling dimension is smaller than two. Such
a field indeed exists: it carries the labels $(r,s) = (3,1)$ and has scaling
dimension $1+\frac{4}{k}$, i.e., it is a relevant field  for all $k$. We note that at $K=\pi$, there also is
a relevant field with labels $(r,s) = (2,1)$ which has scaling dimension
$\frac{3}{8}+\frac{3}{2k}$. As a consequence, a gap is expected to develop if  a
perturbation which staggers the chain is added to the system.

\subsubsection{Stability of the critical phases}
   
We recapitulate that in all three extended critical phases there is no relevant field in the
same symmetry sector as the ground state, which is a requirement for the phases to be
stable. This notion of topological stability will be explained in more detail in the
section~\ref{spin_half_chain} dealing with the anyonic spin-1/2 chains,  
where we show in detail that the critical behavior of those
chains is protected by the topological symmetry.

As was explained above, there is a relevant operator with the same
quantum numbers as the ground state at the superconformal point.
This operator drives the transition from the superconformal point to the Haldane gapped
phase on one side of the phase diagram, and the extended critical region with $Z_2$ sublattice symmetry
on the other side.

\subsection{The gapped Haldane phase}
\label{Haldane}

In addition to the gapless phases that were discussed in detail in the previous section,
the spin-$1$ anyonic chains also exhibits a gapped phase, as can be seen in  Figure~\ref{fig:oddk_phase_diagram}.
The properties of this gapped phase are strikingly similar to the properties of the Haldane phase in the ordinary
bilinear-biquadratic spin-$1$ chain.
For instance, the point $\theta_{2,1}=0$ allows for a straightforward generalization
of the AKLT point of the ordinary SU(2) model. At this AKLT point, the degenerate ground states can
be constructed explicitly (see section~\ref{sec:AKLT-states}). In section \ref{sec:AKLT-states-open},
we discuss  the ground states of the open chain, and find the degeneracy of the anyonic spin-$1$ chain
can be understood in a similar way as the degeneracy of the SU(2) model at the AKLT point.
Before we deal with the ground states at the AKLT point, we first discuss the energy spectrum and the phase
boundaries of the Haldane phase.

\subsubsection{Energy spectrum}
The energy spectrum in the gapped phase is shown in Figures~\ref{Fig:su(2)_5} and \ref{Fig:su(2)_7} for
 coupling parameter $\theta=0$.
It can be seen that there exists a quasiparticle band whose qualitative shape is identical
to the magnon band of triplet excitations of the ordinary AKLT point.
The complete spectrum is shown at angle $\theta_{2,1}=0$: the ground states
occur at momentum $K=0$, and there exists a quasiparticle band  (shown in blue color) and a  continuum
of scattering states (shown in gray color). The quasiparticle band is also displayed for coupling
parameters $\theta_{2,1}$ close to $\theta_{2,1}=0$ (in red for $\theta_{2,1}>0$, in green for $\theta_{2,1}<0$).
It can be seen that when approaching the critical phase with $Z_3$ sublattice symmetry -- i.e., for increasing  $\theta>0$ -- 
 the minimum of the quasiparticle band
moves away from $K=\pi$ towards $K=2\pi/3$ and $K=4\pi/3$.
When decreasing the angle $\theta_{2,1} < 0$, the quasiparticle band 
remains at momentum $K=\pi$, which is consistent with the  $Z_2$ sublattice symmetry
 of the  superconformal critical point.
From a  finite-size scaling analysis of the energy spectra, we confirm that the gapped
phase does indeed extend over a finite range of coupling parameters $\theta$.
Figs.~\ref{Fig:su(2)_5} and \ref{Fig:su(2)_7} 
show that the  size of energy gap (at $\theta_{2,1}=0$) increases from
$\Delta E(k=5)\approx 0.16$ to $\Delta E(k=7) \approx 0.24$.

This behavior suggests that the qualitative shape of the energy spectra
at the AKLT point is preserved for all $k$ with the energy gap at
$\theta_{2,1}=0$ approaching $\Delta E(k\to \infty) \approx 0.41$
\cite{WhiteHuse}.

\subsubsection{Phase boundaries}  

The Haldane phase and the su(2)$_{k-1}\times su(2)_1/su(2)_k$
critical phase are separated by a superconformal critical point, which is
located  at coupling parameter $\theta_{2,1}\approx -0.19\pi$
for both $k=5$ and $k=7$. This is very close to the position of the phase transition
where the Haldane gapped phase gives way for a different phase in the
ordinary SU(2) spin-$1$ chain (see the phase diagram in
figure~\ref{fig:oddk_phase_diagram}), namely $\theta_{2,1} = -\arctan(2/3)$.

The position of the phase boundary at the other end of the gapped phase 
clearly depends on the level $k$. 
Comparing the position of this point for $k=5$ and $k=7$ suggests 
that it moves towards $\theta_{2,1}=\pi/2$ for  increasing $k$.
This scenario  is consistent with the ordinary model, as can be
seen by comparing the phase diagrams of the anyonic and ordinary SU(2)
spin-$1$ chain
(Figs.~\ref{fig:oddk_phase_diagram} and \ref{fig:su2_phase_diagram} respectively).

\subsubsection{Ground states in the periodic chain\\(anyonic equivalent of AKLT point)}
\label{sec:AKLT-states}

\begin{figure}[t]
\begin{center}
\includegraphics[width=\columnwidth]{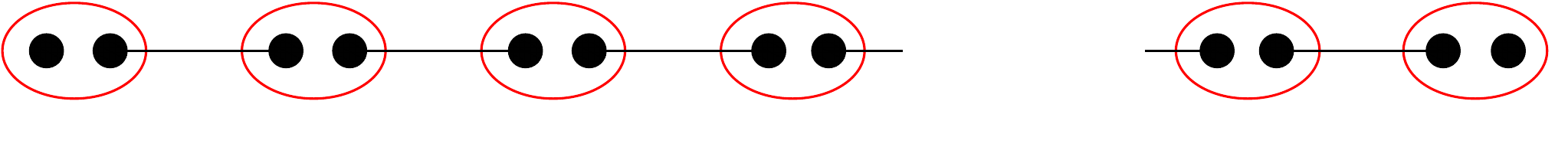}
\caption{The AKLT construction of the valence-bond-solid state on a finite chain of spin-$1$ degrees of freedom.
Each filled circle represents a spin-$1/2$ variable, each dotted ellipse corresponds to
a spin-$1$ particle, and and each line connecting two spin-$1/2$ variables symbolizes
a singlet bond.}
\label{vbs_state}
\end{center}
\end{figure}

In the ordinary SU(2) spin-1 chain, there exists a point within the Haldane gapped phase - the so-called AKLT\cite{AKLT} point -  where the ground state can be obtained exactly.
At the AKLT point, the Hamiltonian penalizes two neighboring spins
who are in the spin-2 channel. To construct the ground state, it is helpful to think of the
spin-$1$'s  as composed of two spin-$1/2$'s 
which are projected onto the spin-$1$ channel.
In the ground state, each of these spin-$1/2$ forms a
singlet with a spin-$1/2$ particle that is associated with a neighboring spin-1, as
depicted in Figure \ref{vbs_state}.
In this situation two neighboring spin-1's will never
combine into an overall spin-2 and, therefore, the state has zero energy. It can be shown that for periodic boundary conditions this ground state is non-degenerate \cite{AKLT}.

At the corresponding point  (angle $\theta_{2,1}=0$) in
the phase diagram of the anyonic chains, the Hamiltonian
(Eq.(\ref{eq:spin1-hamiltonian},\ref{eq:generalprojector}))
 penalizes
two neighboring anyons to fuse in the spin-$2$ channel. As for the ordinary
SU(2) quantum spin model, the ground state can be obtained exactly at this point.
In contrast to the
SU(2) case, there exists a topological symmetry which dictates that the
ground state is degenerate even in the case of periodic boundary conditions (we will
deal with the open chain in the next subsection). One of these degenerate ground
states is easily found, while the others can be obtained by making use of the
topological symmetry operator (see section~\ref{sec:top-sym} for details).

We will present  the simplest case of $k=5$  here, and
give the results for arbitrary $k$ in appendix \ref{app:AKLT-states}.
We start by constructing one zero energy ground state. For $k=5$,
the allowed spins are $0,1,2$, and the fusion rules read
\begin{align*}
0 \times 0 &= 0 & 0 \times 1 &= 1 & 0 \times 2 &=2 \\ 
&& 1 \times 1 &= 0 + 1 + 2 & 1 \times 2 &= 1+2 \\
&& && 2 \times 2 &= 0 + 1 
\end{align*}

In particular, the fusion rule $2\times 1 = 1 +  2$ implies that in the labeling of the
Hilbert space, the assignment $(x_{i-1},x_{i},x_{i+1}) = (2,2,2)$ is allowed. In addition,
$(x_{i-1},x_{i},x_{i+1}) = (2,1,2)$ is allowed as well. Fixing $x_{i-1} = x_{i+1} = 2$, one
finds that the allowed values of $\tilde{x}_i$ in the transformed basis are $\tilde{x}_i = 0,1$,
because $0$ and $1$ are the two possible fusion outcomes of $2\times 2 = 0 +1$. Because
at the AKLT point, only the value $\tilde{x}_i = 2$ is penalized, it follows that the state
$\ket{v_0} = \ket{2,2,\ldots,2}$ is a zero energy ground state (recall that that Hamiltonian is a
positive sum of projectors).

By employing the topological symmetry operators $Y_l$, with $l=1,2$, we can construct
other zero energy ground states. The operators $Y_l$ commute with the Hamiltonian,
thus the states $\ket{v_1} = Y_1 \ket{v_0}$ and $\ket{v_2} = Y_2 \ket{v_0}$ also have zero energy.
It turns out that $\ket{v_0}$ is neither an eigenstate of $Y_1$ nor of $Y_2$.
As a result, the number of ground states is three, which is in accordance with
the number of particle types in the model. We note that $Y_0$ is the identity operator.

The explicit form of the states $\ket{v_1}$ and $\ket{v_2}$ is easily written down. First of all,
the only basis states with non-zero coefficient in $\ket{v_1}$ have $x_i = 1,2$, for all $i$.
Similarly, the only basis states with non-zero coefficient in $\ket{v_2}$ have $x_i = 0,1$, for
all $i$. To specify the coefficients, we introduce the notation $\# l$ which denotes the
number of $i$'s such that $x_i=l$. In addition, $\#(l,m)$ denotes the number
of $i$'s such that $x_i = l$ and $x_{i+1} = m$, where we use periodic
boundary conditions, $x_{L} = x_{0}$.

Then, we have
\begin{align}
\ket{v_1} &= \sum_{x_i \in \{1,2\} } f_1 (\{x_i \} ) \ket{x_0,x_1,\ldots x_{L-1} } \\
f_1 (\{ x_i \} )  &= (-1)^{\# 2}d_1^{-L} d_2^{L/2}
d_1^{3\frac{\#(2,1)}{2}}
d_2^{-\frac{\#(2,1)+\#(2,2)}{2}} 
\nonumber
\end{align}
as well as
\begin{align}
\ket{v_2} &= \sum_{x_i \in \{0,1\} } f_2 (\{x_i \} ) \ket{x_0,x_1,\ldots x_{L-1} } \\
f_2 (\{ x_i \} )  &= (-1)^{\# 1}d_1^{-L/2} d_2^{L/2}
d_1^{\frac{\#1}{2}}
d_2^{-\#1} \ .
\nonumber
\end{align}
Here, $d_1$ and $d_2$ are the quantum dimensions of particles with spin-1 and 2
respectively, and are given by $d_1 = 1+ 2 \sin (3\pi/14)$ and $d_2 = 2 \cos(\pi/7)$, respectively.

We labelled the ground states at the AKLT point
by $\ket{v_l}$ with $l=0,1,2$ for a good reason.
In section~\ref{sec:top-sym}, we explained that the topological symmetry operators
$Y_l$ effectively `add' or fuse a particle of type $l$ to the fusion chain. At the AKLT point,
this notion becomes very explicit. The states $\ket{v_l}$ are thought of as states of the chain
in the $l$ sector. In particular, $\ket{v_0}$ corresponds to the identity sector. Adding a particle
of type $l$, i.e., acting with the operator $Y_l$, gives rise to a state in sector $l$,
or $\ket{v_l} = Y_l \ket{v_0}$. Moreover, if one acts with $Y_l$ on the state $\ket{v_j}$, one
obtains a combination of states, which is given by the fusion rules. In particular,
$Y_1 \ket{v_1} = \ket{v_0} + \ket{v_1} + \ket {v_2}$, $Y_1 \ket{v_2} = \ket{v_1} + \ket{v_2}$
and $Y_2 \ket{v_2} = \ket{v_0} + \ket{v_1}$. Thus,  loosely speaking, the ground states
of the periodic anyonic spin-1 chain at the AKLT point form a representation of the fusion
algebra su(2)$_k$. Because the modular $S$ matrix diagonalizes the fusion rules, one
can easily write down combinations of the ground states which are also eigenstates of
the operators $Y_l$, namely
$\ket{\psi_{\rm AKLT,i}} = \sum_{j=0}^{2}  S_{i,j} \ket{v_{j}}$, where $S_{i,j}$ is
the modular $S$ matrix for (the integer sector of) su(2)$_5$,
and the sum is over integer values.

For the explicit form of the AKLT ground states in the general case su(2)$_k$,
we refer to appendix~\ref{app:AKLT-states}.

\subsubsection{Ground states in the open chain\\ (anyonic equivalent of AKLT point)}
\label{sec:AKLT-states-open}

Before describing the structure of the ground states of the open anyonic chains at the
AKLT point, we  briefly review the physics of the valence bond solid
ground state at the AKLT point ($\theta_{2,1}=0$ in phase diagram
Fig.~\ref{fig:oddk_phase_diagram}) of the ordinary bilinear-biquadratic
spin-$1$ Heisenberg chain \cite{AKLT,Affleck}.
The Hamiltonian at $\theta=0$ consists only of the projector onto a 
total spin-$2$ of two nearest-neighbor spins with
a positive sign. Thus, in the ground state, a total spin-$2$ of two-nearest-neighbor
spins is suppressed. In the usual tensor product basis of local (site) states,
the valence bond solid ground state is given by
\begin{equation}
|\Psi_{ab}\rangle =
\varepsilon^{b_1 a_2}\varepsilon^{b_2 a_3}...\ \varepsilon^{b_{L-1} a_L}
|\psi_{a b_1 }\rangle\otimes |\psi_{a_2 b_2}\rangle\otimes ...\otimes
|\psi_{a_L b}\rangle \ ,
\end{equation}
where  the summation over repeated upper and lower indices is assumed.
The local spin-$1$ state  $|\psi_{a b}\rangle$ is  represented as the symmetric
part of the tensor product of two spin-$1/2$ variables:
\begin{equation}
|\psi_{a b } \rangle= \frac{1}{\sqrt{2}}(|\psi_{a}\rangle\otimes|\psi_{b}\rangle +
|\psi_{b}\rangle\otimes| \psi_{a}\rangle) \ ,
\label{spin_1}
\end{equation}
where $\psi_{a}$ denotes one of the two eigenstates
of the $S^z$ spin-$1/2$ operator, which we label by $a=1,2$.
The antisymmetric tensor $\varepsilon^{ab}$ enforces
a singlet bond of the spin-$1/2$ variables $a_{l+1}$ and $b_l$.
Therefore, the total spin of the two
nearest-neighbor spin-$1$ variables, consisting of four spin-$1/2$
variables which are labeled by $a_l$, $b_l$, $a_{l+1}$, $b_{l+1}$,
can only assume the values $0$ or $1$. 
For a chain with open boundary conditions (see
Fig.~\ref{vbs_state}) the first and the last
spin-$1/2$ variables indexed by $a_1$ and $b_L$ 
do not form a singlet bond.
These two spin-$1/2$ variables can add up to a total spin $0$ or a total spin $1$,
giving rise to a four-fold degeneracy for the spin-$1$ bilinear-biquadratic
chain at the AKLT point with open boundary conditions.

With the results above in mind, we will now consider the
fusion basis of the anyonic spin-$1$ chain, as shown in Figure~\ref{Fig:Spin1-Chain}.
We consider a chain of length $L$ with open boundary conditions in the sense that variables
$x_0$ and $x_{L+1}$ form the ends of the chain.
In analogy with the above discussion, we assume that variables
$x_0$ and $x_{L+1}$ can add up to a total spin of
$x_0\times x_{L+1}$ of $0$ or $1$ in the zero-energy ground states.

For a given choice of $x_0$ and $x_{L+1}$,
we expect that there are no zero-energy ground states if
$|x_{L+1}-x_0|>1$ because the fusion product $x_0\times x_{L+1}$
does not contain $0$ nor $1$ in this case. 
We expect one ground state to be present if $x_0\times x_{L+1}$
contains $0$ or $1$, but not both. Finally, if both $0$ and $1$ appear in the
fusion product $x_0\times x_{L+1}$, we expect two zero-energy ground states.
There is no $S_z$ quantum number in anyonic spin chains 
associated with the `spins', and the state with total spin-$1$ (or better, topological
charge $1$) is thus not degenerate.

The analysis of  the previous subsection is helpful in 
understanding the above discussed results. We found that the
 ground states of the periodic chain have a particular
form; namely,  the only basis-states which have non-zero coefficients in these states
are such that all the $x_i$ take at most two values that have to differ by one.
Thus, there is a ground state with all the $x_i \in \{0,1 \}$, one ground state
with the $x_i \in \{1,2\}$, etc. In addition, the state with all $x_i = (k-1)/2$ is also
a zero energy ground state. 

The ground states of the open chain must be such that the bulk part of these states
does not give an energy contribution. Thus, for a particular choice of boundary
conditions $x_0$ and $x_{L+1}$, one can construct one ground state if $|x_0-x_{L+1}| =1$,
because there is exactly one corresponding zero energy ground state with periodic
boundary conditions. For $x_0 = x_{L+1} = 0$, there is also one zero energy ground state,
while for $x_0 = x_{L+1} > 0$, there are two zero energy ground states.
For $|x_0-x_{L+1}| > 1$, one finds that there are no zero energy ground states.
All of this is in accordance with the considerations above.

\begin{figure}[t]
\begin{center}
\includegraphics[width=\columnwidth]{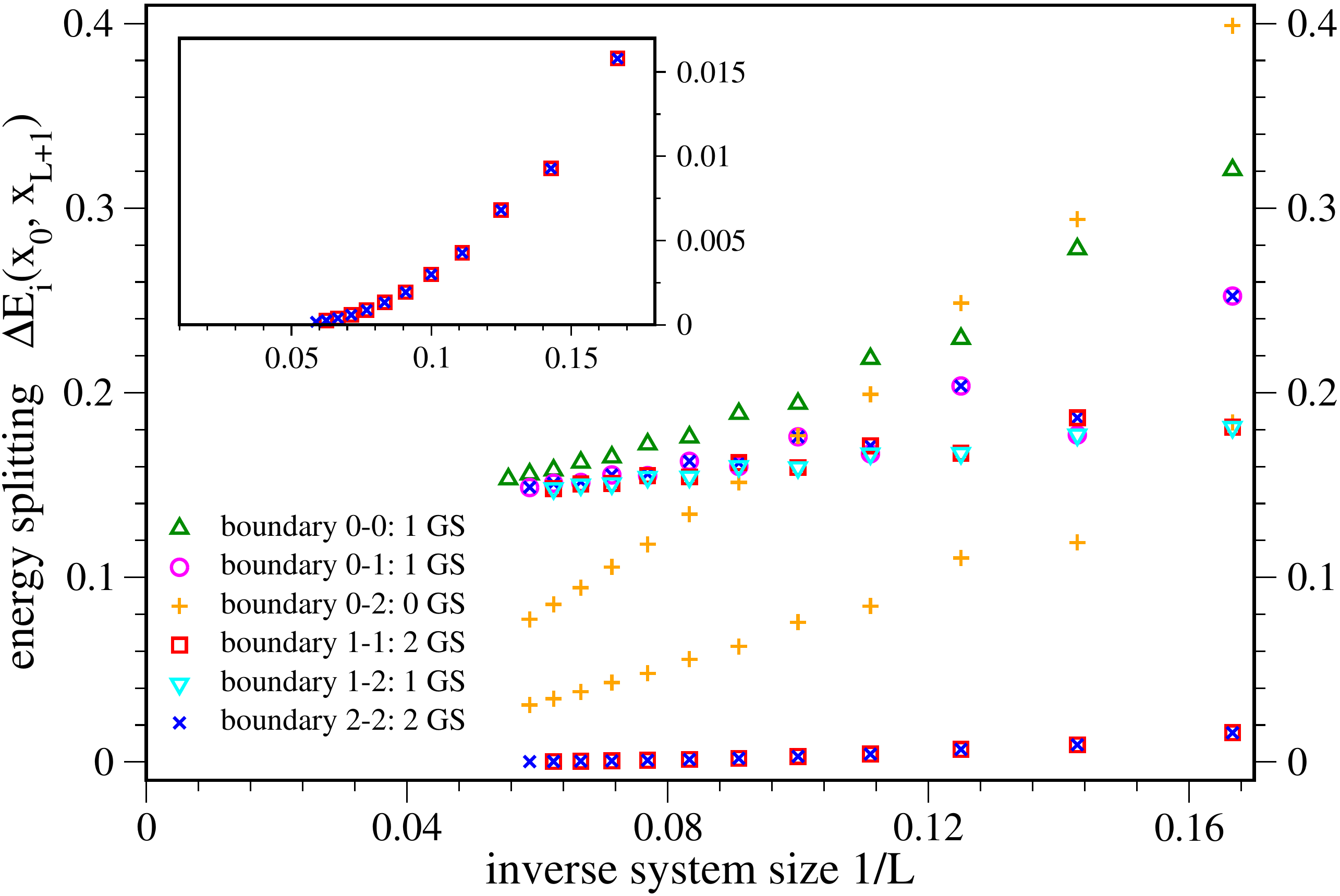}
\caption{The eigenenergies
$\Delta E_i(x_0,x_{L+1}):= E_i(x_0,x_{L+1})-E_0(x_0,x_{L+1})$
($i\ge 1$) of the su(2)$_5$ anyonic spin-$1$ chain with fixed boundaries
$x_0$ and $x_{L+1}$ as a function of $1/L$ at  $\theta_{2,1}=-0.01\pi$.
The legend at the lower left side indicates the values of $x_0$ and $x_{L+1}$.
The energy $E_0(x_0,x_{L+1})$ is the lowest energy and not necessarily
a `ground state energy'.  For $x_0=x_{L+1}=1$, and for $x_0=x_{L+1}=2$,
there are two almost degenerate zero-energy states, and $\Delta E_1(x_0,x_{L+1})$
corresponds to the finite-size splitting of the two ground states that decays exponentially
with system size (see the inset).} 
\label{exp_decay_haldane}
\end{center}
\end{figure}

We computed the ground state degeneracies for all possible choices of fixed boundary
occupations $(x_0,x_{L+1})$ for both the $k=5$ and
the $k=7$ model, and find that the above described picture is indeed the appropriate one.
At the AKLT point $\theta_{2,1}=0$, the ground state energy is independent of
the system size.
In the Haldane gapped phase away from the AKLT point, the ground
state degeneracy is not exact and finite size effects  occur.
In Fig.~\ref{exp_decay_haldane},
we show the lowest energies
$\Delta E_i(x_0,x_{L+1})=E_i(x_0,x_{L+1})-E_0(x_0,x_{L+1})$, $i\ge 1$, of the 
su(2)$_5$ spin-$1$ chain at coupling parameter $\theta_{2,1}=-0.01\pi$.
The energy $E_0(x_0,x_{L+1})$ is the lowest energy of the open chain
with fixed boundary occupations $x_0$ and $x_{L+1}$, and it is not necessarily
a ground state energy. By this we mean that the state is not a perturbation of a
zero energy ground state at $\theta_{2,1} = 0$.
For the boundary condition $x_0=0$, $x_{L+1}=2$, the lowest energy
$E_0(0,2)$ is not a ground state (in the above sense) since both
$\Delta E_1(0,2)$ and $\Delta E_2(0,2)$ approach zero in the limit
$1/L\to 0$, as demonstrated in Fig.~\ref{exp_decay_haldane}.
For the boundary condition $x_0=1$, $x_{L+1}=1$, as well as
$x_0=2$, $x_{L+1}=2$, the ground state is two-fold degenerate,
and the splitting of the two ground state energies at finite system size $L$
decays exponentially in $1/L$, as illustrated in the inset of Fig.~\ref{exp_decay_haldane}.
Again, this is in agreement with the above discussion because
$1 \times 1 = 0 + 1 + 2$ and $2 \times 2= 0 + 1$ (for su(2)$_5$), i.e.
both fusion products allow for a total spin $0$
and a total spin $1$. For all remaining possible boundary conditions,
there is one ground state, as can be seen from
Fig.~\ref{exp_decay_haldane} where $\Delta E_1(x_0,x_{L+1})$ 
approaches a finite energy in the limit $1/L\to 0$.
We also verified this scheme for the su(2)$_7$ model,
and for different values of $\theta_{2,1}$ in the gapped phase.

\section{Anyonic su(2)$_k$  spin-$\bf1$ chains: even $k\geq 6$}

In the previous section, we discussed in detail the odd-$k$ anyonic spin-$1$ chains.
We found that the phase diagram of these models (see Fig.~\ref{fig:su2_phase_diagram}),
bears great resemblance to the 
phase diagram 
of the SU(2) spin-$1$ chain (see
Fig.~\ref{fig:oddk_phase_diagram}).
We observed one striking difference between the ordinary and the anyonic spin-$1$ chains; namely, 
the absence of a (gapped) `dimerized' phase in the
case of the anyonic spin-$1$ chains. In this section, we present our result for the even-$k$ anyonic spin-$1$ chains. For even $k$, the phase
diagram is very similar to the case of odd $k$ with the exception of
an additional gapped phase which resembles the dimerized phase of the SU(2)
spin-$1$ chain. 

In this section, we  focus on the case $k=6$;
however, our analysis for $k=8$ indicates that the case $k=6$ is generic for
$k$ even. The generic structure of the phase diagram for even $k\ge 6$  is
 analogous to the generic structure of the phase diagram for odd $k\ge 5$.
 We note that the case
$k=4$ is special and will be considered in detail in the following section.

The fact that the phase diagrams for $k$ even and odd differ is a very interesting
feature of our model. As far as we are aware, this is the first time that
a dependence on the parity of the level $k$ has been observed. As we will point out in the
discussion, Koo and Saleur \cite{Saleur} considered a closely related loop model which contains
a continuous parameter that plays the role of the discrete level $k$.
The model considered by Koo and Saleur  does not show any sign of the `even-odd' effect we observe. It
would be very interesting to understand the differences and similarities of these two models in
greater detail.

\begin{figure}[t]
\begin{center}
\includegraphics[width=.8\columnwidth]{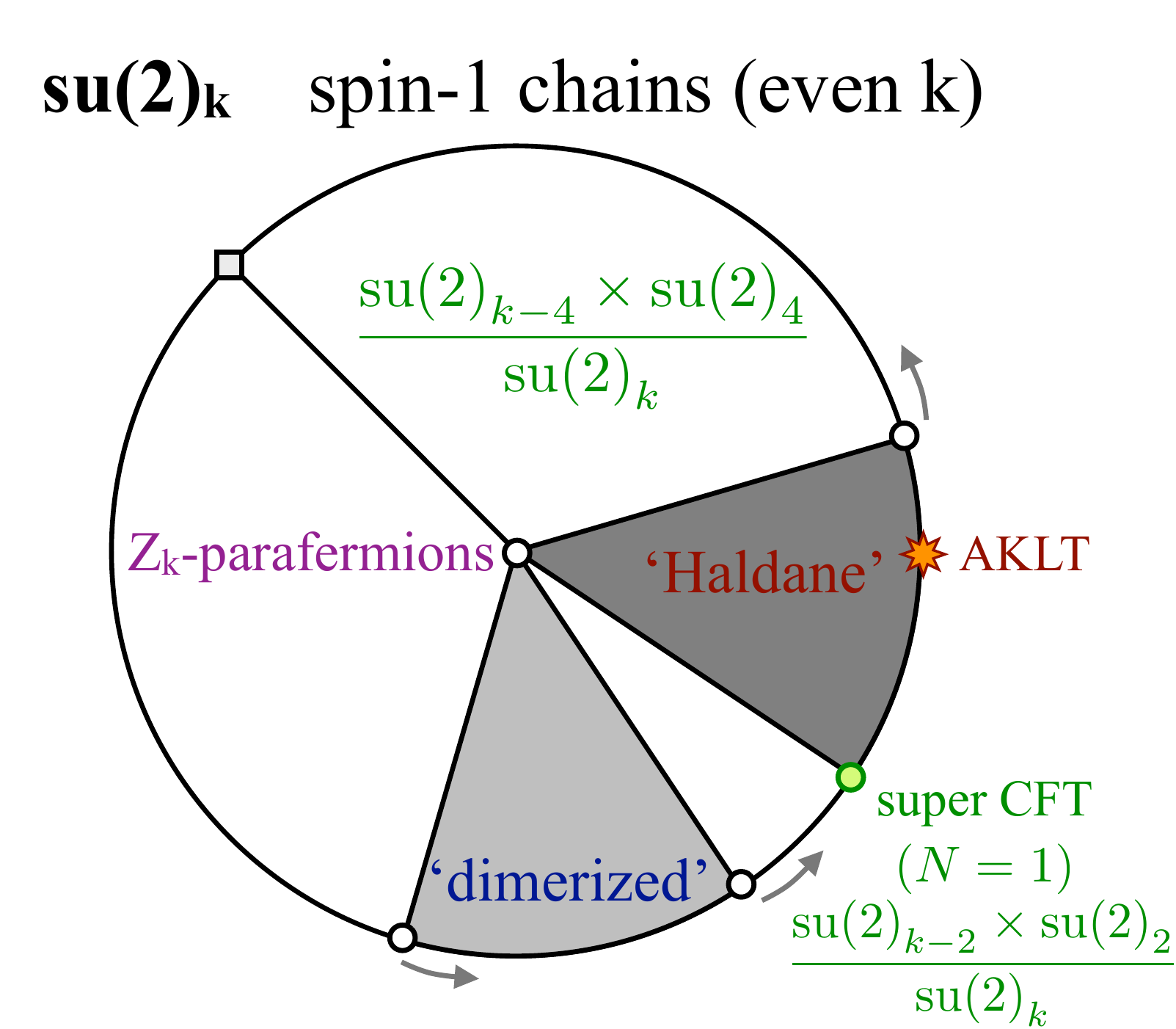}
\caption{(color online)
	      Phase diagram of the even-$k$ anyonic su(2)$_k$ spin-$1$ chain
	      in a projector representation \eqref{eq:spin1-hamiltonian} where  $J_1=-\sin(\theta_{2,1})$ 
	      and $J_2=\cos(\theta_{2,1})$. The locations of the phase boundaries correspond to the
	      case $k=6$. Some of the phase boundaries move with increasing (even) $k$; the arrows
	      indicate the direction of the change.}
  \label{fig:evenk_phase_diagram}
\end{center}
\end{figure}

The phase diagram of the $k=6$ anyonic spin-$1$ chain is presented in Figure
\ref{fig:evenk_phase_diagram}. We will  discuss the similarities and differences of this phase diagram
 to the
phase diagram of the case $k=5$ (Fig. \ref{fig:oddk_phase_diagram}).
The locations of the phase boundaries in Figure \ref{fig:evenk_phase_diagram} correspond to the case $k=6$. As was the case for $k$ odd, we observe that some of the phase boundaries change upon increasing the
value of (even) $k$. The direction of the movement of the phase boundaries is indicated by the arrows in the phase diagram.

Comparing the phase diagrams for odd and even $k$ in Figures
\ref{fig:oddk_phase_diagram} and \ref{fig:evenk_phase_diagram},
we first note that large parts of the phase diagram  have
a similar structure. At angle $\theta_{2,1} = 0$, we encounter a gapped Haldane phase, precisely as in the case of odd $k$. At angle
$\theta_{2,1} \approx - 0.19 \pi$,  there is a phase transition that is described by a $N=1$ supersymmetric minimal model from the Haldane phase into an extended critical region (we will comment on the latter critical region below).
At the other end of the gapped Haldane phase, there is a phase transition at angle
$\theta_{2,1} \approx 0.09 \pi$ (for $k=6$) to a critical region that exhibits 
 a $Z_3$ sub-lattice symmetry and is described by the
coset su(2)$_{4} \times su(2)_{k-4}/su(2)_k$ (we note that the corresponding critical region for odd $k$ is described by the same CFT). This critical
region extends all the way to $\theta_{2,1} = 3\pi/4$ at which point  there is a first order transition to
a critical region with $Z_k$ sublattice symmetry. So far, the phase diagram for
even $k$  has the same structure and phases as the one for odd $k$.

The phase diagrams for odd $k$  versus even $k$  begin to diverge 
at the angle where for $k$ odd, the
critical region with $Z_k$ sublattice symmetry transitions to a critical phase with $Z_2$ sublattice symmetry.
While the former ($Z_k$) critical phase also appears for $k$ even, the latter ($Z_2$ critical phase) does not; rather, there
is  a phase transition at
$\theta_{2,1 } \approx 1.41\pi$ (for $k=6$) to
a gapped phase. This gapped phase is characterized by broken translational invariance, as signified by a zero-energy ground
state at $K=\pi$ present at the angle $\theta_{2,1} = 3\pi/2$. In addition, there are
$(k+2)/2$ degenerate ground states at momentum $K=0$ with topological quantum numbers $(0,1,2,\ldots,k/2)$. The zero energy
ground state at $K=\pi$ is in topological symmetry sector $k/4$. Clearly, the nature of this
`dimerized' gapped  phase differs from the Haldane gapped phase.

Between the `dimerized' gapped  phase and the Haldane gapped phase, we find an
extended critical region.
Due to the rather small extend of this critical region and
the fact that we could not study large enough systems  (the dimension of the Hilbert space increases
with $k$), we have not been able to determine which CFT describes this extended critical
region.

It is interesting to note that the structure of the phase diagram for even $k$ bears closer
resemblance 
to the phase diagram of the SU(2) bilinear-biquadratic spin-$1$ chain,
(see Figure \ref{fig:su2_phase_diagram}) than to the  phase diagram for odd-$k$ anyonic spin-$1$ chains.
In particular, both the phase diagrams of the ordinary SU(2) spin-$1$ chain and the even-$k$ anyonic spin-$1$ chain exhibit 
dimerized phases in the area surrounding the angle $\theta_{2,1} = 3\pi/2$.
It appears   that for 
increasing even $k$,
the phase diagram of the anyonic chain gravitates towards the phase diagram of the SU(2) chain. Our results  for  the $k=8$
anyonic chain are consistent with this picture.

 The phase diagram for the $k=6$ anyonic spin-$1$ chain 
displays a unique feature; namely, its structure is symmetric in the line through the points
$\theta_{2,1} = 3\pi/4,7\pi/4$. The underlying reason is that the fusion rules of the su(2)$_k$
theory are symmetric under the exchange $j \leftrightarrow k/2-j$, where the labels $j$ take the
values $j=0,1/2,\ldots,k/2$. In the case of $k=6$, this symmetry exchanges anyon spins $1\leftrightarrow 2$. The
location of the symmetry points follow from our parametrization of the hamiltonian, as given in
equation \eqref{eq:spin1-hamiltonian}. We  point out that this symmetry only relates the
{\em sets of energy eigenvalues}, but not the possible degeneracy of the levels or their angular
momenta.

For example, the energy levels  levels
at the point $\theta_{2,1} = \pi$ - where the system is described by the $Z_6$
parafermion theory - are identical to those at angle $\theta_{2,1} = \pi/2$. At the latter point, the   system is described by the coset su(2)$_2 \times su(2)_4/su(2)_6$, 
which for $k=6$ corresponds to the $Z_6$ parafermions. We note that the  momenta of the states
are not identical.

Similarly, the energies of the levels in the dimerized  gapped phase  are the same as the energies of the levels in the Haldane phase, even
though the nature of these gapped phases is very different. We will return to this issue below. Finally, we note that the phase transition from
the dimerized phase to the
critical region in between the dimerized phase and the Haldane phase is given by an
$N=1$ supersymmetric model. As far as we can tell from our numerics, this is only true
for the case $k=6$. For $k=8$ and higher, we have not been able to determine the CFT
describing this phase transition.


\section{Anyonic su(2)$_k$  spin-$\bf1$ chains: $k=4$}
\label{spin1_k4_chain}

Having discussed the anyonic spin-$1$ models for odd $k\geq 5$  and
even $k\geq 6$, we finally turn our attention to the  remaining case
$k=4$. We already pointed out in the introduction that  the phase diagram for $k=4$
has a different structure than the phase diagrams for other values of $k$. The underlying reason
is that the spin-$1$ particle is special in this case. The symmetry of the fusion
rules under the exchange $j\leftrightarrow k/2-j$ implies that 
 $j=1$ is mapped onto itself for $k=4$. In addition, $k=4$ is the lowest $k$ for which a
general fusion rule $1\times 1 = 0 + 1 + 2$ applies. We refer to the discussion in section \ref{sec:discussion}
for more details.

\begin{figure*}[t]
\begin{center}
\includegraphics[width=\linewidth]{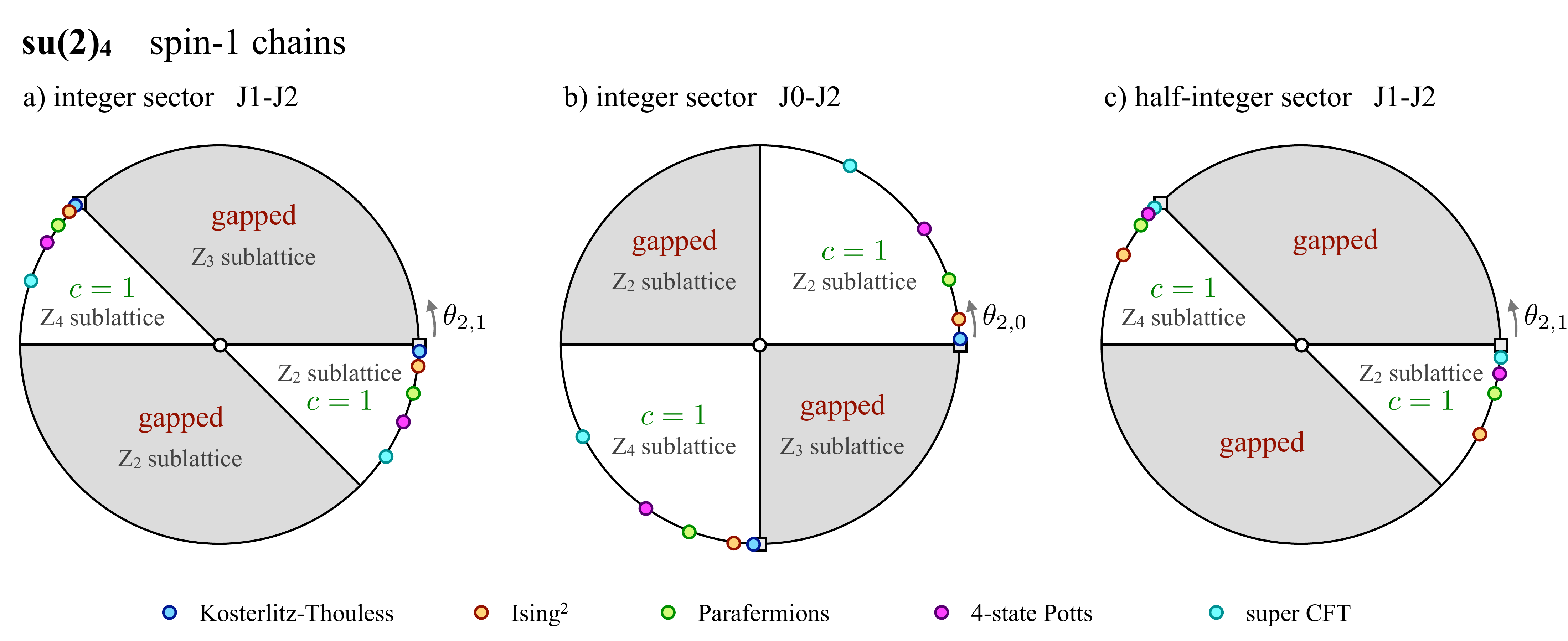}
\caption{(color online)
	     Phase diagrams of the su(2)$_4$ spin-$1$ chain in the integer sector and half-integer sector and different
	      projector representations. The colored circles indicate special points in the $c=1$ gapless phase that can
	      be matched to the labelled conformal field theories.}
  \label{PD_spin1_k4}
\end{center}
\end{figure*}

\subsection{Hilbert space and Hamiltonian}
\label{spin1_k4_hamiltonian}

The basis of the su(2)$_4$ spin-$1$ chain is
depicted in Fig.~\ref{Fig:Spin1-Chain}.
Each labeling
$\{x_i\}_{i=0,...,L-1}\in \{0,\frac{1}{2},1,\frac{3}{2},2\}$ that 
 satisfies the fusion rules at the vertices
corresponds to a different basis state.
In fact, the Hilbert space of the su(2)$_4$ spin-$1$ chain
splits into two independent sectors:
 the fusion rules impose that
the local basis elements are either all integer valued or all
half-integer valued. We shall use the following terminology:
\begin{itemize}
\item Integer sector (IS): $\{x_i\}_{i=0,...,L-1}\in \{0,1,2\}$.
\item Half-integer sector (HIS): $\{x_i\}_{i=0,...,L-1}\in \{\frac{1}{2},\frac{3}{2}\}$.
\end{itemize}
We will only consider periodic boundary conditions for the su(2)$_4$ chain,
i.e., $x_{L}=x_{0}$.

We find that the differences in behavior between the IS and HIS su(2)$_4$ spin-$1$
chains are rather subtle. We will first describe
the behavior of the model in the IS sector, followed by a discussion of the
HIS sector.

As a first minor difference, we note that
the number of states in the HIS is given by $2^{L}+\delta_{L,0}$, where $L$ is the length of 
the chain. In the IS sector, however, the
number of states is $2^L+1$ when $L>0$ is even and $2^L-1$ when $L$ is odd.
The additional state in the even-$L$ IS 
occurs at momentum $K=\pi$,  while the additional state in the odd-$L$ HIS occurs at momentum $K=0$.
Those are the only differences; the remaining $2^L$ ($2^L-1$) states where $L$ even (odd)
have the same momenta in the integer and half-integer sectors.

As we did for $k\geq 5$, we represent the Hamiltonian of the su(2)$_4$ spin-$1$ chain
in terms of the
projectors onto the $1$ and $2$ channels with couplings $J_1$ and $J_2$, respectively.
These couplings are parametrized by an angle
$\theta_{2,1}$ where $J_2 = \cos\theta_{2,1}$ and $J_1 = -\sin\theta_{2,1}$. Explicitly,
the Hamiltonians read
\begin{align}
\label{k4_spin1_IS}
H^{(k=4)}_{\rm IS} &= \sum_i \cos\theta_{2,1} P^{(2)}_{i,{\rm IS}} - \sin\theta_{2,1} P^{(1)}_{i,{\rm IS}}
\\
\label{k4_spin1_HIS}
H^{(k=4)}_{\rm HIS} &= \sum_i \cos\theta_{2,1} P^{(2)}_{i,{\rm HIS}} - \sin\theta_{2,1} P^{(1)}_{i,{\rm HIS}} \ .
\end{align}

The explicit form of the projectors are given in appendix~\ref{spin1_chain_k=4}.

\subsection{Phase diagram in the integer Hilbert space sector (IS)}		

The phase diagram of the IS su(2)$_4$ spin-$1$ chain (Hamiltonian given in eq.~\eqref{k4_spin1_IS}) is shown in the
left most panel of Fig.~\ref{PD_spin1_k4}.
The phase diagram consists of two extended gapped phases which are separated
by two extended gapless regions. The two phase transitions between the
gapped phase with a $Z_3$-sublattice structure and the two gapless regions
are first order. However, the phase transitions into the gapped
phase with a $Z_2$-sublattice structure are continuous.

The critical behavior of the gapless regions is
described by the $Z_2$ orbifold theory of the u(1)-compactified boson with central
charge $c=1$. Interestingly, the compactification radius varies continuously
as a function of $\theta_{2,1}$ in the gapless regions. We found it difficult to determine
the range of compactification radii which are realized in the model. The reason is
that the finite size data makes it difficult to  determine the location of the transition between the gapped phase
with the $Z_2$-sublattice structure and the critical regions. We will devote a separate subsection \ref{su2k4transitions}
to the issue of the location of these phase boundaries, dealing with the IS and the HIS
at the same time.

\subsubsection{Gapped phases (IS)}

{\em Gapped phase $\theta_{2,1}=\pi/2$  --}
The gapped phase containing the point $\theta_{2,1}=\pi/2$
extends from $\theta_{2,1}= 0$ to $\theta_{2,1} = 3\pi/4$. These
phase boundaries are easy to locate because the transitions to the gapless
regions are first order transitions, as we will show in section \ref{su2k4transitions}.

This gapped phase has a $Z_3$ sublattice symmetry, which results in a
three-fold degenerate ground state for system sizes
 that are a multiple of $3$. These ground states occur at momenta $K=0,2\pi/3,4\pi/3$ and
their exact form can be established throughout the whole gapped phase.

At angle $\theta_{2,1} = \pi/2$, the Hamiltonian can be solved exactly. At this point,
the Hamiltonian reduces to the equal sum of two projectors, namely
onto the spin-$0$ and spin-$2$ channels (in addition, there is also a constant term $-L$).
Throughout the region $0\leq \theta_{2,1} \leq 3\pi/4$, the Hamiltonian is a sum of two projectors
with positive coefficients (for a matrix representation of the Hamiltonian see
appendix~\ref{spin1_chain}).
The three degenerate ground states are build from the basis state of the
form $| 1 1 a_1\, 1 1 a_2\, 1 1 a_3\, \ldots, 1 1 a_{L/3}\rangle$ and its two
translations, where the $a_i$ represent the states
$|a\rangle_{3i} = (|0\rangle - |2\rangle)/\sqrt{2}$ at position $3i$.
These three states can easily be combined to form three momentum eigenstates.
These ground states have energy $-L$ and are eigenstates of the
two projectors with eigenvalue $0$. The latter explains that these ground states
persist throughout the whole gapped phase.

{\em Gapped phase $\theta_{2,1}=3\pi/2$ --}
In the gapped phase surrounding the point $\theta_{2,1}=3\pi/2$, the spectrum exhibits a $Z_2$ sublattice symmetry
and a cosine-shaped quasi-particle dispersion. For even-$L$ system sizes, the ground
state is threefold degenerate, with the ground states occurring at momenta $K=0,0,\pi$. Two of the three ground states at angle
$\theta_{2,1} = 3\pi/2$ consist of states of the form $| 1 b_1\, 1 b_2 \ 1 b_3 \, \ldots, 1 b_{L/2} \rangle$ and its translation by
one site, where $| b \rangle_{2i} = (|0\rangle + |2\rangle)/\sqrt{2}$ at site $2i$.
These two state can be combined to form the two ground states at momenta $K=0,\pi$. The state
$| 1 1 1 \ldots 1\rangle$ is the third ground state and has momentum $K=0$. For odd system
sizes, this state is the only ground state of the system.\\

\begin{table*}[t]
\begin{tabular}{c||c|c|c|c|c|c|c||c||c|c|c|c} 
\hspace{1cm}$p$   &$1$ & $2$  & $3$&  $4$ & $6$ &$9$ &$10$& Y & IS & IS& HIS&HIS  \\
 & KT& (Ising)$^2$&pCFT& Potts&  sCFT& & &Top. & $Z_2$ & $Z_4$&$Z_2$ & $Z_4$ \\
$h+\bar{h}$&  & & & & && & &$K$ & $K$ & $K$ & $K$ 
 \\ \hline
\raisebox{-2mm}{\rule{0cm}{5mm}}
$0$ & $0$& $0$& $0$& $0$ & $0$& $0$&$0$ & $y_0$ & $0$ & $0$ & $0$&$0$  \\
\hline 
\raisebox{-2mm}{\rule{0cm}{5mm}}
$\frac{1}{8}$ & $\frac{1}{8}$& $\frac{1}{8}$&$\frac{1}{8}$& $\frac{1}{8}$& $\frac{1}{8}$ & $\frac{1}{8}$&$\frac{1}{8}$& $y_{1/2}$& $0$ & $\frac{\pi}{2}$ & $0$ & $\frac{\pi}{2}$  \\
\hline 
\raisebox{-2mm}{\rule{0cm}{5mm}}
$\frac{1}{8}$ &$\frac{1}{8}$& $\frac{1}{8}$&$\frac{1}{8}$&$\frac{1}{8}$&$\frac{1}{8}$&$\frac{1}{8}$&$\frac{1}{8}$& $y_{1/2}$ & $\pi$ &  $\frac{3\pi}{2}$ &  $\pi$ &  $\frac{3\pi}{2}$  \\ 
\hline 
\raisebox{-2mm}{\rule{0cm}{5mm}}
$\frac{9}{8}$&$\frac{9}{8}$&$\frac{9}{8}$&$\frac{9}{8}$&$\frac{9}{8}$&$\frac{9}{8}$&$\frac{9}{8}$&$\frac{9}{8}$& $y_{1/2}$& $0$   &  $\frac{\pi}{2}$  &  $0$   &  $\frac{\pi}{2}$\\
\hline 
\raisebox{-2mm}{\rule{0cm}{5mm}}
\raisebox{-2mm}{\rule{0cm}{5mm}}
$\frac{9}{8}$ &$\frac{9}{8}$&$\frac{9}{8}$&$\frac{9}{8}$&$\frac{9}{8}$&$\frac{9}{8}$&$\frac{9}{8}$&$\frac{9}{8}$&$y_{1/2}$ & $\pi$ &  $\frac{3\pi}{2}$  & $\pi$ &  $\frac{3\pi}{2}$\\
\hline 
\raisebox{-2mm}{\rule{0cm}{5mm}}
$2$  & $2$  &$2$ & $2$ &  $2$ & $2$&$2$& $2$&$y_0$  &  $0$  &  $0$       &  $0$  &  $0$   \\
\hline 
\raisebox{-2mm}{\rule{0cm}{5mm}}
$\frac{1}{2p}$& - &$\frac{1}{4}$&$\frac{1}{6}$ & $\frac{1}{8}$& $\frac{1}{12}$&$\frac{1}{18}$&$\frac{1}{20}$ &$y_1$ & $\pi$  & $\pi$ & $0$ & $\pi$ \\ 
\hline 
\raisebox{-2mm}{\rule{0cm}{5mm}}
$\frac{4}{2p}$&-&-&$\frac{2}{3}$&$\frac{1}{2}$& $\frac{1}{3}$&$ \frac{2}{9}$&$\frac{1}{5}$& $y_1$ &  $0$& $0$    & $ 0$ & $0$  \\
\hline  
\raisebox{-2mm}{\rule{0cm}{5mm}}
 $\frac{9}{2p}$& -&-&-&$\frac{9}{8}$& $\frac{3}{4}$&$\frac{1}{2}$&$\frac{9}{20}$ &$y_0$ & $\pi$ & $\pi$ & $0$ & $\pi$\\ 
\hline 
\raisebox{-2mm}{\rule{0cm}{5mm}}
$\frac{16}{2p}$& -& -&-&-& $\frac{4}{3}$&$\frac{8}{9}$&$\frac{4}{5}$&   $y_1$ & $0$ & $0$ & $0$ & $0$ \\
\hline  
\raisebox{-2mm}{\rule{0cm}{5mm}}
$\frac{25}{2p}$&- &-&-&-&$\frac{25}{12}$&$\frac{25}{18}$&$\frac{5}{4}$& $y_1$ &$\pi$& $\pi$& $0$ & $\pi$ \\ 
\hline 
\raisebox{-2mm}{\rule{0cm}{5mm}}
 $\frac{36}{2p}$&- &-&-&-& - & $2$ &$\frac{9}{5}$  &$y_0$ & $0$ &$0$ &$0$ &$0$   \\
\hline 
\raisebox{-2mm}{\rule{0cm}{5mm}}
$\frac{p}{2}$&$\frac{1}{2}$   &$1$  &$\frac{3}{2}$  &$2$   & $3$ &$9/2 $ &$ 5  $  &$y_0$,$y_1$ &$0$  &$0$  &$0$ & $\pi$ \\
\hline  
\raisebox{-2mm}{\rule{0cm}{5mm}}
 $\frac{p}{2}$&$\frac{1}{2}$&$1$    &$\frac{3}{2}$
 &$2$  & $3$  & $ 9/2$&$ 5 $ &$y_0$,$y_1$ &$0,\pi$  &$\pi$ &$\pi$  & $0$,$\pi$ \\ 
\end{tabular}
\caption{The scaling dimensions ($h+\bar{h}$) of the operators of the 
$Z_2$ orbifold of the compactified boson on a circle of radius $R=\sqrt{2p}$ for
some integer $p$.
The following abbreviations are used:
sCFT = the (minimal) superconformal CFT with central charge $c=1$,
Potts = $4$-state Potts CFT,
pCFT = $Z_4$ parafermion CFT,
(Ising)$^2$ = square of the Ising CFT,
KT = Kosterlitz-Thouless theory, equivalent to the compactified boson theory u(1)$_8$.
We also list the numerically observed topological quantum numbers
($Y$-symmetry: $y_0=y_2=2$, $y_{1/2} = y_{3/2} = 0$, $y_1=-1$) and momentum quantum
numbers $K$ at which the fields appear in the various critical regions.
The symmetry sectors of the fields with scaling dimensions $p/2$ depend on $p$.
This is a consequence of the fact that the field with scaling dimension
$(p-1)^2/2p$ at radius $p$ corresponds to the field with dimension $p/2$ at radius $p-1$.
}
 \label{comp_boson_table}
\end{table*}

\subsubsection{Gapless phases (IS)}\label{gapless_spin1_k4_IS}

The critical behavior of the su(2)$_4$ spin-1 chain is particularly interesting. We find that the
critical behavior depends continuously on the angle describing the interaction. At particular
values of the angle $\theta_{2,1}$, the behavior matches
particular CFTs with central charge $c=1$. In particular, these CFTs are 
the $Z_2$ orbifolds of a boson compactified on a circle of radius $R=\sqrt{2p}$.
For $p$ integer, these are rational conformal field theories \cite{ginsparg,dijkgraaf}, described in  detail in the appendix \ref{App:orbifolds}. In this section,
we will limit the discussion to the most prominent features of these theories.
In section \ref{su2k4transitions}, we will point out the particular orbifold theories that are
realized in the su(2)$_4$ spin-1 chain.

To identify the critical theories describing the critical behavior as a function of the
angle, we employ the standard technique of first shifting the spectrum such that the
ground state has zero energy, followed by a rescaling of the energy to elucidate the
conformal nature of the spectrum.

By means of this procedure, we identified several of the $c=1$ orbifold theories. These theories `share' several operators that appear
in the spectrum throughout the critical region. These operators are the
ground state with $h_0 = \bar{h}_0 = 0$, two twist fields $\sigma_{1,2}$ with
scaling dimension $h_\sigma+\bar{h}_\sigma = 1/8$, two twist fields $\tau_{1,2}$ with
scaling dimension $h_\tau+\bar{h}_\tau = 9/8$, a field $\Theta$ with scaling dimension
$2$, and finally, two fields $\Phi_{1,2}$ with scaling dimension $p/2$. For $p=1$, the fields
just described exhaust the full list, but in general, there are $p-1$ additional fields $\phi_\lambda$
with scaling dimension $\frac{\lambda^2}{4p}$. These fields, as well as the associated momenta
and topological symmetry
sectors  are given in table~\ref{comp_boson_table}.
We checked that the assignments of the topological symmetry sectors are compatible with the
fusion rules of the orbifold CFTs (for details, see appendix~\ref{App:orbifolds}).
For various values of $p$, the orbifold theories are also known under specific names, such as 
the Kosterlitz-Thouless theory ($p=1$),
the theory of two decoupled Ising models ($p=2$),
the $Z_4$ parafermion CFT ($p=3$),
the $4$-state Potts model ($p=4$),
and the superconformal minimal model with $c=1$ ($p=6$).

\begin{figure*}[tb]
  \begin{center}
  \includegraphics[width=.48 \linewidth]{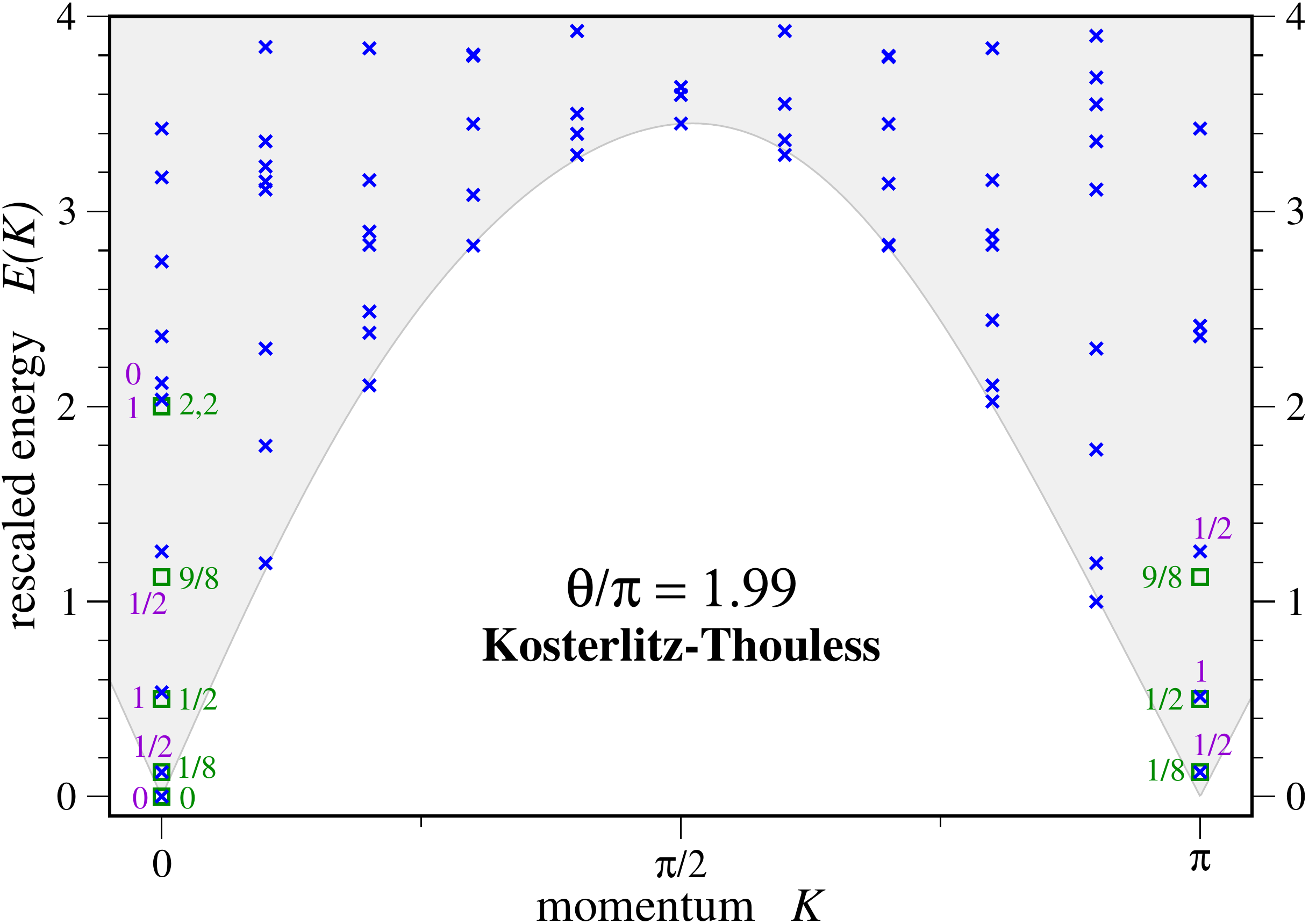} 
  \hskip 0.03 \linewidth
  \includegraphics[width=.48 \linewidth]{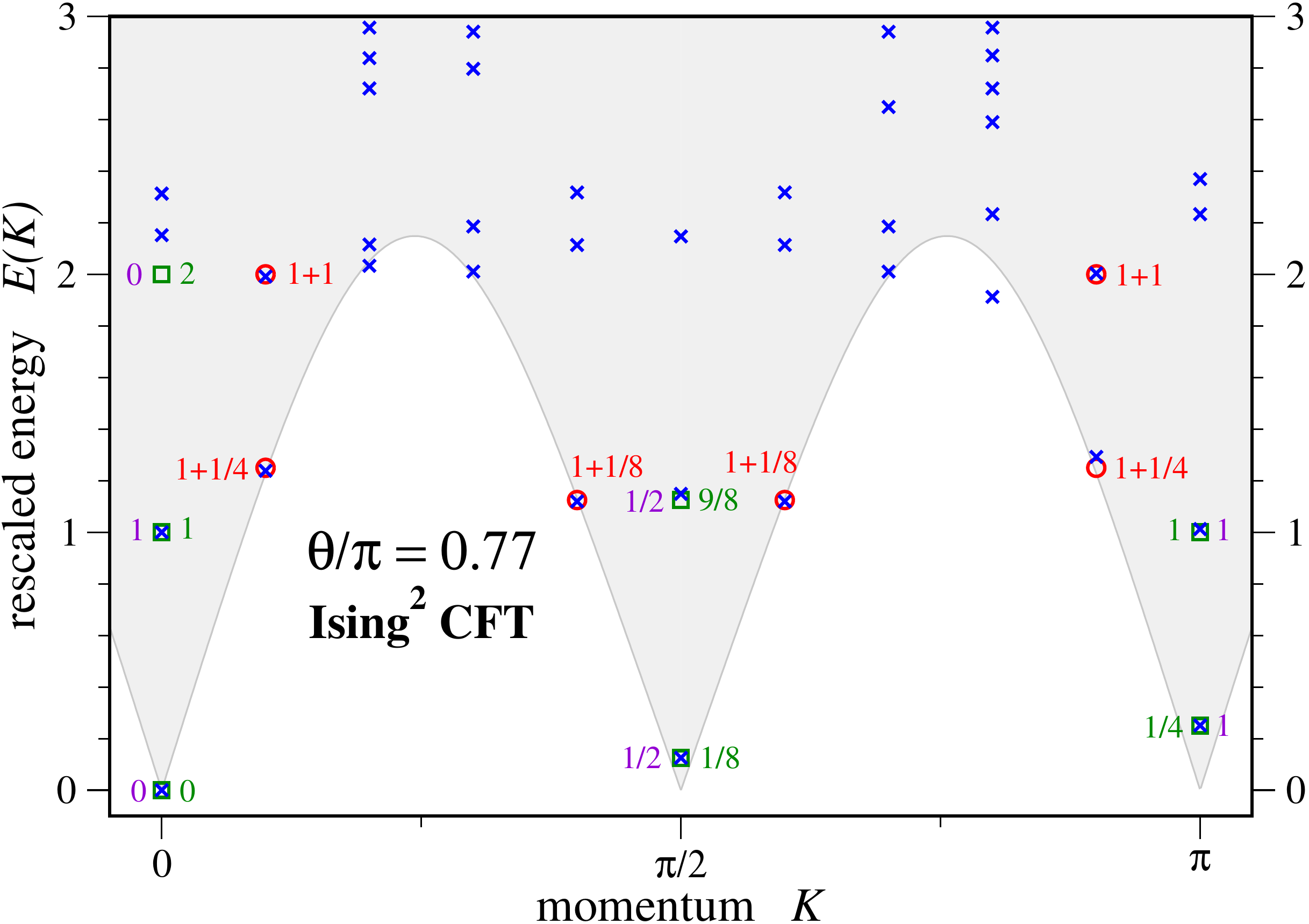} 
  \vskip 5mm
  \includegraphics[width=.48 \linewidth]{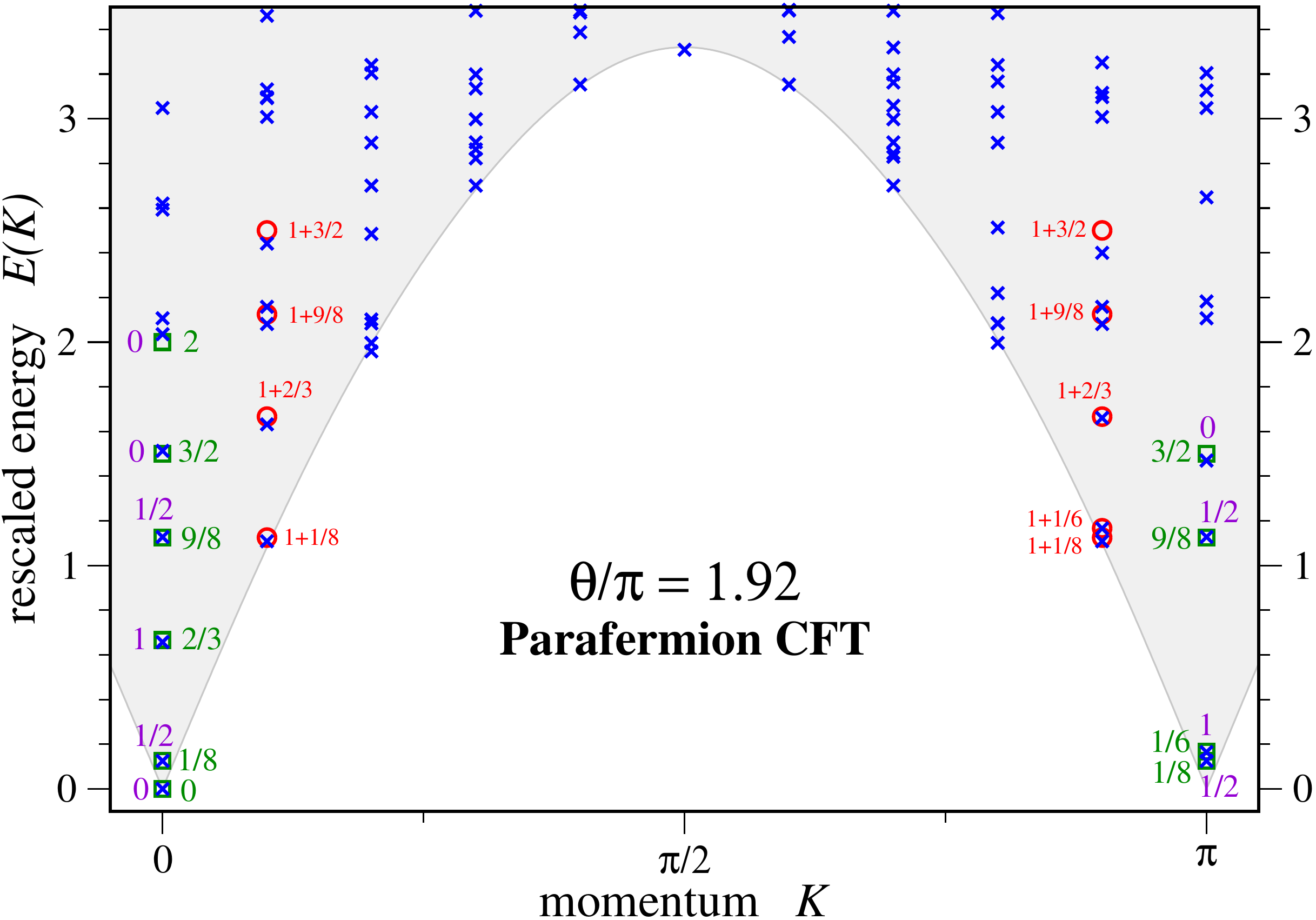} 
  \hskip 0.03 \linewidth
  \includegraphics[width=.48 \linewidth]{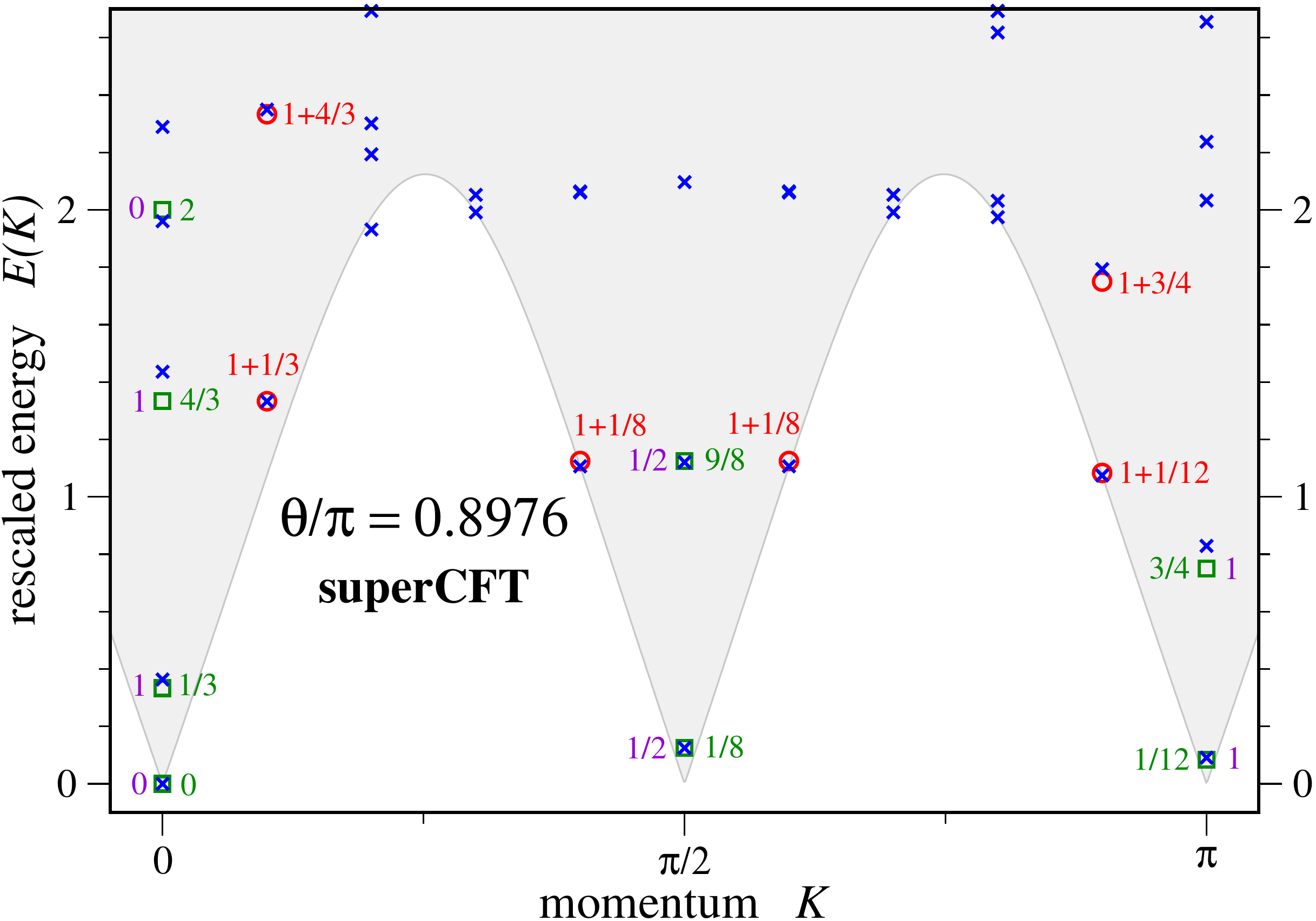} 
     \caption{(color online)
                   {\bf The su(2)$_4$ chain -- integer sector:} Energy spectra at various  in the gapless phases
                   of the phase diagram in Fig.~\ref{PD_spin1_k4}a).
                   The energy spectra have been rescaled to match the conformal field theory prediction 
                   given in Eq.~\eqref{CFT_energy_levels}. 
                   Green squares indicate the location of the primary fields, red circles the descendant fields.
                   The topological symmetry sector is indicated by the violet index. 
                   Data shown are for system size L = 20.
                   }
     \label{Fig:su(2)_4-IS}
  \end{center}
\end{figure*}

We identified several of the $c=1$ $Z_2$ orbifold theories, including the ones
with $p=1,2,3,4,5,6$. In the left side of Figure~\ref{Fig:su(2)_4-IS}, we show the energy spectra associated with the $p=1$  and $p=3$ orbifold theories in the $Z_2$ critical region.
 In the right side of Figure~\ref{Fig:su(2)_4-IS}, we display   the energy spectra associated with the $p=2$  and $p=6$ orbifold theories in the $Z_4$ critical region.

\begin{table}[ht]
\begin{tabular}{r|c|l|l}
$p$ & theory & \multicolumn{1}{c|}{$Z_4$} & \multicolumn{1}{c}{$Z_2$} \\
\hline
$1$ & Kosterlitz-Thouless & $0.755\pi^*$ & $-0.01 \pi$ \\
$2$ & Ising$^2$                  & $0.77\pi$   & $-0.04 \pi$ \\
$3$ & parafermion             & $0.80\pi$ & $-0.08 \pi$ \\
$4$ & 4-state Potts            & $0.83\pi$   & $-0.13 \pi$ \\
$5$ & & $0.88\pi$ & $-0.17\pi$ \\
$6$ & superCFT               & $0.92\pi$ & $-0.20 \pi$ \\
$7$ & & $0.96\pi$ & $-0.23\pi^*$ \\
$8$ & & $0.98\pi$ & $-0.24\pi^*$
\end{tabular}
\caption{The approximate locations of some of the critical
theories of the su(2)$_4$ spin-1 chain \eqref{k4_spin1_IS}
in the integer sector (IS) are listed for both the $Z_2$ and the $Z_4$ critical regions.
The angles without asterisk are obtained directly from exact diagonalization for $L=20$,
i.e. we matched the momentum resolved spectrum to the CFT.
The angles with an asterisk were obtained by using the relations between the angles
$\theta_2$ and $\theta_4$, as explained in the text.
We only list those values of $p$ for which we could match the CFT description
beyond any doubt.
}
\label{Tab:su2_4-IS}
\end{table}

In table~\ref{Tab:su2_4-IS}, we list the locations of some of the critical points as
extracted from the numerical data. The procedure we followed to obtain these locations
will be described in more detail in section \ref{su2k4transitions}.
The location $\theta \approx - 0.20 \pi$ of case $p=6$ - the superconformal
theories - is very close to the location of the superconformal point
for the su(2)$_k$ spin-1 chains with $k\geq 5$, namely
$\theta_{2,1} \approx -0.19 \pi$.

The location of the superconformal point
in the $Z_4$ critical region is $\theta_{2,1} \approx 0.92\pi$. In general, the relation
between critical angle in the $Z_2$ critical region (which we will for now denote by
$\theta_2$, similarly, $\theta_4$ denotes the angle in the $Z_4$ critical region)  is
\begin{align}
\theta_4 &= \pi - \tan^{-1} (1 + \tan \theta_2) & 
\theta_2 &= - \tan^{-1} (1 + \tan \theta_4) \ .
\label{eq:th2-vs-th4}
\end{align}

The spectra in Figure~\ref{Fig:su(2)_4-IS} illustrate the different sublattice
symmetry for the two gapless regions. In these spectra, we also indicate the
topological symmetry sectors of some of the low-lying states. In the
case of su(2)$_4$, the topological symmetry operator $Y$ has three distinct
eigenvalues, which are given by $y_0 = y_2 = 2$, $y_{1/2} = y_{3/2} = 0$ and
$y_1 = -1$. We will thus use the labels $y=0,1/2,1$ for these sectors. 

The presence of the different critical models with the same central charge
$c=1$ indicates the presence of a marginal operator that drives the
`transition' between the different critical theories and that gives rise to continuously
varying critical exponents. Indeed, all the orbifold models share a marginal
operator $\Theta$ with scaling dimension $2$ whose  topological
symmetry coincides with that of the ground state. It is this operator which is responsible for the
critical region with continuously changing exponents.
It proved difficult to locate the phase transition between the
critical regions and the gapped phase around $\theta_{2,1}=3\pi/2$.
One reason might be that the transition to the gapped phase is also driven
by a marginal operator, which allows for large
finite size effects that thwarts the localization of these critical points.


\subsection{Phase diagram in the half-integer Hilbert space sector (HIS)}	

The behavior of the su(2)$_4$ spin-$1$ chain in the half-integer sector
mimics very closely that of the integer sector. The phase diagram is presented in
the rightmost panel in Fig~\ref{PD_spin1_k4}. The phase boundaries are
located at the same positions, but the details of the observed
phases differ slightly. In the following discussion of the HIS su(2)$_4$ spin-$1$ chain, we will emphasize the differences between the two 
sectors. 


\subsubsection{Gapped phases (HIS)}

As was already noted above, there are some differences in the dimensions of the  Hilbert spaces  in the IS and the HIS, respectively. As a consequence, the IS and HIS models have  different sublattice
structures in the gapped phases. Namely, in the half-integer sector, the
ground state occurs at momentum $K=0$, and it is non-degenerate.
All other features of the gapped phases in the half-integer sector are very similar to those  observed in the integer sector. 

{\em Gapped phase $\theta_{2,1}=\pi/2$  --} 
In gapped phase that surrounds the angle $\theta_{2,1}$, the ground state is non-degenerate and occurs at momentum $K=0$ (there is no sublattice structure).
The model can be solved at angle  $\theta_{2,1} =\pi/2$:
the ground state can be expressed as follows,
$$
| GS \rangle = \sum_{x_{i}=1/2,3/2} (-1)^{\# (3/2,3/2)} | x_0,x_2,\ldots, x_{L-1} \rangle \ .
$$
Here,  $\#(3/2,3/2)$ denotes the number of times the sequence
$(x_{i},x_{i+1})=(3/2,3/2)$ occurs in the state $|x_0,x_2,\ldots,x_{L-1}\rangle$ (note that periodic boundary conditions impose $x_{L}=x_0$). As was the case for the ground state(s) at
$\theta_{2,1} = \pi/2$ in the IS, this state is in fact the ground state throughout the whole gapped
phase, i.e. for angle $0\leq \theta_{2,1} \leq 3\pi/4$.

{\em Gapped phase $\theta_{2,1}=3\pi/2$  --}
In the gapped phase that surrounds the angle $\theta_{2,1}=3\pi/2$, the ground state is non-degenerate and occurs at momentum $K=0$ (there is no sublattice symmetry).
At $\theta_{2,1} = 3\pi/2$, the ground state is given by
$$
| GS \rangle = \sum_{x_{i}=1/2,3/2} (-1)^{\# (1/2,3/2)} | x_0,x_2,\ldots, x_{L-1} \rangle \ .
$$
All basis states contribute to the ground state. The sign of a term is 
given by the number of times the sequence $(x_{i},x_{i+1})=(1/2,3/2)$ occurs in the
basis state $|x_0,x_2,\ldots,x_{L-1}\rangle$
(periodic boundary conditions are assumed).

\begin{figure*}[t]
  \begin{center}
  \includegraphics[width=.48 \linewidth]{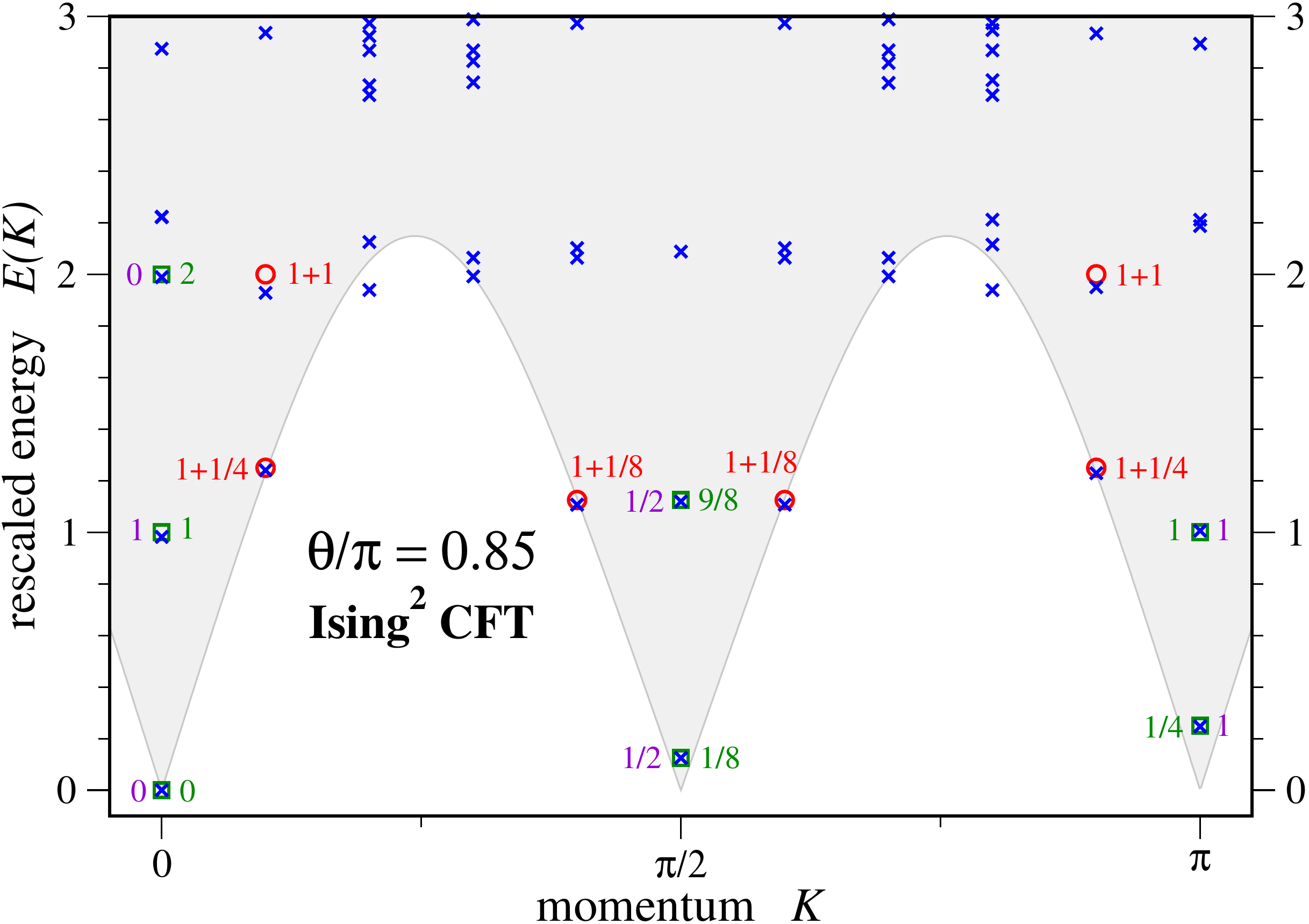} 
  \hskip 0.03 \linewidth
  \includegraphics[width=.48 \linewidth]{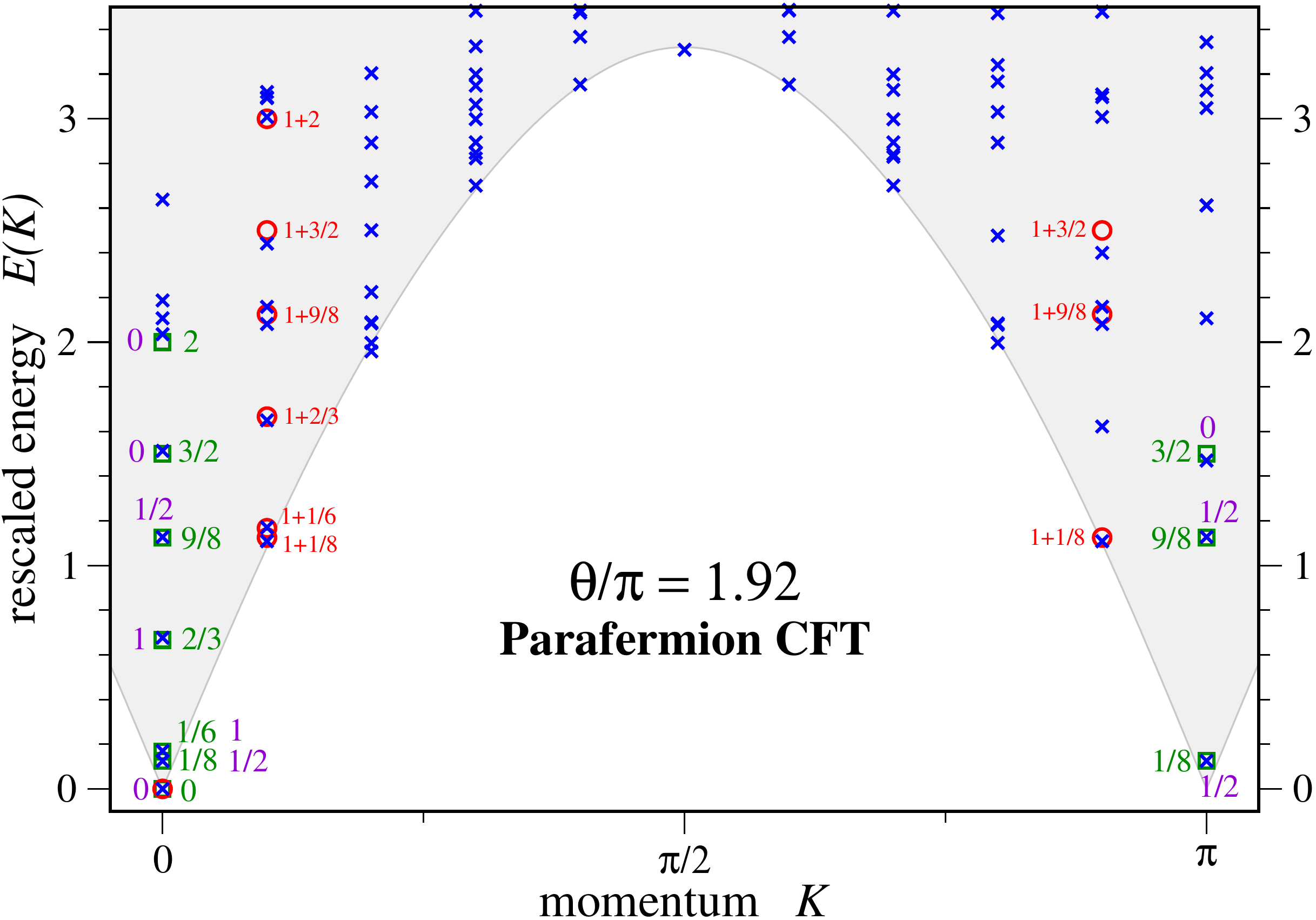} 
    \caption{(color online)
    	         {\bf The su(2)$_4$ chain -- half-integer sector:} 
	           Energy spectra at various points in the gapless phases
                   of the phase diagram displayed in Fig.~\ref{PD_spin1_k4}c).
                   The energy spectra have been rescaled to match the conformal field theory prediction 
                   given in Eq.~\eqref{CFT_energy_levels}. 
                   Green squares indicate the location of the primary fields, red circles the descendant fields.
                   The topological symmetry sector is indicated by the violet index. 
                   Data shown are for system size L = 20.}
     \label{Fig:su(2)_4-HIS}
  \end{center}
\end{figure*}
    
\subsubsection{Gapless phases (HIS)}

As in the integer sector, the phase diagram in the half-integer
sector has two extended regions where the model is critical. The criticality
is again described by $Z_2$ orbifold models.  
We identified the orbifold models
with parameters $p=2,\ldots,9$. Some of the critical angles are given in
table~\ref{Tab:su2_4-HIS}.

\begin{table}[ht]
\begin{tabular}{r|c|l|l}
$p$ & theory & \multicolumn{1}{c|}{$Z_4$} & \multicolumn{1}{c}{$Z_2$} \\
\hline
$2$ & Ising$^2$                  & $0.852\pi$   & $-0.148 \pi$ \\
$3$ & parafermion             & $0.795\pi$ & $-0.078 \pi$ \\
$4$ & 4-state Potts            & $0.774\pi$   & $-0.046 \pi$ \\
$5$ & & $0.766\pi$ & $-0.030\pi$ \\
$6$ & superCFT               & $0.761\pi$ & $-0.020 \pi$ \\
$7$ & & $0.758\pi^*$ & $-0.015\pi$ \\
$8$ & & $0.756\pi^*$ & $-0.011\pi$ \\
\end{tabular}
\caption{The approximate locations of some of the critical
theories of the su(2)$_4$ spin-1 chain
in the half integer sector (eq \eqref{k4_spin1_HIS})
for both the $Z_2$ and the $Z_4$ critical regions.
The angles without asterisk are obtained directly from exact diagonalization for $L=20$ by matching the momentum resolved spectrum to the CFT. The angles
with an asterisk were obtained by using the relations between the angles
$\theta_2$ and $\theta_4$, as given in eq.~\eqref{eq:th2-vs-th4}.
We only list those values of $p$ for which we were able to match the CFT description
beyond any doubt.
}
\label{Tab:su2_4-HIS}
\end{table}

The difference between the two gapless regions in the half-integer sector 
lies in the momentum quantum numbers, as indicated
in Table~\ref{comp_boson_table}. The topological symmetry sectors in the HIS coincide with those
 found in the IS  (see Table~\ref{comp_boson_table}).
While this is  to be expected for topological
quantum numbers, it nevertheless shows that our results are consistent.

A major distinction between the IS and the HIS  phase diagram of the su(2)$_4$ spin-$1$ chain  is the order of the orbifold theories.
By comparing the leftmost and the rightmost panels of Fig.~\ref{PD_spin1_k4}, it can be
seen that in the integer sector, the orbifold theories appear in ascending $p$ order when moving
away from the first order transition points,
while in the half-integer sector, the orbifold theories appear in descending $p$ order when moving
away from gapped phase I.
We exploit this result in locating the position of the critical endpoint of one of the the gapped phases (see following subsection). 

\subsection{The location of the phase boundaries}

\label{su2k4transitions}

To locate the boundaries of the gapped and critical regions of the su(2)$_4$ spin-1
chain, we consider the ground state energy as a function of the interaction
angle $\theta$. The analysis is most easily carried out
by using an alternative parametrization of the Hamiltonian. Two spin-1 anyons can
fuse into either a spin-0, a spin-1 or a spin-2 anyon; therefore we can write the Hamiltonian in terms
of projectors onto the spin-$2$ and spin-$0$ channels, instead of the
spin-$2$ and spin-$1$ channels as we did in eq.(\eqref{k4_spin1_IS}). By making use of the relation
$\idop = P^{(0)} + P^{(1)} + P^{(2)}$, we find that the Hamiltonian
\begin{equation}
H^{(k=4)}_{\rm J2-J0} = \sum_i \cos\theta_{2,0} P^{(2)}_{i} - \sin\theta_{2,0} P^{(0)}_{i}
\label{k4_spin1_J2-J0}
\end{equation}
is related to the Hamiltonian of equation \eqref{k4_spin1_IS}
\begin{equation}
\label{k4_spin1_J2-J1}
H^{(k=4)}_{\rm J2-J1} = \sum_i \cos\theta_{2,1} P^{(2)}_{i} - \sin\theta_{2,1} P^{(1)}_{i}
\end{equation}
via
\begin{equation}
\tan \theta_{2,1} = -\frac{\tan\theta_{2,0}}{1+\tan\theta_{2,0}} \ ,
\end{equation}
up to an unimportant shift in energy.

\begin{figure}[h]
\includegraphics[width=8.5cm]{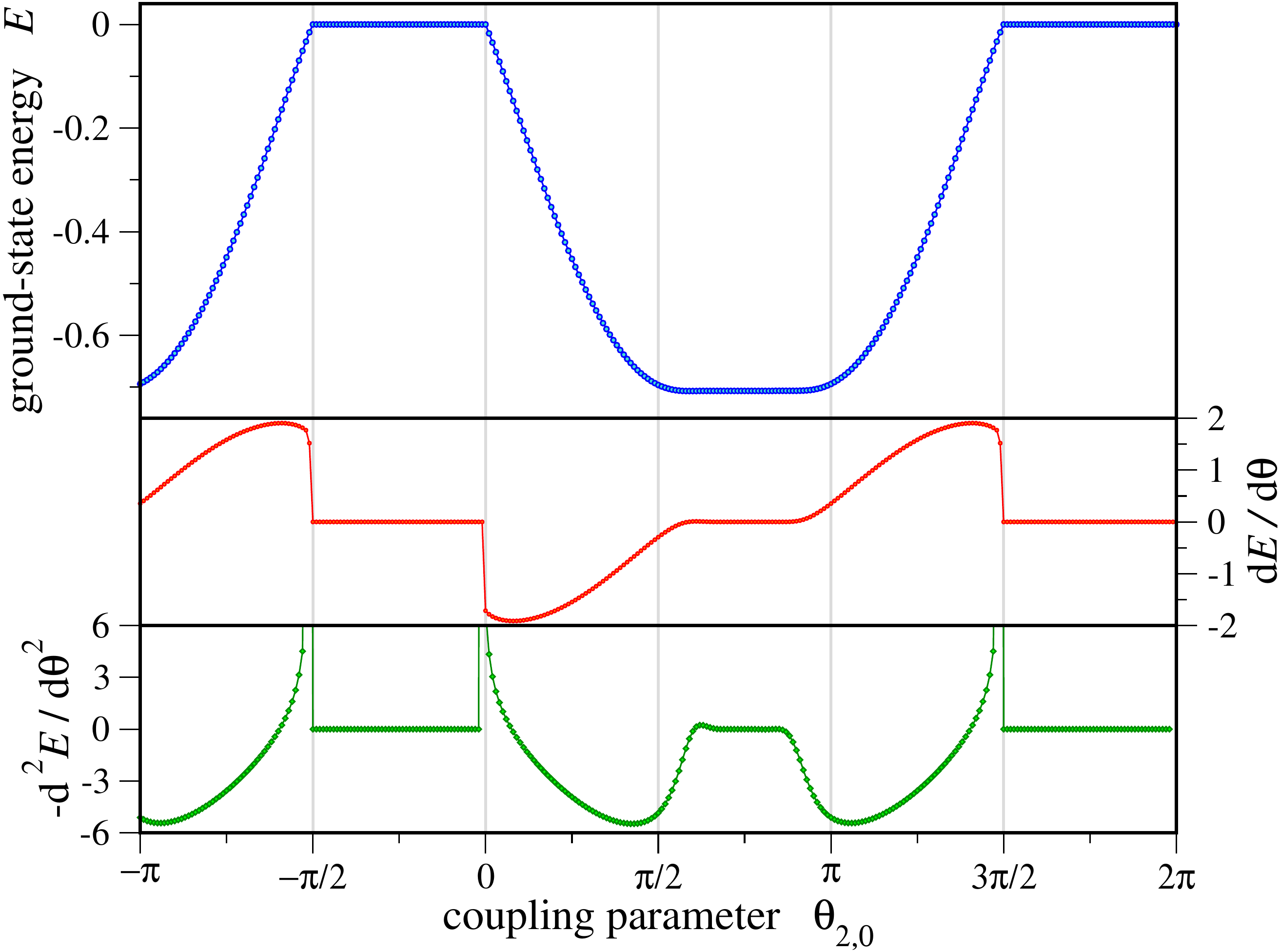}
\caption{(color online) The ground-state energy per site (upper panel) and its first and second derivative (middle and lower panel, respectively) of the IS su(2)$_4$ chain. Data shown is shown for system size $L=18$.}
\label{Fig:GS-energy-spin-1-IS}
\end{figure}

The ground state energy as a function of the angle $\theta_{2,0}$ is given in
figure~\ref{Fig:GS-energy-spin-1-IS} for a chain of size $L=18$.
The kinks in the ground state energy  indicate that there are
 two first order phase transitions. These first order
phase transitions mark the boundaries of the gapped phase located at  $-\pi/2 < \theta_{2,0} < 0$ in the new angle variable
$\theta_{2,0}$ ($0<\theta_{2,1}<3\pi/4$
in terms of the original variable $\theta_{2,1}$, see phase
diagram in Figure~\ref{PD_spin1_k4}).

\begin{figure}[t]
\includegraphics[width=8.5cm]{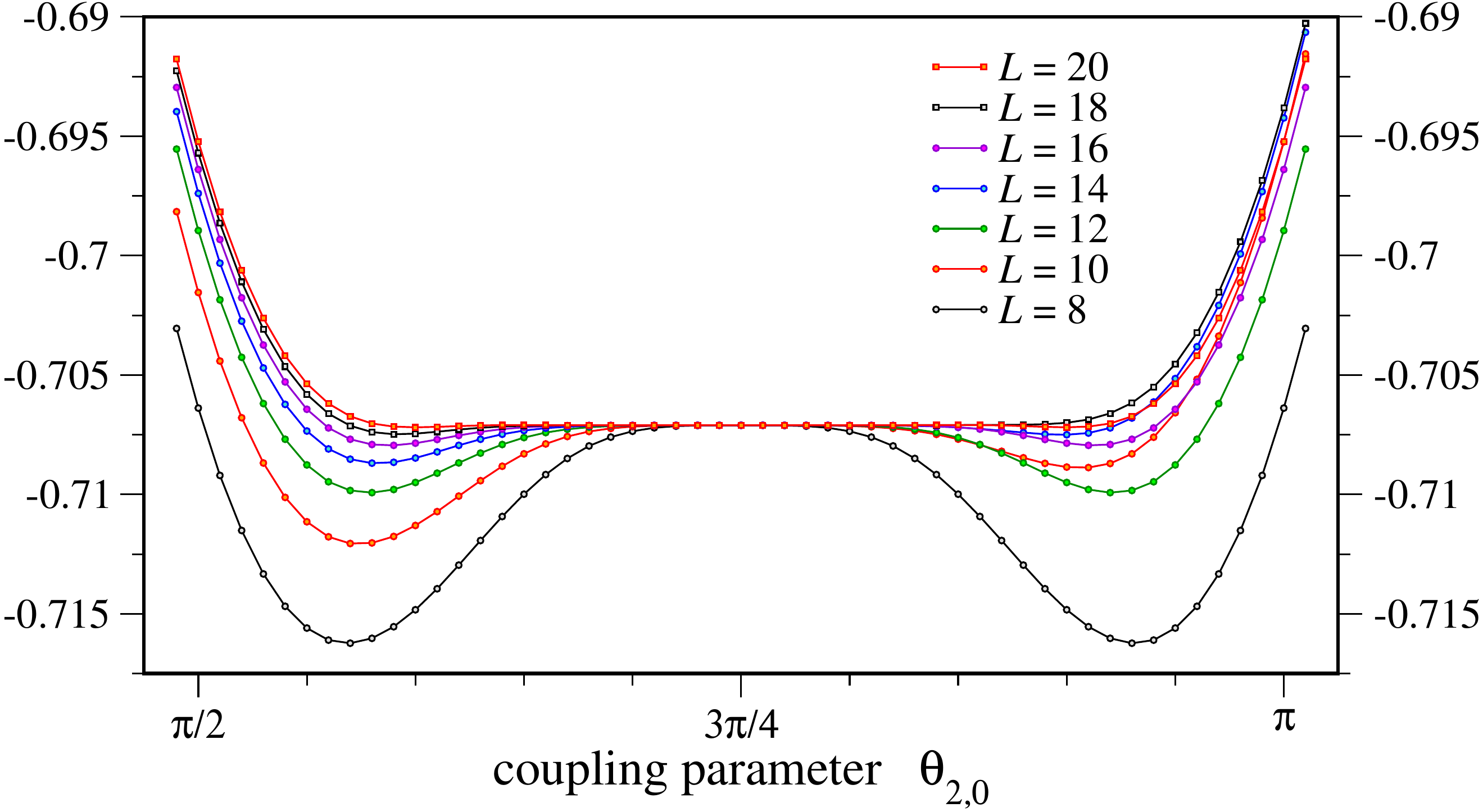}
\caption{(color online) The ground-state energy per site of
the IS su(2)$_4$ chai for system sizes
$L=8,\ldots,20$, in steps of two.}
\label{Fig:GS-energy-spin-1-IS-Ldep}
\end{figure}

To identify the location of the continuous transition between the other gapped phase and the 
neighboring gapless phases,  we 
plot the first and second derivatives of the ground state energy per site.
From these derivatives,  it can be concluded that these transitions are
roughly located at $\theta_{2,0} = \pi/2$ and $\theta_{2,0} = \pi$. In terms
of the original variable $\theta_{2,1}$, these locations correspond to
$\theta_{2,1} = -\pi/4$ and $\theta_{2,1} = \pi$. This conclusion is corroborated by
figure~\ref{Fig:GS-energy-spin-1-IS-Ldep}, in which we plot the
ground state energy in the gapped phase surrounding the angle
$\theta_{2,1} = 3\pi/2$, i.e. $\theta_{2,0} = 3\pi/4$
for system sizes ranging from $L=8$ to $L=20$.

In order to locate the phase boundaries, we
 also considered the structure of the orbifold CFTs describing
 the gapless phases (we refer to Appendix~\ref{App:orbifolds} for more details on the orbifold CFTs).
We know that throughout the critical region,  two fields  with scaling dimension $h+\bar{h} = 1/8$ and two
fields with scaling dimension $h+\bar{h} = 9/8$ must appear. In addition, there are several fields with scaling dimensions
$\lambda^2/(2p)$ ($\lambda=1,\ldots,p-1$) for some value of $p$.
Depending on the sector (IS or HIS), and depending on the 
critical region, these fields appear at different  momenta, as detailed 
 in table~\ref{comp_boson_table}. This table also includes our numerical results for the topological
symmetry sectors of the various fields.

The structure of the critical theories describing the critical region allows us to
numerically determine the value of $p$ as a function of the angle $\theta_{2,0}$. Moreover, in doing so, we will gain insight into the locations of the
phase boundaries. We proceed as follows: We first shift the spectrum such
that the ground state has energy zero, and we rescale the spectrum such that the
two degenerate lowest fields with topological eigenvalue $y=0$ have energy $1/8$.
Since these fields are always among the low-lying fields, finite size effects
 are insignificant.
 After shifting  and rescaling  the energy,
we focus on the two states corresponding to the fields with scaling dimensions
$1/(2p)$ and $4/(2p)$. By equating the numerical energies to the $p$-dependent
predictions from the conformal field theory, we obtain a numerical
estimate of $p$ as a function of the interaction angle. We note 
one has to be watchful of level crossings when using this procedure.

\begin{figure}[t]
  \begin{center}
  \includegraphics[width=\columnwidth]{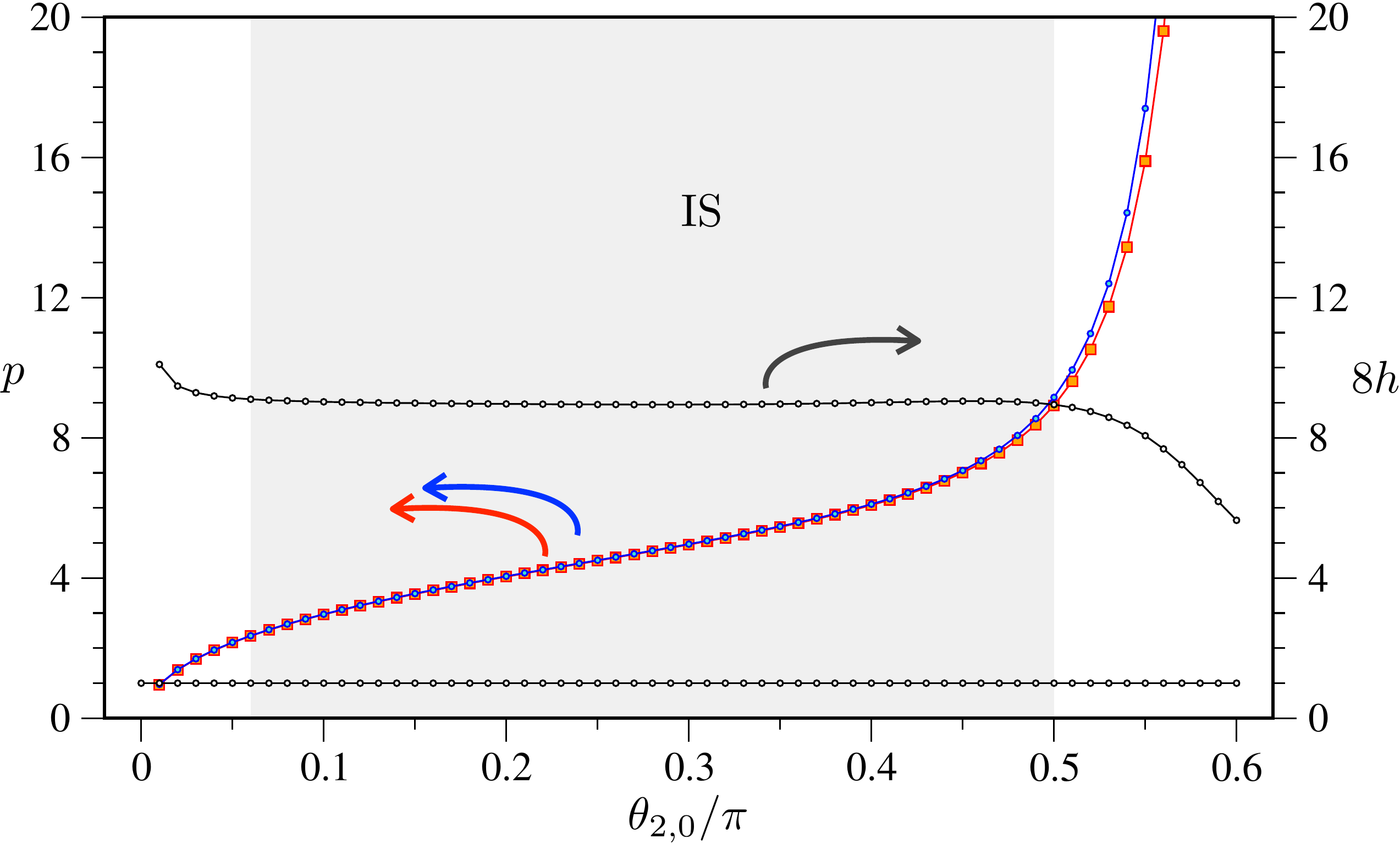}
  \caption{Numerical estimate (system size $L=20$) of parameter $p$ from the eigenenergies that are associated 
  with operators with scaling dimensions
   $1/2p$ (red squares) and $4/2p$ (blue dots) in the IS gapless phase
   with twist operators at $k=0$ and $k=\pi$.
  The parameter $p$ is about $1$ at $\theta_{2,0}=0$, and grows to about $9$ at
  $\theta_{2,0} = \pi/2$. For angles  $\theta_{2,0} >\pi/2$, the estimates of $p$ obtained from the
 two operators start to deviate.
  The black dots correspond
  the scaling dimension of the fields with dimensions $h=1/8$ and $h=9/8$ multiplied by eight, as
  obtained from exact diagonalization. The shaded region indicates the range of $\theta_{2,0}$
  for which the latter dimension lies between $8.9<8h<9.1$. 
 }
  \label{AT_int_Z2}
  \end{center}
  \end{figure}

\begin{figure}[t]
  \begin{center}
 \includegraphics[width=\columnwidth]{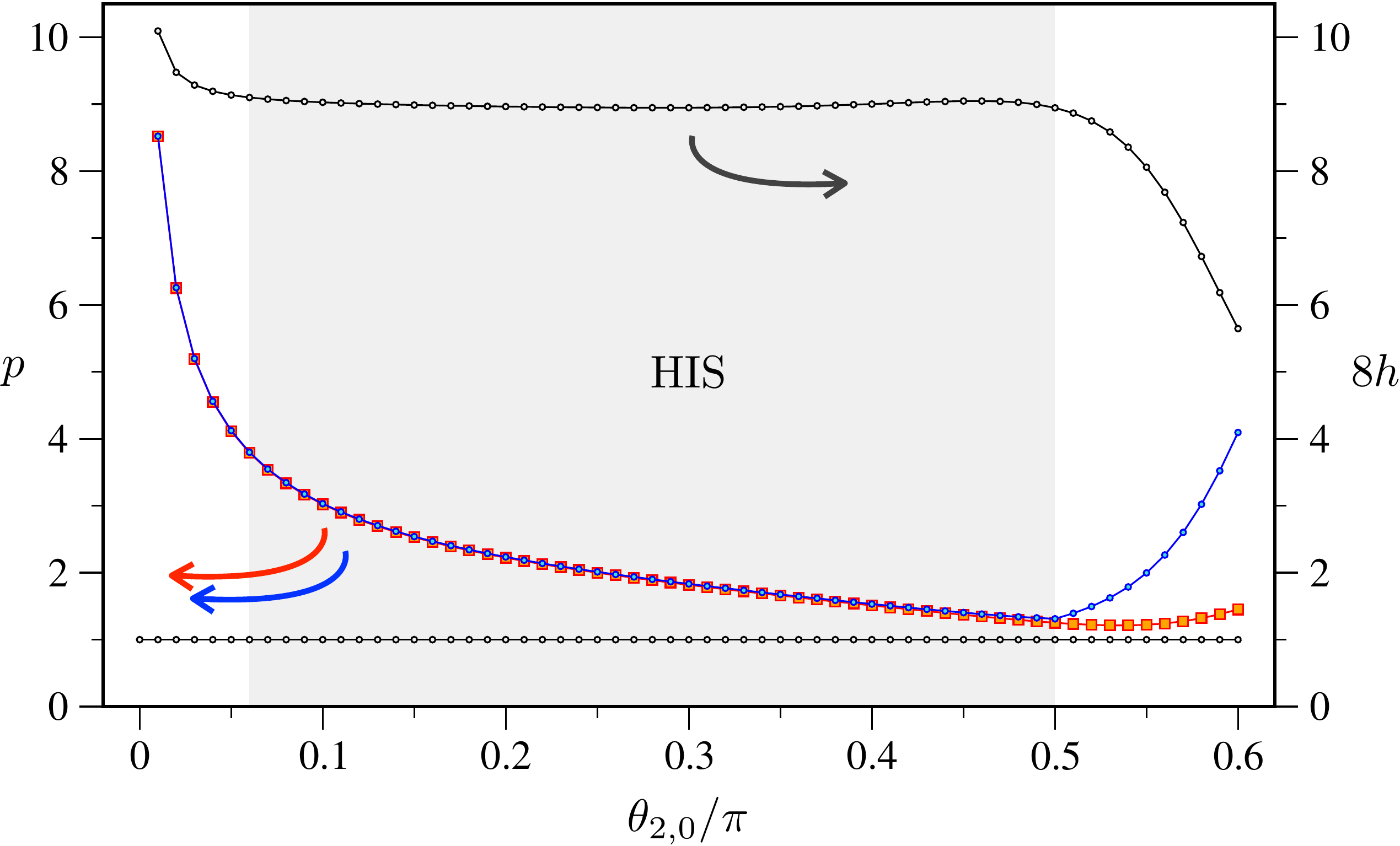}
  \caption{Numerical estimate (system size $L=20$) of parameter $p$ from the eigenenergies that are associated 
  with operators with scaling dimensions
   $1/2p$ (red squares) and $4/2p$ (blue dots) in the HIS gapless phase. The black dots correspond
   to the scaling dimension of the fields with dimensions $h=1/8$ and $h=9/8$ multiplied by eight, as
   obtained from exact diagonalization. The shaded region indicates the range of $\theta_{2,0}$
  for which the latter dimension lies between $8.9<8h<9.1$. 
    }
  \label{AT_half_Z2}
  \end{center}
  \end{figure}

In Figs.~\ref{AT_int_Z2} and \ref{AT_half_Z2}, we display the numerically obtained values for $p$ as a function of the angle for system size $L=20$. In these Figures,
we also show the energy of the state corresponding to the field with scaling dimension $9/8$.
 The range of angles $\theta_{2,0}$ over which the field with scaling dimension $9/8$ is constant is shaded in Fig.~\ref{AT_int_Z2}: the shaded region includes all angles for which the energy associated with the field multiplied by eight takes values between $8.9$ and $9.1$.
It is immediately apparent that the two independent numerical
estimates  of $p$ agree very well in the range $0<\theta_{2,0}<\pi/2$. This applies
 to both integer and half-integer sector.  In addition, the energy
of the state corresponding to the field with scaling dimension $9/8$ agrees very
well with the prediction over this range. Thus our numerical data is consistent with
the picture that the $Z_2$ critical region extends over the range
$0\leq \theta_{2,0} \leq \pi/2$, giving  way at $\theta_{2,0}=\pi/2$ to the gapped
phase with a $Z_2$ sub-lattice structure.

We do not include similar Figures for
the $Z_4$ critical region, but  note that they show very similar behavior. This leads to the location
of the boundaries of this critical region being $\theta_{2,0} = \pi$ and
$\theta_{2,0} = 3\pi/2$, the latter being the location of the first order transition.

We also studied  the various
values of $p$ which are realized in the su(2)$_4$ spin-1 model. 
 The region close to the first order transitions is most suitable for identifying the various orbifold models because of drastic changes in the spectrum in this region.
 Since the  $Z_2$ orbifold CFTs appear in opposite order in the 
integer and half-integer sectors, respectively, 
  both the low- and the
high-$p$ orbifold CFTs are observed near the first order phase transitions at $\theta_{2,0} = 0$.
In the IS, we identified the $p=1$ orbifold CFT (see Figure~\ref{Fig:su(2)_4-IS}), which 
 suggests that the gapless phases in the IS su(2)$_4$ chain include the orbifold CFTs starting at 
the lowest integer value $p=1$. 
In contrast, in the HIS, we were able to match the spectrum for the $p=2$ CFT, but we did not find evidence that the $p=1$ model exists in the phase diagram.
Moreover, we found that the values of $p$ in the HIS increase quite rapidly when decreasing $\theta_{2,0}$
to zero. 
We were  able to identify the orbifold CFTS  up to 
$p=9$. The reason is that the
abundance of low-lying (primary) fields in the high-$p$ orbifold theories  
requires large system sizes to identify these CFTs with sufficient accuracy.

Further insight into the critical phases can be gained by considering the  topological sectors of the various operators in
the spectra for the integer values of $p$ (see Table~\ref{comp_boson_table}). All orbifold CFTs (i.e., all $p$) include  a marginal operator with conformal dimension $h=2$. This marginal operator has
 momentum $K=0$ and topological quantum number $y_0$, i.e., it has the same
quantum numbers as the ground state. 
It is this marginal operator which causes the
continuously varying critical behavior within  the gapless phase. With increasing
 $p$, the number of  fields whose scaling dimensions are smaller than two
 increases. However, these fields are not relevant because their topological and/or momentum quantum numbers differ from those of the ground state.
The lowest-$p$ orbifold CFT for which there  exists  an additional marginal operator
with the same quantum numbers as the ground state is $p=9$. 
For general $p$, this
operator has scaling dimension $h+\bar{h} = 36/(2p)$ (see Table~\ref{comp_boson_table}).
The existence of an additional relevant operator for $p>9$ (as $36/2p$ surpasses two for $p>9$)
 suggests  that the range of $p$ values which are realized in our model is $p=1,2,\ldots,9$.
Moreover, if the marginal operator which first appears for $p=9$ is indeed the operator which is
driving the phase transition, it is not surprising that the location of the continuous phase
transition is hard to determine.

In conclusion, we  provide evidence that the orbifold CFTs with 
$p=2,3,\ldots,9$ are realized in both IS and HIS of the su(2)$_4$ spin-$1$ anyon chain, while
the  $p=1$ orbifold CFT appears only in the IS. 
We note that the su(2)$_4$ anyonic spin-$1$ chain has some similarities with the
one-dimensional quantum Ashkin-Teller model\cite{Kohmoto}.
The one-dimensional quantum Ashkin-Teller model, which is an
anisotropic version of the two-dimensional Ashkin-Teller model\cite{Ashkin_Teller}, 
also has a line of critical
points on its self-dual line, realizing the orbifold CFTs with $p=1,2,3,4$, in addition to two
gapped phases, one of which has a $Z_2$ sub-lattice structure.

\section{Anyonic su(2)$_k$ spin-$1/2$ chains}
\label{spin_half_chain}

\begin{figure*}[ht]
\begin{center}
\includegraphics[width=\columnwidth]{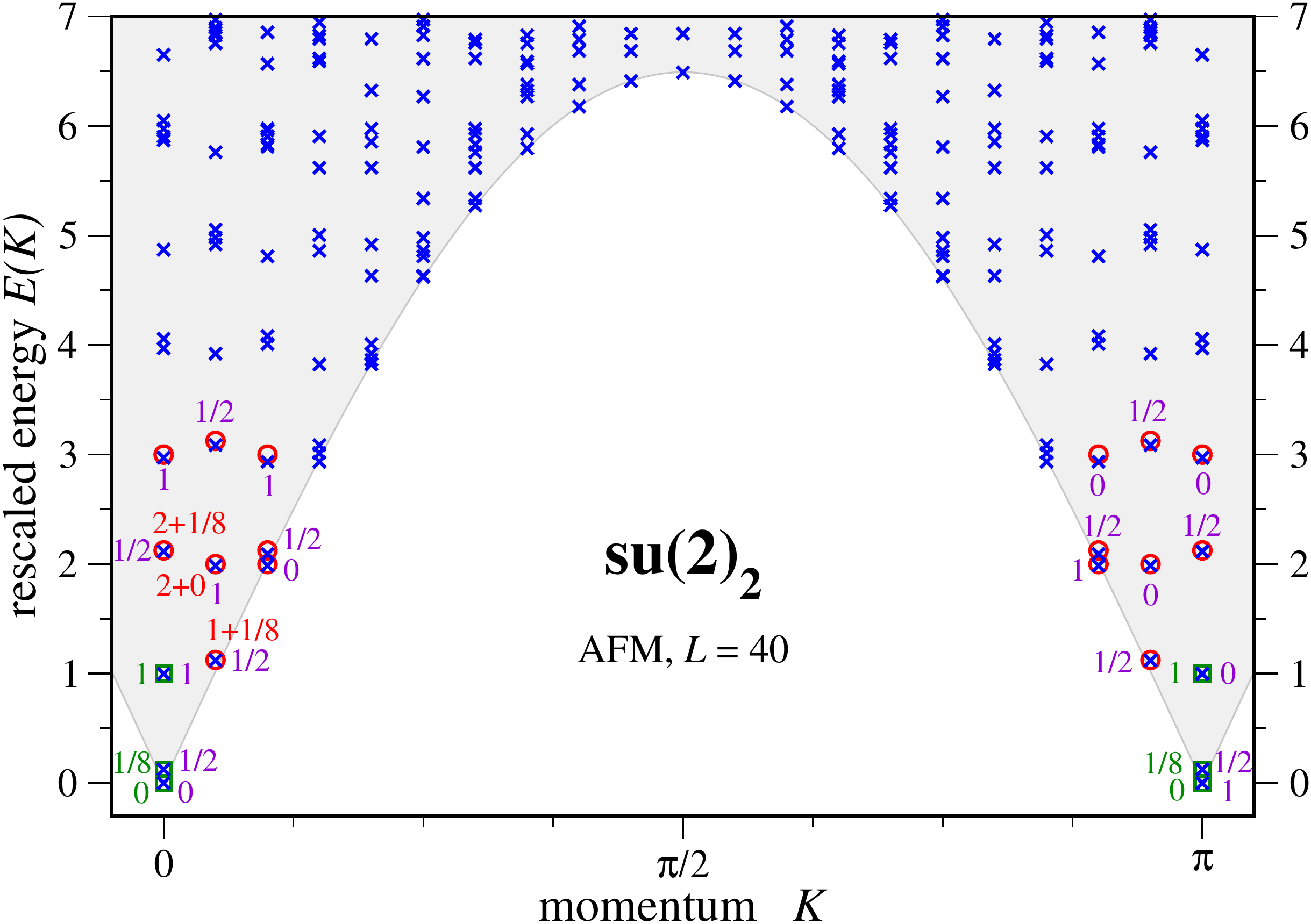}
\includegraphics[width=\columnwidth]{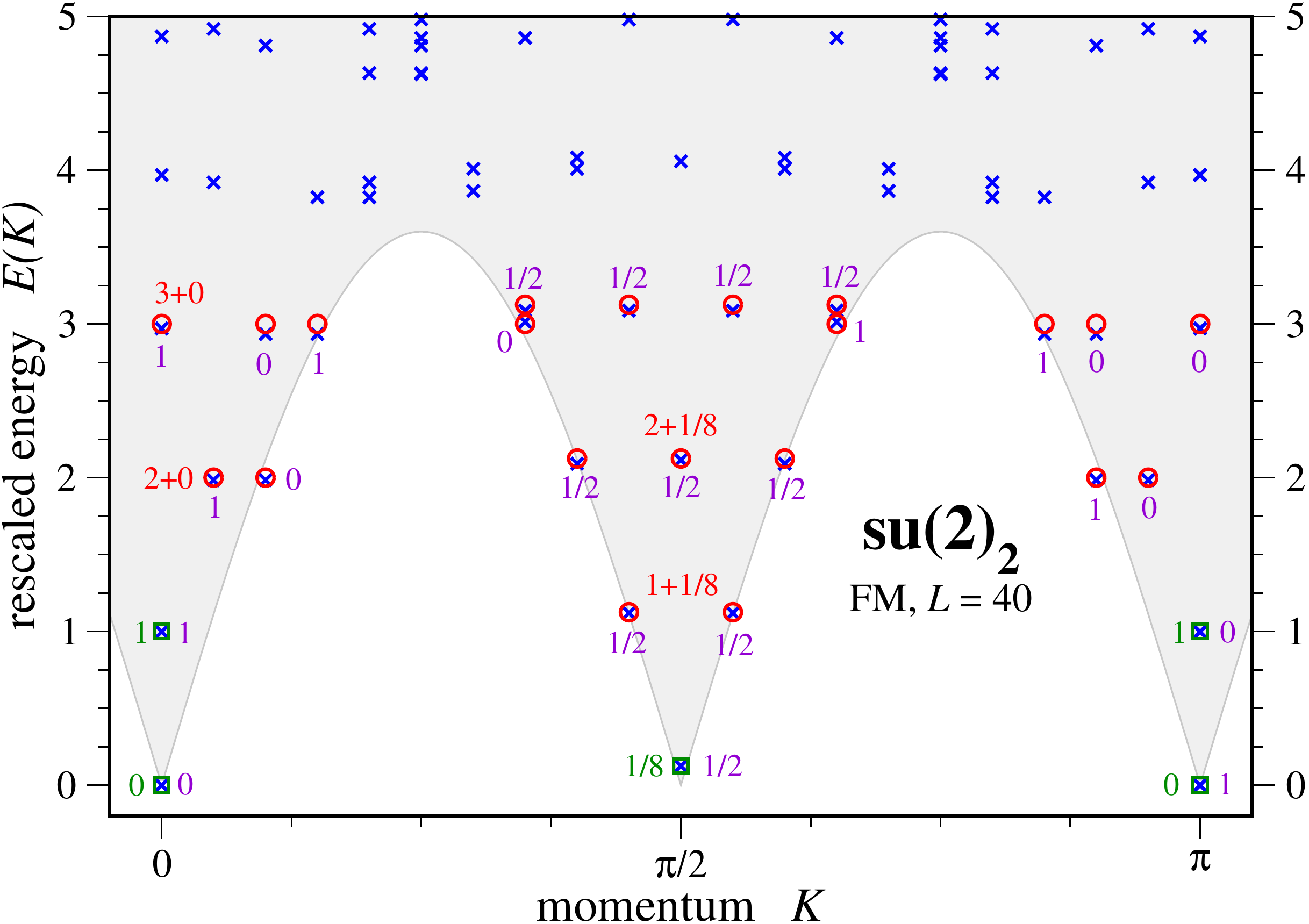}

\caption{(color online)
	      {\bf The su(2)$_2$ spin-1/2 chain:}  Energy spectra have been rescaled to 
	     match the conformal field theory prediction given in Eq.~\eqref{CFT_energy_levels}. 
              Green squares indicate the location of the primary fields, red circles specify the descendant fields.
              The topological symmetry sector is indicated by the violet index. 
	     Data shown is for system size $L=40$.}
\label{AFM_k2}
\end{center}
\end{figure*}
\begin{figure*}[ht]
\begin{center}
\includegraphics[width=\columnwidth]{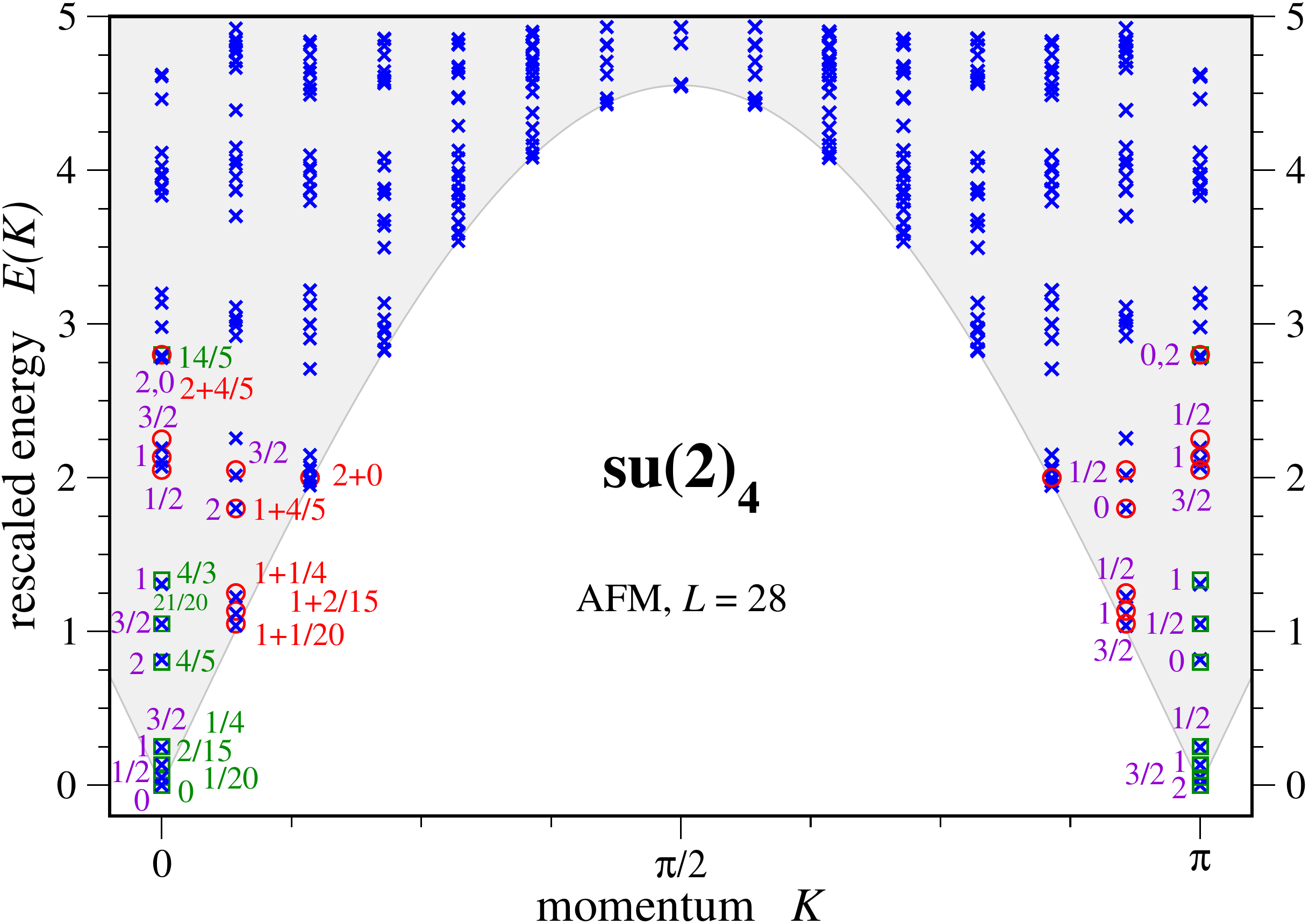}
\includegraphics[width=\columnwidth]{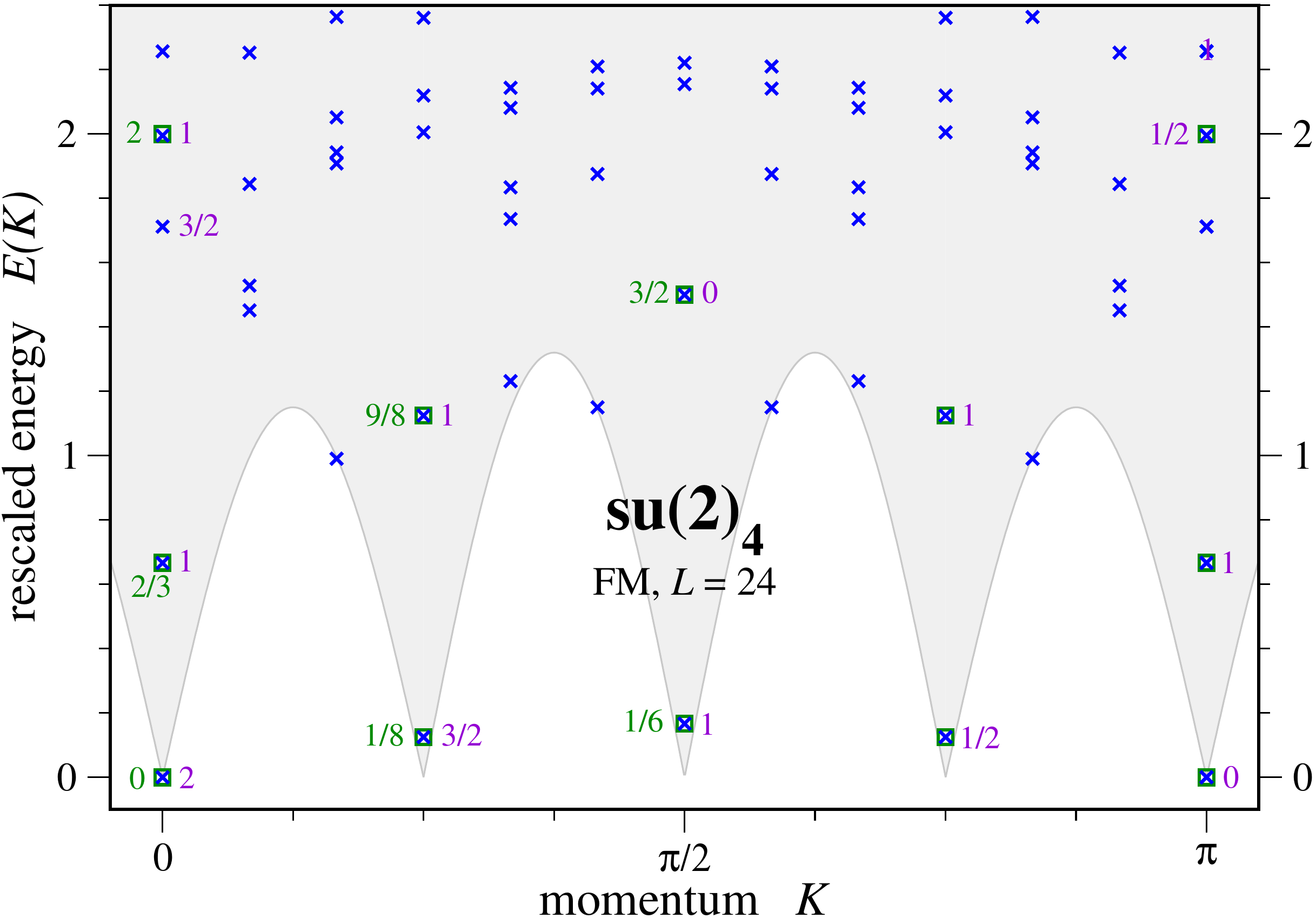}
\caption{(color online)
	      {\bf The su(2)$_4$ spin-1/2 chain:}  Energy spectra have been rescaled to 
	     match the conformal field theory prediction given in Eq.~\eqref{CFT_energy_levels}. 
              Green squares indicate the location of the primary fields, red circles specify the descendant fields.
              The topological symmetry sector is indicated by the violet index. 
	     Data shown is for system size $L=28$ and $L=24$.}
\label{AFM_k4}
\end{center}
\end{figure*}

In this section, we discuss the results of our study of the su$(2)_k$ spin-$1/2$ anyonic spin chains
for $k=2,4,5$. The case $k=3$ is the original `Golden Chain'
model, which marked the beginning of the study of anyonic quantum spin chains \cite{Feiguin_07}.
In the latter publication, it was established numerically that for both antiferromagnetic as
well as ferromagnetic interactions, the system is critical, and that the system can be described by
the tri-critical Ising model and  the $Z_3$ parafermion cft  (three-state Potts model criticality), respectively\cite{Feiguin_07}. 

In addition, it was shown that the model can be
mapped onto an exactly solvable model, namely a particular `restricted solid-on-solid'
(RSOS) model\cite{Feiguin_07}. This mapping is applicable to arbitrary $k$, and thus the critical
behavior of the spin-$1/2$ anyonic chains is described by the $k$-critical Ising
model for AFM interactions and  $Z_k$-parafermions for FM interactions\cite{Feiguin_07}. 

Finally, it was conjectured in Ref.~(\onlinecite{Feiguin_07}) that the criticality of these spin-$1/2$
anyonic chains is not merely due to a fine tuning of parameters, but is in fact protected
by a non-local, topological symmetry of the model. This implies that the model remains gapless if a 
perturbation  which preserves both the spatial and the  topological symmetry is added to the model.
This property is essential for the nucleation of a new topological liquid as a result of interactions between
anyons\cite{su2k_short}.

In this section, we consider the topological symmetry properties of the
 su(2)$_k$ spin-$1/2$ chains
and explain why the criticality  is topologically protected for all finite $k$. Explicit Hamiltonians  are given in appendix~\ref{spin12_chain}. 

\begin{figure*}[t]
\begin{center}
\includegraphics[width=\columnwidth]{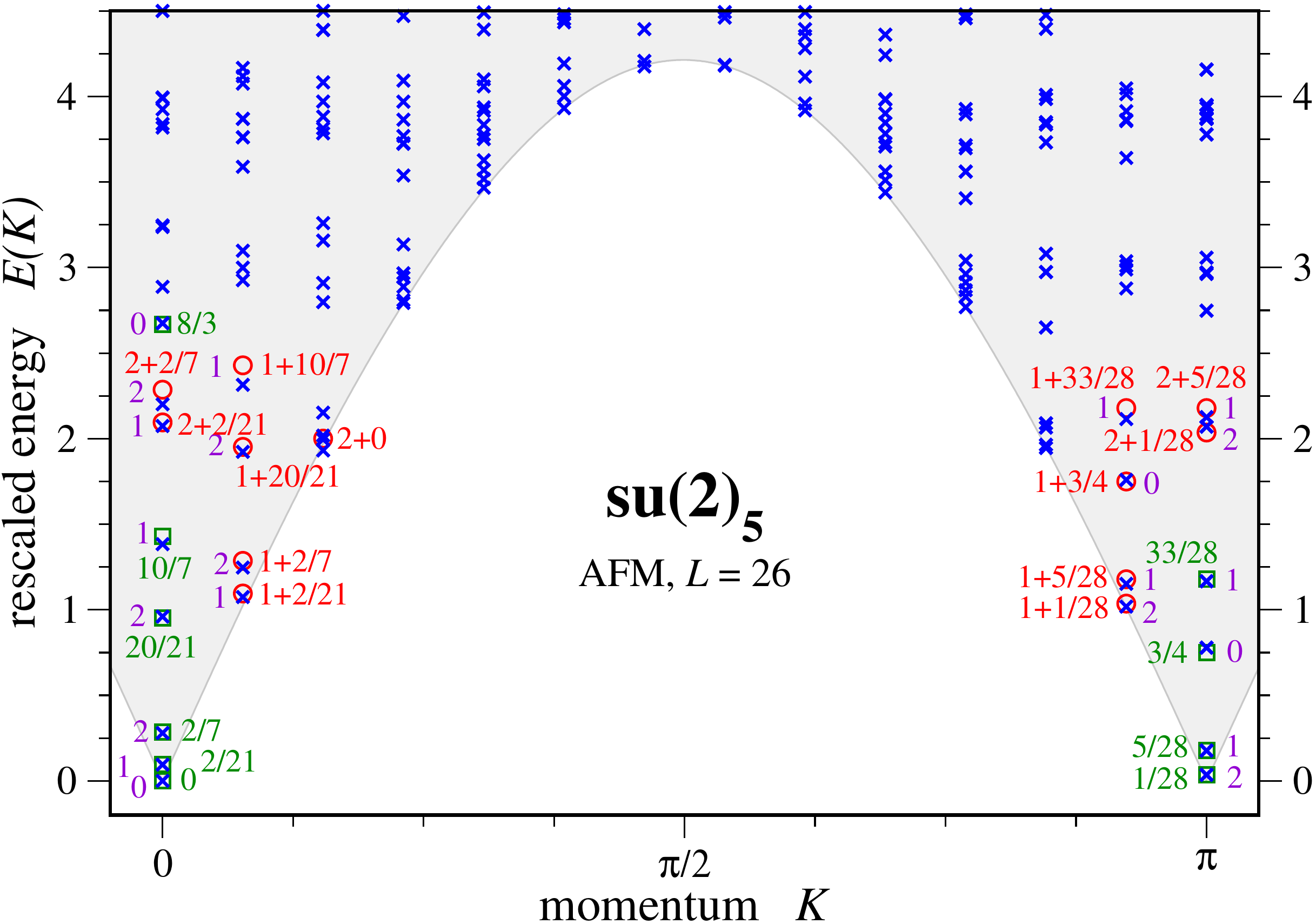}
\includegraphics[width=\columnwidth]{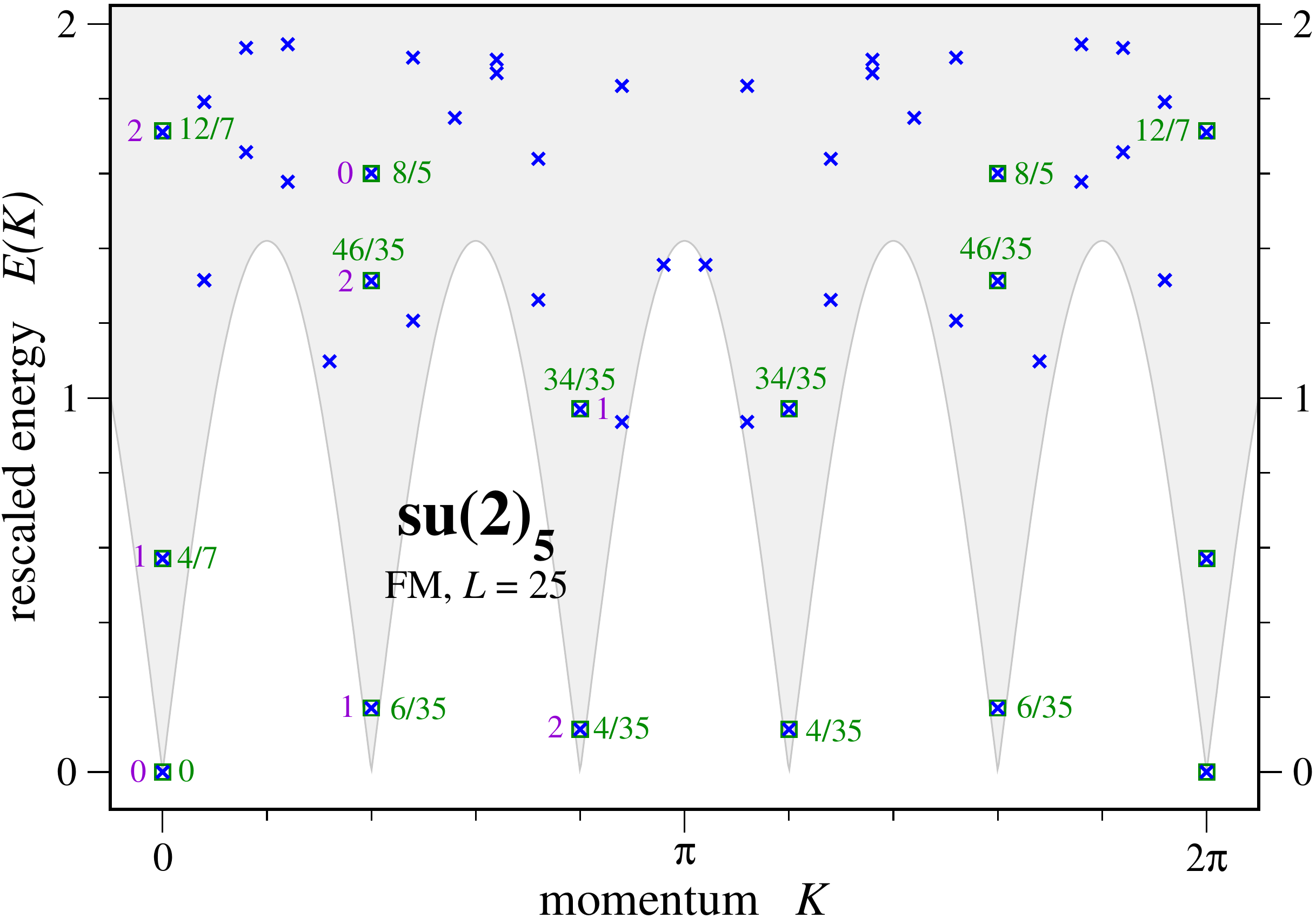}
\caption{(color online)
	     {\bf The su(2)$_5$ spin-1/2 chain:} Energy spectra have been rescaled to 
	     match the conformal field theory prediction given in Eq.~\eqref{CFT_energy_levels}. 
              Green squares indicate the location of the primary fields, red circles the descendant fields.
              The topological symmetry sector is indicated by the violet index. 
	     Data shown is for system size $L=26$ and $L=25$, respectively.}
\label{AFM_k5}
\end{center}
\end{figure*}

The numerically obtained spectra for both AFM and FM interactions are given
in Figures \ref{AFM_k2}, \ref{AFM_k4} and \ref{AFM_k5} for $k=2$, $k=4$ and
$k=5$. The spectra were obtained by exact diagonalization of the Hamiltonian,
followed by  shifting and rescaling of the spectrum in order to match the conformal
field theory predictions. 

The numerical results confirm that the spin-$1/2$ su(2)$_k$
chains are described by the $k$-critical Ising model for AFM interactions and the
$Z_k$-parafermion CFT for FM interactions. 
Details of these CFTs are given in  appendices
\ref{app:minmod} and \ref{app:zk-pf}.

In the remainder of this section, we discuss 
the assignment of  topological
symmetry sectors  to  the states
in the energy spectra, as indicated in  Figures \ref{AFM_k2}, \ref{AFM_k4} and
\ref{AFM_k5}. The topological symmetry sectors  were obtained
by acting with the operator $Y$ on the eigenstates.  Because $Y$ commutes with both
the Hamiltonian and the momentum operator, and because most states are non-degenerate,
it follows that the eigenstates of the Hamiltonian (in the momentum
representation) are also eigenstates of the topological operator $Y$.

We begin the analysis with a general observation.
A topological symmetry sector is assigned to each state in the spectrum. Moreover, each state  is associated with a field in the
conformal field theory describing the critical behavior of the chain. These conformal fields
 satisfy certain fusion rules, which, generally, are
 different from the fusion rules of the anyons themselves (typically, the number of conformal fields differs from the number of types of anyons).
 As a result, the topological symmetry sectors must be associated with the conformal fields in a  manner that both 
  the su(2)$_k$ fusion rules of the anyons and
fusion rules of the conformal fields are satisfied.
For the case of the su(2)$_k$ spin-1/2 anyonic chain, this constraint is obeyed for the following reason:
the relevant critical theories are so-called coset theories, which contain a su(2)$_k$
theory and other theories such as u(1).
This implies that the fields in the
critical theory inherit su(2)$_k$ topological symmetry labels; thus, 
the topological symmetry sectors can be assigned in a consistent manner.

\subsection{The ferromagnetic case}

The ferromagnetic su(2)$_k$ spin-$1/2$ anyon chain is described by the 
 coset
theory su(2)$_k/u(1)_{2k}$ (details  can be found in appendix~\ref{app:zk-pf}).
  The fields in this
conformal field theory carry two labels, a su(2)$_k$ label $l$, and an u(1)
label $m$, where
$l=0,1,\ldots,k$, $m=0,1,\ldots 2k-1$ and  $l+m= 0 \bmod 2$.
Under fusion of two fields with labels $(l_1,m_1)$ and $(l_2,m_2)$, the
labels $m_1$ and $m_2$ are added modulo $2k$, while the labels $l_1$ and
$l_2$ satisfy the fusion rules of su(2)$_k$. Thus, the fields $(l,m)$ can be assigned
a topological label $l$, and this assignment automatically
obeys the correct fusion rules.

The momentum quantum numbers of the fields cannot be predicted from the conformal field
theory itself. 
Different realizations of a particular
CFT may vary in the assignment of momenta to conformal fields. For example,
the su(2)$_2$ spin-1/2 chain is described by the Ising CFT for both AFM and
FM interactions, but the states corresponding to the $\sigma$, or $j=1/2$ field
occur at different momenta, as  illustrated in Figure~\ref{AFM_k2}.

We first consider the case of $k$ even. The Hilbert space is given
by labelings of the the fusion chain as displayed in Figure~\ref{Fig:Spin1-Chain}, where, in the 
case of the spin-$1/2$ anyon chain,
 the `incoming' labels are spin-$1/2$ anyons. As a consequence, 
the labels $x_i$ alternate between integer and half-integer values. Thus, there
are two decoupled `sectors': In one sector, the labels of the odd sites correspond to integer-spin anyons,
while in the other sector, the odd sites correspond to half-integer-spin anyons.
Because of this, each field in the conformal field theory will appear twice in the
spectrum, once at momentum $K$ and once at momentum $K+\pi$.

\begin{table}[ht]
\begin{tabular}{r | c c c c c c c c }
& $m:$\\
& $0$ & $1$ & $2$ & $3$ & $4$ & $5$ & $6$ & $7$ \\
\hline
l: $0$ & $0$ & x & $\frac{3}{2}$ & x & $2$ & x & $\frac{3}{2}$ & x\\
$1$ & x & $\frac{1}{8}$ & x & $\frac{9}{8}$ & x & $\frac{9}{8}$ & x & $\frac{1}{8}$\\
$2$ & $\frac{2}{3}$ & x & $\frac{1}{6}$ & x & $\frac{2}{3}$ & x & $\frac{1}{6}$ & x\\
$3$ & x & $\frac{9}{8}$ & x & $\frac{1}{8}$ & x & $\frac{1}{8}$ & x & $\frac{9}{8}$\\
$4$ & $2$ & x & $\frac{3}{2}$ & x & $0$ & x & $\frac{3}{2}$ & x
\end{tabular}
\caption{Scaling dimensions in the $Z_4$ parafermion
model.}
\end{table}

As was already discussed above, for the case of even $k$, 
the topological sector $y$ of a field labeled by
$(l,m)$ is determined by $l$, namely $y = l/2$.
Our numerical results show that the momenta of the fields 
are either given by $K(m) = \frac{m \pi}{k}$,
or  by $K(m) = \pi + \frac{m\pi}{k}$,
as can be seen in the right hand side panel of Figure~\ref{AFM_k4}.
To establish that the FM spin-$1/2$ chain is stable under
perturbations preserve both spatial and topological symmetry, we need to 
show that
 there are no relevant operators with the same momentum and topological quantum
numbers as the ground state.

In the case of odd $k$, anyon spins $j$ are automorph to anyon spins
$k/2-j$ (see Appendix A), and therefore the labels of the conformal fields are
 given by $(l,m)$ where both $l$ and $m$ are even. 
 The topological
sectors are given by $l/2$, and the momenta of the fields
are given by $K(m)  = \frac{m\pi}{k}$.

From the above discussed relations between
 field labels $(l,m)$ and the quantum numbers (topological sectors and momenta),
it becomes apparent that 
each momentum and each topological sector  appears at most once.
This implies that
the critical behavior is indeed stable to perturbations which preserve
both spatial and topological symmetry. 

\subsection{The anti-ferromagnetic case}

In this section, we show that  
the criticality of the anti-ferromagnetic su(2)$_k$ spin-$1/2$ chain is stable under perturbations that do not break
the symmetries of the model.
The model is described by the
$k$-critical Ising model, which can be formulated in terms of a
coset-model su(2)$_1 \times su(2)_{k-1}/su(2)_k$ (some details of this
coset model can be found in appendix~\ref{app:minmod}).
The conformal fields in this CFT are labeled by $(r,s)$, where
the $r$ label ($1\leq r \leq k$) is associated with su(2)$_{k-1}$,
 while $s$ ($1\leq s \leq k+1$) is associated with su(2)$_k$.
 There is also a label associated
with su(2)$_1$, however,  this label is fixed by the constraint $t = r+s \bmod 2$.

The topological sectors are given by $(s-1)/2$. Since $s$ is the conformal label associated with the
 denominator su(2)$_k$ of the coset, 
the fusion rules of the coset CFT are consistent with the fusion rules
associated with the topological sectors.

\begin{table}[ht]
\begin{tabular}{r | c c c c c}
& $s:$\\
& $1$ & $2$ & $3$ & $4$ & $5$ \\
\hline
r: $1$  & $0$ & $\frac{1}{4}$ & $\frac{4}{3}$ & $\frac{13}{4}$ & $6$
\raisebox{-2mm}{\rule{0cm}{5mm}} \\ 
$2$ & $\frac{4}{5}$ & $\frac{1}{20}$ & $\frac{2}{15}$ & $\frac{21}{20}$ & $\frac{14}{5}$
\raisebox{-2mm}{\rule{0cm}{5mm}} \\
$3$ & $\frac{14}{5}$ & $\frac{21}{20}$ & $\frac{2}{15}$ & $\frac{1}{20}$ & $\frac{4}{5}$
\raisebox{-2mm}{\rule{0cm}{5mm}} \\
$4$ & $6$ & $\frac{13}{4}$ & $\frac{4}{3}$ & $\frac{1}{4}$ & $0$
\raisebox{-2mm}{\rule{0cm}{5mm}}
\end{tabular}
\caption{Scaling dimensions for the tetra-critical Ising model.}
\end{table}

In the case of even $k$, all  fields appear twice in the spectrum,
(once at momentum $K$ and once at $K+\pi$) as a result of   
 the `doubling' of the
Hilbert space.
Our numerical calculations yield the following. The topological
sector of each field is determined by $s$, namely $y=(s-1)/2$. The momentum
of a  field labeled by $(r,s)$  is given by either $K = (r+s \bmod 2) \pi$ or by $K = (r+s+1 \bmod 2)\pi$; the system size determines which one of the two possibilities occurs (
we verified this behavior for $k=2,4$).

For odd $k$, the association of field labels $(r,s)$ with topological and momentum sectors
coincides with that for even $k$. 
However, only odd values of $s$ appear, due to the above mentioned automorphism of anyon spins.
These results were verified for  $k=3$ in Ref. (\onlinecite{Feiguin_07}), and
 for $k=5$ in this study (see 
figure~\ref{AFM_k5}).

To confirm that the criticality of the AFM spin-$1/2$ chains is stable under perturbations 
which preserve the spatial and topological symmetries of the model,
we have to analyze the scaling dimensions of the fields which have the same topological quantum number as the ground state. The ground state has label $s=1$ (i.e., topological sector $y=0$).
The scaling dimensions of the fields with label
$s=1$ are given by $2h = (r^2(k+2)-2(k+1)r+ k)/(2(k+1))$, which for
$r\geq 1$ increases monotonically. 
The most relevant field in the same momentum
sector as the ground state thus carries the labels $(r,s) = (3,1)$, and has
scaling dimension $2h = 2+\frac{4}{k+1}$, which is irrelevant for $k$ finite, and
becomes marginal in the limit $k\rightarrow \infty$. Again, we conclude that the
AFM spin-$1/2$ chains are stable with respect to perturbations which preserve both topological and translational symmetry.

When breaking the spatial symmetry of the model by dimerizing the system, 
the most relevant field has labels $(r,s) = (2,1)$ and thus a scaling dimension
$2h = (k+4)/(2(k+1))$ that is relevant for all $k$. Therefore, a perturbation which
breaks translational symmetry may open up a gap.

\section{Discussion}
\label{sec:discussion}

\begin{figure*}[ht]
\begin{center}
\includegraphics[width=\linewidth]{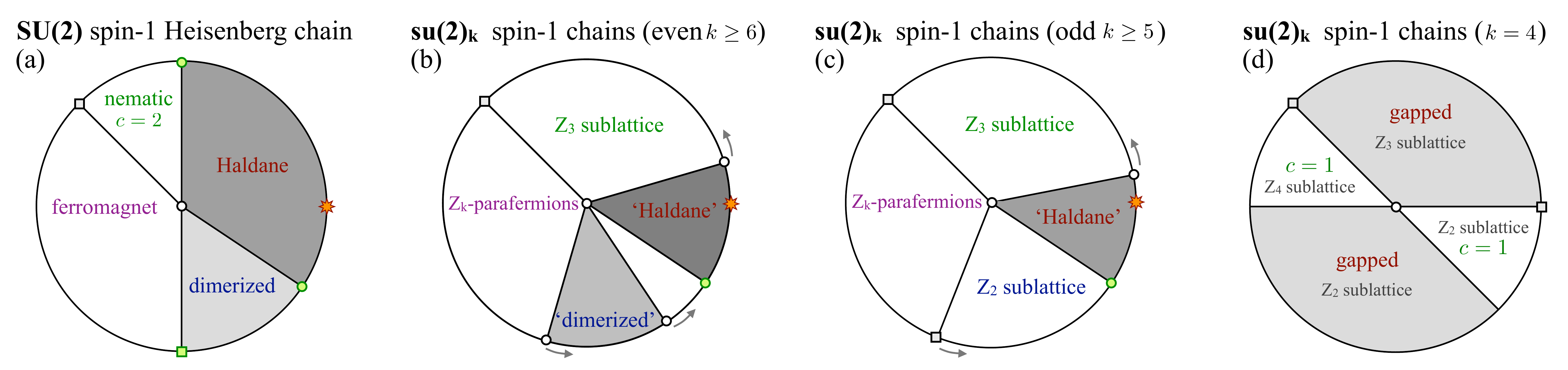}
\end{center}
\caption{%
Phase diagrams of the various spin-$1$ models considered in the paper:
(a) the bilinear-biquadratic spin-$1$ Heisenberg chain,
(b) the generic even $k\geq 6$ anyonic spin-$1$ chain,
(c) the generic odd $k\geq 5$ anyonic spin-$1$ chain,
(d) the special case $k=4$.}
\label{fig:overview}
\end{figure*}

The anyonic analogs of the SU(2) Heisenberg spin-$1$ model  have a
rich structure, as can be seen from the phase diagrams of
 the ordinary bilinear-biquadratic spin-$1$ model, the generic {\em even}
$k\geq 6$ anyonic model, the generic {\em odd} $k\geq 5$ anyonic model, and
 the special case $k=4$ (displayed side by side in Fig.~\ref{fig:overview}).

The distinct nature of the phase diagram for $k=4$ originates
in the symmetry of the
fusion rules of the su(2)$_k$ theory  under the
exchange $j \leftrightarrow \frac{k}{2}-j$ which, for $k=4$, maps $j=1$ onto
itself. It is also the lowest value of $k$ for which a generic fusion rule 
$1\times 1 = 0 + 1 + 2$ applies (compare with $k=3$ where
$1\times 1 = 0 + 1$), thus making it possible to define an anyonic spin-$1$ model.
Moreover,   the central charge of the defining su$(2)_4$ algebra is an integer 
($c=2$), and the  quantum dimensions of the su$(2)_4$ anyons
are all integers or square roots of integers (we discuss the various anyon
models in more detail in Appendix~\ref{app:su2k-anyons}).
We note that  fusion models with such quantum dimensions
typically do not permit `universal quantum computation',
a property which requires a `fine tuning' of the braid properties\cite{freedman02a,freedman02b}.
Models analogous to the case $k=4$ have been studied
from the integrability point of view \cite{finch,Verbus}.

Upon increasing the level $k$, the su(2)$_k$ anyon model increasingly resembles the
ordinary SU(2) spin algebra. In terms of the quantum group language, the limit
$k\rightarrow\infty$ corresponds to $q\rightarrow 1$, where $q=e^{\pi i/(k+2)}$.
For $q=1$, the quantum group reduces to the ordinary SU(2) algebra.
One would therefore intuitively expect that the  phase diagram of the generic $k$ case has
the same structure as the phase diagram of the SU(2) bilinear-biquadratic spin-$1$ chain.
The numerics presented in the paper shows that this is indeed the case for both even and
odd $k$, with one notable exception:
for even $k$, we find a gapped dimerized phase that is separated from the Haldane gapped phase by an
extended critical region; in contrast, for odd $k$, we observe only an extended
critical region but no dimerized phase.
 The fact that the anyonic
spin-$1$ models behave differently for even and odd $k$ is very interesting in its
own right. To the best of our knowledge, this is the first time that such an
`even-odd' effect in the level $k$ has been observed.

 In the following, we discuss some of the differences between the cases of odd $k$ 
versus even $k$.
It is instructive to consider the model for the lowest (generic) value of even $k$, i.e., $k=6$.
 From the symmetry of the fusion
rules under the exchange $j \leftrightarrow \frac{k}{2} - j$ it follows that $j=1$ and $j=2$
are exchanged. This implies a  `symmetry' in the phase diagram of the $k=6$ model
under exchange of the projectors $P^{(1)}$ and $P^{(2)}$.
The parametrization chosen in this paper, 
$H = \sum_{i} \cos\theta_{2,1} P^{(2)}_i - \sin\theta_{2,1} P^{(1)}$,
renders  the phase diagram symmetric in the line through the points
$\theta_{2,1} = 3\pi/4$ and $\theta_{2,1} = 7\pi/4$.
It is important to realize that this `symmetry' only applies the values of the energies which
appear in the spectra, but not to the momenta and the degeneracies of the energy
levels. In particular, the gapped dimerized phase is the `mirror phase' of the gapped Haldane phase.
We also note that the same mechanism resulted in a symmetric phase diagram for $k=4$,
if plotted in terms of the projectors $P^{(0)}$ and $P^{(2)}$.

For even $k=6$, there is an extended critical region between these two gapped phases; however, we were not able
to determine its precise critical behavior.
For $k=8$, the extend of this critical region is smaller, and it is therefore not inconceivable
that for very large even $k$, this critical region will shrink to a single critical point separating
the two gapped phases, as is the case for the SU(2) spin-$1$ bilinear-biquadratic model.

As indicated by the above terminology, the gapped phase around the angle $\theta_{2,1} = 0$ is the
anyonic analogue of the Haldane gapped phase\cite{Haldane}. The ground states
 at $\theta_{2,1} = 0$ can be obtained exactly, and they are the anyonic
analogues of the AKLT state\cite{AKLT}. 
In section~\ref{sec:AKLT-states-open}, we studied this `AKLT' point of our anyonic models
with open boundary conditions. We obtained edge states similar to the ones observed in the
SU(2) case. In the case of periodic boundary conditions, we find a $k+1$-fold
degenerate ground state (one for each topological sector), occurring at momentum $K=0$.
Although the Haldane phases of the SU(2) and anyonic models share many properties,
they differ in their degeneracy for periodic boundary conditions. Therefore, it is interesting
to investigate in which way the underlying quantum group symmetry changes the
classification of gapped phases in one-dimensional spin systems\cite{Chen11}.

The dimerized  gapped phase of the anyon model  exhibits exactly the same values of the
energy levels as the gapped Haldane phase, as pointed out above. Nevertheless, this gapped
phase is of a different nature. At $\theta_{2,1} = 3\pi/2$, there is a (unique) zero
energy state at momentum $K=\pi$, i.e., the phase is dimerized - like the 
corresponding phase in the SU(2) spin-$1$ model.
 In addition, there is a set of degenerate zero
energy ground states at momentum $K=0$, where the number of states depends on
$k$. 

Almost two decades ago, Koo and Saleur \cite{Saleur} considered a spin-$1$ type loop model that was
based on the  `fused Potts model'. The underlying algebra of their model is
the Birman-Murakami-Wenzl (BMW) algebra,
 which replaces the
Temperley-Lieb \cite{tl71} algebra that  appears in the study of the Potts model
in its loop representation. For details on the BMW algebra, we refer to \cite{fk10}.
The Koo-Saleur model contains a {\em continuous} parameter $Q$,
which is closely related to the discrete level $k$ in the anyon models we consider (see below). 
More specifically, the model Koo and Saleur consider is
\begin{equation}
H_{\rm KS} = \sum_i (Q-1)(\sin\omega-\cos\omega) P^{(0)}_i - (Q-2)\cos\omega P^{(1)} \ .
\end{equation}
The projectors $P^{(0)}$ and $P^{(1)}$ project two neighboring spin-1 loops onto the
spin-0 and spin-1 channels, respectively (see \cite{fk10} for explicit expressions of these
projectors in terms of the BMW algebra). The number of Potts states $Q$  is related to the quantum dimension of the spin-1/2 anyons $d_{1/2}$ (or the parameter
$d$ appearing in the Temperley-Lieb algebra) via
$Q= d_{1/2}^2 = 4\cos(\pi/(k+2))^2$, and thus $Q=1,2,3,4$ corresponds to
$k=1,2,4,\infty$. In particular, the case $Q=4$ corresponds to the ordinary
SU(2) spin-$1$ chain. 
We note that the anyonic chains can only be defined for
integer $k\geq 4$, and recall that  we parametrized the anyonic
spin-$1$ model as $H= \sum_i \cos(\theta_{2,1}) P^{(2)}_i - \sin(\theta_{2,1}) P^{(1)}$.
By making use of the relation $\id = P^{(0)} + P^{(1)} + P^{(2)}$, one finds the following
relation between the parameters of the models
\begin{align}
\cos\theta_{2,1} &= -(Q-1)(\sin\omega-\cos\omega) \nonumber\\
\sin\theta_{2,1} &= -\cos\omega + (Q-1)\sin\omega \ .
\end{align}

Despite the similarities between the model of Koo and Saleur and our anyonic model,
they behave rather differently.
The phase boundaries between the various phases observed in the Koo-Saleur model
depend smoothly on the continuous parameter $Q$, while the phase diagrams of the
anyonic spin-1 models depend on whether $k$ is even or odd. In addition, the Koo-Saleur
model displays non-unitary critical behavior, while the critical behavior of the
anyon models is described by unitary CFTs.
The explanation for this difference in behavior should be sought in the
representations used in the two models. 
In the Koo-Saleur model, a representation which essentially behaves like a SU(2)
representation is used (which permits to define the model as a function of the continuous
parameter). In the anyonic version, the truncated su(2)$_k$ representations play a central role.
For a related discussion in the general context of loop models, we refer to\cite{fendley02,fendley06}.
 
These observations suggest that a deeper investigation into the differences and similarities of
the two models is warranted, especially because the Koo-Saleur model exhibits various integrable
points\cite{Saleur}. One of the integrable points identified in \onlinecite{Saleur} corresponds to the
supersymmetric critical point forming the boundary of the Haldane phase. The location of this
integrable point, in terms of the parameters used in this paper, is
$\tan \theta_{2,1} = -\frac{1}{2}\frac{d_1+1}{d_1}$, where $d_1 = 1 + 2 \cos \left( 2\pi / (k+2) \right)$
(see \cite{nissinen-up}).
For $k\geq4$, this location depends only weakly on $k$, namely, for $k=4$, one obtains
$\theta_{2,1} = -\arctan(3/4) \approx -0.2048 \pi$, while in the limit $k \rightarrow \infty$, one obtains
$\theta_{2,1} = -\arctan(2/3) \approx -0.1872 \pi$. The location of the critical end point of the Haldane
phase we obtained in this paper are consistent with the location of this integrable point.

To solve the anyonic spin-1 chain at this integrable point, one approach is to map the model
to a fused RSOS model, as studied in\cite{date86,Date} (see also
\onlinecite{pasquier88,gepner92-up}).
This subject will be described in a separate publication\cite{nissinen-up}.

  
\acknowledgments
We acknowledge insightful discussions with P. Fendley, P. Finch, J. Nissinen, and H. Saleur. 
We thank the Aspen Center for Theoretical Physics, where parts of this manuscript were written,
for hospitality and support under Grant No. NSF 1066293.
A.W.W. L. was supported, in part, by NSF DMR-0706140. C.G. was supported, in part, by NSERC-163953. S.T. was supported, in part, by SFB TR 12 of the DFG.


\appendix

\section{su(2)$_k$ anyons}
\label{app:su2k-anyons}

In this appendix, we briefly review of the properties of 
su(2)$_k$ anyons - the building blocks of the 
anyonic chains considered in this paper - for arbitrary level $k\geq 2$.
We explicitly discuss the levels $k=2,\ldots,7$.
For a general discussion of anyon models,
see e.g.  Refs.~\onlinecite{Kitaev06,ZHWang,bonderson-thesis}.

The anyons of the su(2)$_k$ theories are closely related to ordinary
SU(2) spin degrees of freedom; thus we label the 
anyons by their `generalized angular momenta', or simply `spin' value
$j=0,\frac{1}{2},1,....,\frac{k}{2}$.
We note that  in the su(2)$_k$ theory, there is a maximum
allowed value of the `spin', namely $k/2$, a feature not present 
for ordinary SU(2) spins.

Ordinary spins can be combined using tensor products. In general,
combining two spins gives rise to several different spins.
 An analogous
phenomenon occurs if we combine two anyons of the su(2)$_k$ theory. In the
following, we will assume that $k$ is fixed, but arbitrary; i.e.,  
the anyons combined belong to the same theory.
 The rules for combining two anyons - also denoted as `fusion rules' -
 are closely related to the SU(2) tensor products,
namely
\begin{equation}
j\times j' = \sum_{j''=|j-j'|}^{\min(j+j',k-j-j')} j'' \, .
\label{fusion_rules}
\end{equation}
The only difference to the case of ordinary SU(2) spins is the cutoff in the upper
limit of the sum in Eq. \eqref{fusion_rules}. The cutoff is the result
 of the finite number of types of anyons in the su(2)$_k$ theories. The fusion rules  in Eq. \eqref{fusion_rules} are associative.

The fusion rules can be represented in terms of the fusion matrices
$N_j$ which, in the case of su(2)$_k$ anyons, have entries
(the so-called fusion coefficients)
$N_{j,j'}^{j''}=1$ if and only if the fusion of labels $j$ and $j'$
gives rise to the label $j''$, and zero otherwise. In general, fusion
coefficients bigger than one are possible, but they do not appear in the context
of this paper.

Fusion is commutative and associative, i.e., fusing several anyons in different order gives
rise to the same result. This implies that the fusion matrices $N_j$ 
commute and that they can be diagonalized simultaneously. 
Diagonalizing the fusion matrices yields the quantum dimensions $d_j$,
\begin{equation}
N_j \ {\bf d}  = d_j \ {\bf d}\, ,
\end{equation}
where ${\bf d}$ is a vector whose components are the quantum dimensions $d_j$.
The total quantum dimension $\mathcal{D}$ is defined as
\begin{equation}
\mathcal{D} = \sqrt{\sum_j d_j^2 }\, .
\end{equation}
For su(2)$_k$ anyons, the quantum dimensions are given by
\begin{eqnarray}
d_0 &=& 1\ , \nonumber \\
d_{1/2} &=& 2\cos\left ( \frac{\pi}{k+2}\right )\, , \nonumber \\
d_{j} &=& d_{1/2}d_{j-1/2} - d_{j-1}, \quad j\ge 1\ .
\label{quant_dim}
\end{eqnarray}
Explicitly, one obtains
\begin{equation}
d_j = {\sin\left(\frac{(2j+1)\pi}{k+2}\right)}/{\sin\left(\frac{\pi}{k+2}\right)} \ ,
\end{equation}
where we note that the dimensions $d_j$  depend on the level $k$, which we have suppressed
in the notation.

The matrix which diagonalizes the fusion rules is called the
modular $S$-matrix. Its entries for the su(2)$_k$ theories are
given by
\begin{equation}
S_{j,j'} (k) = \sqrt{\frac{2}{k+2}} \sin \left( \frac{(2j+1)(2j'+1)\pi}{k+2} \right) \ .
\end{equation}

For odd $k$,  there exists an automorphism relating anyons with spin $j$ to anyons with spin
 $\frac{k}{2} - j$. The automorphism thus relates integer and half-integer spins, 
 reducing the study of odd-$k$ anyon systems to 
only integer (or only half-integer) anyon spins (it also means that there are only
 $k/2$ distinct anyon types for odd $k$).
In this manuscript, we consider anyons with integer spin when studying odd-$k$ systems.

\begin{figure}[t]
\begin{center}
  \includegraphics[width=.6\linewidth]{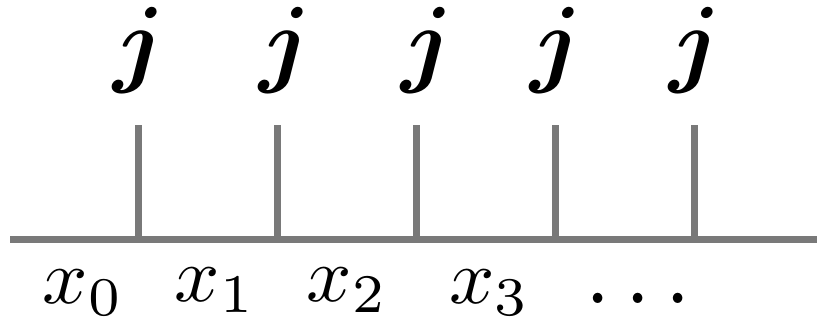}
\caption{Basis (fusion diagram) of a chain of spin-$j$ anyons ($j=1/2$ in the case of the
su(2)$_k$ spin-$1/2$ chain, 
 discussed in section~\ref{spin_half_chain} and $j=1$ in the case of the 
 su(2)$_k$ spin-$1$  chain, discussed in  section~\ref{bil_biq_chain}).}
\label{anyon_chains_general}
\end{center}
\end{figure}

 The Hilbert space of a multi-anyon system is non-local, and it can be represented by a 
a trivalent graph with each line segment representing an anyonic degree of freedom.
Such a graph is called a fusion diagram. The labeling of the segments has to be such
that the fusion rules are obeyed at all the vertices. In Figure \ref{anyon_chains_general}, 
we display the fusion diagram
that defines the Hilbert space of the models studied in this paper.

Each distinct labeling of the fusion diagram defines a basis state
$|\psi\rangle = |x_0,x_2,...,x_{L-1}\rangle$. We define the basis states
 $|\psi\rangle$ to be orthogonal, i.e., 
the inner product of two basis states is one if the
labels of the two states are identical, and	zero otherwise.
The number of basis states in a chain of spin-$j$ anyons of length $L$  grows 
asymptotically as $d_j^L$, where $d_{j}$ is the quantum dimension of
the anyon of type $j$. It is important to note that $d_{j}$ generally is not
an integer, as would be the case for ordinary SU(2) spins. This means that
it is not possible to associate a local Hilbert space with each anyon, and that the total Hilbert space
is not a simple tensor product of local Hilbert spaces.
 It also implies that there are no `internal $s_z$ quantum numbers'  in anyonic Hilbert spaces. The reason  behind all these features is that
the fusion rules enforce non-local constraints on the possible labelings of the
fusion digrams.

In order to define Hamiltonians acting on anyonic Hilbert spaces, 
the anyonic analog to the 
 $6j$-symbols for ordinary spin degrees of freedom has to be considered.
The anyonic version of  the $6j$-symbols 
is the so-called $F$-transformation, which relates the two different ways three anyon spins,
$j_1$, $j_2$, $j_3$, can fuse into a fourth anyon spin $j_4$. The $F$-matrix can be defined
as a result  
of the associativity of the fusion rules:
\begin{equation}
\includegraphics[width=5.5cm]{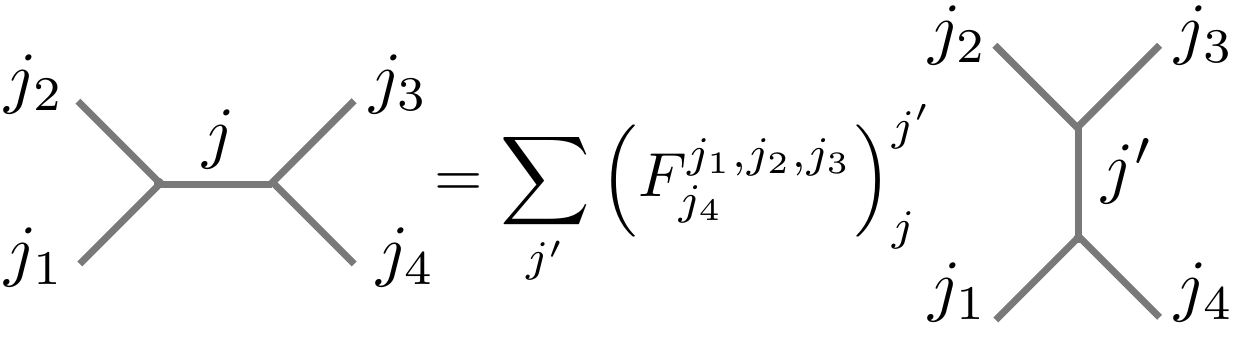}\, .
\label{F_matrix}
\end{equation}
In the case of su(2)$_k$, the $F$-matrices are uniquely determined by 
a consistency relation, namely the pentagon equation, and by imposing unitarity.
A useful expression (for general $k$) can be found in \cite{kr88},		
and is given in the appendix~\ref{app:f-symbols}.
 
A further basis transformation of interest is the so-called $S$-transformation which 
relates the `flux' of anyon spin $j$ through a loop of anyon spin $l$
to the case without anyon loop by
\begin{equation}
\includegraphics[width=3.8cm]{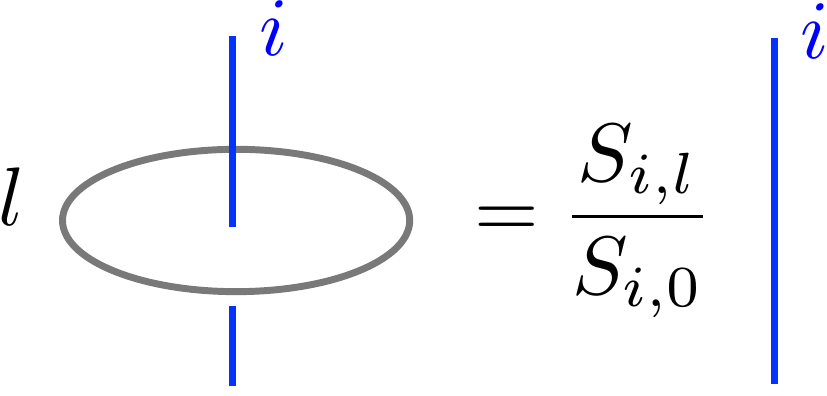}\, .
\label{S_eq}
\end{equation}
The matrix elements of this transformation are the elements of the
modular $S$-matrix \cite{ZHWang,Kitaev06}.

In the following, we give matrix representations of  some of the above discussed properties of a model of
 su(2)$_k$ anyons. Upper indices in round brackets denote the level $k$.
 
\noindent
{\it Level $k=2$: Ising anyons --}
This class of anyons comprises the spin-$0$ anyon, 
the Ising anyon (spin-$1/2$) with non-Abelian braiding
properties, and the  fermion (spin-$1$). The non-trivial
fusion rules are given by
\begin{align}
\frac{1}{2} \times \frac{1}{2} &= 0 + 1 \ ,&
\frac{1}{2} \times 1 &= \frac{1}{2} \ ,&
1 \times 1 = 0 \ .
\label{fusion_rules_k2}
\end{align}
The corresponding quantum dimensions are given by
\begin{align}
d^{(2)}_{0}=d^{(2)}_1 &=1\ , &
d^{(2)}_{\frac{1}{2}} &= \sqrt{2},  
\end{align}
and the $S$-matrix takes the explicit form
(the entries are ordered according to ascending value of the anyon spins)
\begin{equation}
S^{(2)} = \frac{1}{2}
\begin{pmatrix}
1 & \sqrt{2} & 1 \\
\sqrt{2} & 0 & -\sqrt{2} \\
1 & -\sqrt{2} & 1 \\
\end{pmatrix} \ .
\end{equation}

\noindent
{\it Level $k=3$: Fibonacci anyons --}
This class of non-Abelian anyons exhibits only two distinct particles, with spins
$0$ and $1$ (the Fibonacci anyon) respectively
(the spins $\frac{1}{2}$ and $\frac{3}{2}$ are automorph to spins $1$ and $0$, respectively).
Thus, there is only one non-trivial fusion rule,
\begin{equation}
1 \times 1 = 0 + 1
\end{equation}
and the quantum dimensions are given by
\begin{align}
d^{(3)}_0&=1 \ , & d^{(3)}_1 &=   (1+\sqrt{5})/2 \ .
\end{align}
Using the notation $\phi  = (1+\sqrt{5})/2$,
the $S$-matrix reads
\begin{equation}
S^{(3)} =\frac{1}{\sqrt{2 + \phi}}
\begin{pmatrix}
1 & \phi \\ \phi & -1 
\end{pmatrix} \ .
\end{equation}

\noindent
{\it Level $k=4$ --}
The $k=4$ anyon model contains five anyon spins, namely
$j=0,\frac{1}{2},1,\frac{3}{2},2$. The fusion rules are given by\\
\begin{equation}
\begin{tabular}{c|cccc}
$\times$ & $\frac{1}{2}$ & $1$ & $\frac{3}{2}$ & $2$ \\
\hline
$\frac{1}{2}$ & $0+1$ & $\frac{1}{2} + \frac{3}{2}$ & $1 + 2$ & $\frac{3}{2}$ \\
$1$ & & $0+1+2$ & $\frac{1}{2} + \frac{3}{2}$ & $1$ \\
$\frac{3}{2}$ & & & $0+1$ & $\frac{1}{2}$ \\
$2$ & & & & $0$
\end{tabular}
\end{equation}
The (non-trivial) quantum dimensions can be obtained from Eq.~(\ref{quant_dim}),
\begin{align}
d^{(4)}_{0} = d^{(4)}_{2} &= 1 \ , &
d^{(4)}_{\frac{1}{2}} = d^{(4)}_{\frac{3}{2}} &= \sqrt{3} \ , & 
d^{(4)}_1 &= 2  \ .
\end{align}
Finally, the $S$-matrix takes the form
\begin{equation}
S^{(4)} = \frac{1}{2\sqrt{3}}
\begin{pmatrix}
1 & \sqrt{3} & 2 & \sqrt{3} & 1 \\
\sqrt{3} & \sqrt{3} & 0 & -\sqrt{3 }& -\sqrt{3} \\
2 & 0 & -2 & 0 & 2\\
\sqrt{3} & -\sqrt{3} & 0 & \sqrt{3} & -\sqrt{3} \\
1 & -\sqrt{3} & 2 & -\sqrt{3} & 1 \\
\end{pmatrix}
\label{Smatrix_k4}
\end{equation}

\noindent
{\it Level $k=5$ --}
This class of non-Abelian anyons gives rise to three distinct anyon particles with
spins $0$, $1$ and $2$ (which are automorph to the spins
$\frac{5}{2}$, $\frac{3}{2}$ and $\frac{1}{2}$, respectively).
The non-trivial fusion rules are given by 
\begin{align}
1 \times 1 &= 0 + 1 + 2\ , &
1 \times 2 &= 1 + 2\ , & 
2 \times 2 &= 0 + 1\ .
\end{align}
The quantum dimensions take the following values,
\begin{align}
d^{(5)}_{0} &= 1\ , \nonumber\\
d^{(5)}_{1} &= (d_{2}^{(5)})^2-1 = 1+2\cos(2\pi/7)\ , \nonumber\\
d_2^{(5)} &= 2\cos(\pi/7) \ .
\end{align}
The $S$-matrix of the su(2)$_5$ theory  is given by
\begin{equation}
S =  \frac{1}{\mathcal{D}^{(5)}}
\begin{pmatrix}
1 & d^{(5)}_1 & d^{(5)}_2 \\
d_1^{(5)} & -d_2^{(5)} & 1\\
d_2^{(5)} & 1 & -d_1^{(5)} 
\end{pmatrix}
\end{equation}
where $\mathcal{D}^{(5)}$ denotes the total quantum dimension
$\mathcal{D}^{(5)} = \sqrt{1+(d_1^{(5)})^2+(d_2^{(5)})^2}$ of the su(2)$_5$ theory (restricted to the 
integer `spins').

\noindent
{\it Level $k=6$ --}
The anyon model with $k=6$ has seven anyons labeled by
$j=0,\frac{1}{2},1,\frac{3}{2},2,\frac{5}{2},3$. The fusion rules read
\begin{equation}
\begin{tabular}{c|cccccc}
$\times$ & $\frac{1}{2}$ & $1$ & $\frac{3}{2}$ & $2$ & $\frac{5}{2}$ & $3$\\
\hline
$\frac{1}{2}$ & $0+1$ & $\frac{1}{2} + \frac{3}{2}$ & $1 + 2$ &
$\frac{3}{2} + \frac{5}{2}$ & $2 + 3$ & $\frac{5}{2}$ \\
$1$ & & $0+1+2$ & $\frac{1}{2} + \frac{3}{2} + \frac{5}{2}$ & $1 + 2 + 3$ &
$\frac{3}{2} + \frac{5}{2}$ & $2$ \\
$\frac{3}{2}$ & & & $0+1+2+3$ & $\frac{1}{2}+\frac{3}{2}+\frac{5}{2}$ &
$1+2$ & $\frac{3}{2}$ \\
$2$ & & & & $0 + 1 + 2$ & $\frac{1}{2}+\frac{3}{2}$ & $1$ \\
$\frac{5}{2}$ & & & & & $0+1$ & $\frac{1}{2}$ \\
$3$ & & & & & & $0$ \\
\end{tabular}
\end{equation}
The quantum dimensions can be obtained from Eq.~(\ref{quant_dim}),
\begin{align}
d^{(6)}_{0} &= d^{(6)}_{3} = 1 &
d^{(6)}_{\frac{1}{2}} &= d^{(4)}_{\frac{5}{2}} = \sqrt{2+\sqrt{2}} \ , \nonumber \\ 
d^{(6)}_{1} &= d^{(6)}_{2} = 1+\sqrt{2}  & 
d^{(6)}_{\frac{3}{2}} &= \sqrt{2} \sqrt{2+\sqrt{2}}\ .
\end{align}
The entries of the $S$-matrix are given by
$S_{i,j} = \sqrt{\frac{2}{k+2}} \sin(\frac{(2i+1)(2j+1)\pi}{(k+2)})$, for $i,j=0,1/2,1,\ldots,k/2$, with
$k=6$.

\noindent
{\it Level $k=7$ --}
Finally, we provide some details of the $k=7$ model, which contains four  distinct
anyons with spins $0$, $1$, $2$ and $3$.  
The fusion rules are 
\begin{equation}
\begin{tabular}{c|ccc}
$\times$ & $1$ & $2$ & $3$ \\
\hline
$1$ & $0+1+2$ & $1+2+3$ & $2+3$ \\
$2$ & & $0+1+2+3$ & $1+2$\\
$3$ & & & $0+1$\\
\end{tabular}
\end{equation}
and the quantum dimensions are given by
\begin{align}
d^{(7)}_{0} &= 1\ , & d^{(7)}_{1} &= 1+2\cos(2 \pi/9)\ , \nonumber\\
d^{(7)}_{2} &= 1+2\cos(\pi/9) \ , & d^{(7)}_{3} &= 2\cos(\pi/9) \ .
\end{align}
The entries of the $S$-matrix are given by
$S_{i,j} = \sqrt{\frac{4}{k+2}} \sin(\frac{(2i+1)(2j+1)\pi}{(k+2)})$, for $i,j=0,1,\ldots,(k-1)/2$, with
$k=7$.

\section{$F$-matrices of the su(2)$_k$ theories}
\label{app:f-symbols}

In this section, we give an explicit expression for the $F$-symbols, following Ref.\cite{kr88}
We begin with some preliminary notation.
The $q$-numbers are defined as
$\qnum{n} =\sum_{i=1}^{n} q^{\frac{n+1}{2}-i} = \frac{q^{\frac{n}{2}}-q^{-\frac{n}{2}}}{q^{\frac{1}{2}}-q^{-\frac{1}{2}}}$.
The $q$-factorials are defined as $\qnum{n}! = \qnum{n}\qnum{n-1}\cdots \qnum{1}$,
for integer $n>0$, and $\qnum{0}! = 1$.
The labels of the anyons $a,b,\ldots$ take the values $0,1/2,1,\ldots$.
The quantum dimensions are
$d_j = \qnum{2j+1} = \sin\left(\frac{(2j+1)\pi}{k+2}\right)/\sin\left(\frac{\pi}{k+2}\right) = d_{k/2-j}$.
Moreover, we define 
\begin{equation}
\Delta (a,b,c) = \sqrt{\frac{\qnum{a+b-c}!\qnum{a-b+c}!\qnum{-a+b+c}!}{\qnum{a+b+c+1}!}}
\end{equation}
where  
$a\leq b+c$, $b\leq a+c$, $c\leq a+b$ and $a+b+c = 0 \bmod 1$.
\begin{widetext}
Using the above  introduced notation, the $F$-symbols can be written  as follows\cite{kr88}
\begin{equation}
\begin{split}
\left(\fs{a}{b}{c}{d}\right)^{e}_{f} &= 
(-1)^{a+b+c+d} \Delta(a,b,e)\Delta(c,d,e)\Delta(b,c,f)\Delta(a,d,f)
\sqrt{\qnum{2e+1}}\sqrt{\qnum{2f+1}}\\
&\sideset{}{'}\sum_{n}
\frac{(-1)^{n}\qnum{n+1}!}
{\qnum{a+b+c+d-n}!\qnum{a+c+e+f-n}!\qnum{b+d+e+f-n}!}\\
&\times \frac{1}
{\qnum{n-a-b-e}!\qnum{n-c-d-e}!\qnum{n-b-c-f}!\qnum{n-a-d-f}!} \ ,
\end{split}
\end{equation}
where the sum over $n$ runs over (non-negative) integers such that
$$\max(a+b+e,c+d+e,b+c+f,a+d+f)\leq n \leq \min(a+b+c+d,a+c+e+f,b+d+e+f) \ ,$$
which guarantees that the arguments of the $q$-factorials are non-negative integers. 
\end{widetext}


\section{Microscopic models}
\label{app:hamiltonians}

\subsection{Basis and Hamiltonian}
We consider a chain of spin-$j$ anyons, using the basis displayed in 
 Fig.~\ref{anyon_chains_general}.
We  fix the spin-$j$ anyon to be either a spin-$1/2$ anyon,
or a spin-$1$ anyon; however, the Hamiltonian defined below can be generalized to any value  $j\in \{0,1/2,...,k/2\}$.
Throughout most of this paper, we apply  periodic boundary conditions, \ie, $x_{L}=x_0$, where $L$ denotes
the number of anyonic quasiparticles in the chain.
 \begin{figure}[t]
\begin{center}
\includegraphics[width=7cm]{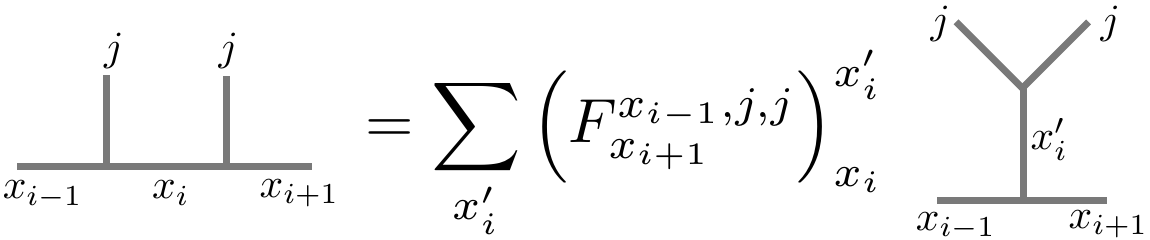}
\caption{Basis transformation used to  obtain the fusion product of two neighboring spin-$j$ anyons.}
\label{basis_chain_heis}
\end{center}
\end{figure}

We consider interactions between nearest neighboring spin-$j$ anyons. In the case of
$j=1/2$ (the `su(2)$_k$ spin-$1/2$ chain'), two neighboring spin-$1/2$ anyons may fuse into a
spin-$0$ or a spin-$1$ anyon. 
In contrast, for the case of $j=1$ (the `su(2)$_k$ spin-$1$ chain'), 
two neighboring spin-$1$ anyons may fuse into 
a spin-$0$, a spin-$1$, or a spin-$2$ anyon (for $k\ge 4$).
In order to obtain the fusion product
of two nearest-neighbor spin-$j$ anyons in the basis shown in
Fig.~\ref{anyon_chains_general}, an $F$-transformation has to be
performed, as illustrated in Fig.~\ref{basis_chain_heis}.
 Consequently, the projector
  onto a particular fusion channel $l$ is 
composed of two $F$-transformations. This projector, denoted by $P_i^{(l)}$,  penalizes the fusion of anyons at positions 
$i$ and $i+1$  into an 
$l$-anyon, and it is defined as follows,
\begin{equation}
\begin{split}
&P_i^{(l)} \ket{x_0,...,x_{i-1},x_i,x_{i+1},...,x_{L-1}} = \\
&\sum_{x_i'} (F^{x_{i-1},j,j}_{x_{i+1}})_{x_i}^l (F^{x_{i-1},j,j}_{x_{i+1}})_{x_i'}^l
\ket{x_0,...x_{i-1},x_i',x_{i+1},...,x_{L-1}}
\end{split}
\label{projector}
\end{equation}
We note that this definition utilizes  that  $F=F^{-1}$ for su(2)$_k$. The
explicit form of the local projectors for the systems studied in this paper 
is given in appendix \ref{app:f-symbols}.
The  Hamiltonians discussed in the following section are composed of the sum  of the local
projection operators $P_i^{(l)}$ onto the fusion product $l$ of
two nearest neighbor spin-$j$ anyons

\subsection{Hamiltonian of the su(2)$_k$ spin-$1/2$ chain}
\label{spin12_chain}

The Hamiltonian of the su(2)$_k$ spin-$1/2$ chain is given by 
\begin{equation}
H = J \sum_i  P^{(0)}_i \ ,
\label{anyon_chain1/2}
\end{equation}
where the projector $P_i^{(0)}$ is defined in Eq.~(\ref{projector})
(note that $l=0$ and $j=\frac{1}{2}$), and the coupling constant
takes the values $J=\pm 1$. 
In analogy to the `ordinary' Heisenberg spin-$1/2$ chain,
we denote the case $J=-1$ in Hamiltonian (\ref{anyon_chain1/2})
as antiferromagnetic (AFM) coupling while $J=1$ is ferromagnetic (FM) coupling.

In the following, we present matrix representations of the Hamiltonians of the spin-$1/2$ anyon chains for 
$k=2,3,4,5$. The matrix formulation for the su(2)$_3$ spin-$1/2$ chain
 was first introduced in \cite{Feiguin_07} (see also \cite{anyon_review}).
Local basis elements are labeled by $x_i$, where 
$x_i \in \{ 0,1/2,1,\ldots, k/2\}$. The order of anyon spins in the matrix representation
 is ascending. 
We also  introduce the
operators $n_i^{j}$ acting on local state $|x_i\rangle$: 
  $n_i^{j}|x_i\rangle = e |x_i\rangle$ where the eigenvalue $e=1$  if the local basis element $x_i=j$,
and $e=0$ otherwise.

From the definition of the summands of the Hamiltonian (eq.\ref{projector}),
 it is apparent that the matrix representation of the projector $P_i^{0}$ 
depends on the basis elements $x_{i-1}$, $x_i$ and $x_{i+1}$.
In particular, non-trivial contributions to $P_i^0$ exist only for certain
values  of $x_{i-1}$ and $x_{i+1}$, and thus each contribution to the projector will 
be proportional to $n^j_{i-1}n^{j'}_{i+1}$, for some values of $j$ and $j'$. By specifying both
$j$ and $j'$, the possible values of $x_i$ are fixed by the fusion rules. If there is
only one value $x_i$ can take (for given $j$ and $j'$), we omit the identity operator that is applied to 
basis element $|x_i\rangle$.
If there is more than one
possible value of $x_i$, we specify the matrix assigning the correct
energies. 

In the case of even-$k$ spin-$1/2$ chains,  the fusion rules
Eq.~(\ref{fusion_rules_k2}) impose that the values of the local basis
elements $x_{i}$  alternate between integer and half-integer values.
For the odd-$k$ spin chains (both spin-$1/2$ and spin-$1$ chains), we only consider the integer anyon spin subspace (recall the automorphism
 that applies to odd $k$ anyons, see appendix~\ref{app:su2k-anyons}).
For the even-$k$ spin-$1$ chains, the Hilbert space splits into two disjoint
sectors, the integer sector (all $x_i$ take integer values), and the half-integer sector (all 
$x_i$ assume half-integer values). 

\begin{widetext}
\subsubsection{su(2)$_2$ spin-$1/2$ chain}
The Hamiltonian eq.~\eqref{anyon_chain1/2} takes a rather
simple form in the case of  su(2)$_2$, namely
\begin{equation}
H^{(k=2)} = 
J\sum_{i} n_{i-1}^0 n_{i+1}^0 + n_{i-1}^1 n_{i+1}^1 +
+\frac{1}{2}n_{i-1}^{1/2} n_{i+1}^{1/2}
\begin{pmatrix}
1 & -1 \\ -1 & 1
\end{pmatrix}_{i} \ .
\end{equation}

\subsubsection{su(2)$_3$ spin-$1/2$ chain}
The Hamiltonian for the $k=3$ spin-$1/2$ chain is given by
\begin{equation}
H^{(k=3)} = 
J\sum_{i} n_{i-1}^0 n_{i+1}^0 + \frac{1}{d^2} n_{i-1}^1 n_{i+1}^1
\begin{pmatrix}
1 & -\sqrt{d} \\ -\sqrt{d} & d
\end{pmatrix}_i \  ,
\end{equation}
where
$d=d_1 = (1+\sqrt{5})/2$.

\subsubsection{su(2)$_4$ spin-$1/2$ chain}
In the case $k=4$, the local basis elements  alternate between
integer spin, $x_i \in \{ 0,1,2\}$ and half integer spin, $x_{i+1} \in \{ 1/2,3/2\}$.
The Hamiltonian takes the following form
\begin{equation}
H^{(k=4)} = 
J\sum_{i} n_{i-1}^0 n_{i+1}^0 + n_{i-1}^{2} n_{i+1}^{2}
+\frac{1}{2}n_{i-1}^{1} n_{i+1}^{1}
\begin{pmatrix}
1 & -1 \\ -1 & 1
\end{pmatrix}_{i} 
+ \frac{1}{3}n_{i-1}^{1/2} n_{i+1}^{1/2}
\begin{pmatrix}
1 & -\sqrt{2} \\ -\sqrt{2} & 2
\end{pmatrix}_{i}
+ \frac{1}{3} n_{i-1}^{3/2} n_{i+1}^{3/2}
\begin{pmatrix}
2 & -\sqrt{2} \\ -\sqrt{2} & 1
\end{pmatrix}_{i}
\end{equation}

\subsubsection{su(2)$_5$ spin-$1/2$ chain}
Using the notation
$d_1 = 1+2\cos(2\pi/7)$ and $d_2 = 2 \cos(\pi/7)$, the Hamiltonian reads
\begin{equation}
H^{(k=5)} = 
J\sum_{i} n_{i-1}^0 n_{i+1}^0 +
 \frac{1}{d_1 d_2} n_{i-1}^{1} n_{i+1}^{1}
\begin{pmatrix}
d_1 & -\sqrt{d_1 d_2} \\ -\sqrt{d_1 d_2} & d_2
\end{pmatrix}_{i} 
+ \frac{1}{d_2^2} n_{i-1}^{2} n_{i+1}^{2}
\begin{pmatrix}
1 & -\sqrt{d_1} \\ -\sqrt{d_1} & d_1
\end{pmatrix}_{i}
\end{equation}

\subsection{Hamiltonian of the su(2)$_k$ spin-$1$ chain}
\label{spin1_chain}
We define the Hamiltonian of the su(2)$_k$ spin-$1$ chain as follows,  
\begin{equation}
H= J_1 \sum_i P^{(1)}_i+J_2\sum_i P^{(2)}_i\, .
\label{anyon_chain1}
\end{equation}
The projectors $P^{(1)}_i$ and $P^{(2)}_i$ are defined in Eq.~(\ref{projector}), where $l=1$ and $l=2$, respectively.
This Hamiltonian is the su(2)$_k$ anyonic equivalent of  the bilinear-biquadratic spin-$1$
chain.
Throughout the paper, we  parametrize the Hamiltonian by the angle $\theta$ as follows:
 $J_1 =-\sin(\theta_{2,1}) $, $J_2= \cos(\theta_{2,1})$.

The Hamiltonian Eq.~(\ref{anyon_chain1}) is defined for levels $k\ge 4$, in
which case the fusion of two spin-$1$ anyons may result in  a spin-$0$,
a spin-$1$ or a spin-$2$ anyon (for level $k=3$, spins $\frac{1}{2}$ and $1$ are automorph, \ie, the spin-$1$ chain is 
equivalent to the spin-$1/2$ chain; moreover, the fusion rules imply $1\times 1 = 0 +1$.
For $k=2$, the fusion of two spin-$1$ particles is trivial, $1\times 1 = 0$).

\subsubsection{The su(2)$_4$ spin-$1$ chain}
\label{spin1_chain_k=4}

We shall now present a matrix representation of the Hamiltonian of the su(2)$_4$ spin-$1$ chain, using 
 the same notation as  in the previous
subsection.
In the integer sector (IS), the projectors onto the different channels can be written as follows,
\begin{align}
P^{(0)}_{i,{\rm IS}} &= n^{0}_{i-1}n^{0}_{i+1} + n^{2}_{i-1}n^{2}_{i+1}  +
\frac{1}{4}n^{1}_{i-1} n^{1}_{i+1}
\begin{pmatrix} 1 & -\sqrt{2} & 1 \\ -\sqrt{2} & 2 & -\sqrt{2} \\ 1 & -\sqrt{2} & 1 \end{pmatrix}_i \\
P^{(1)}_{i,{\rm IS}} &= n^{0}_{i-1}n^{1}_{i+1} + n^{1}_{i-1}n^{0}_{i+1} +
n^{1}_{i-1}n^{2}_{i+1} + n^{2}_{i-1}n^{1}_{i+1}  +
\frac{1}{2}n^{1}_{i-1} n^{1}_{i+1}
\begin{pmatrix} 1 & 0 & -1 \\ 0 & 0 & 0 \\ -1 & 0 & 1 \end{pmatrix}_i  \\
P^{(2)}_{i,{\rm IS}} &= n^{0}_{i-1}n^{2}_{i+1} + n^{2}_{i-1}n^{0}_{i+1}  +
\frac{1}{4}n^{1}_{i-1} n^{1}_{i+1}
\begin{pmatrix} 1 & \sqrt{2} & 1 \\ \sqrt{2} & 2 & \sqrt{2} \\ 1 & \sqrt{2} & 1 \end{pmatrix}_i 
\end{align}

In the half-integer sector (HIS),
we can write the projectors as follows,
\begin{align}
P^{(0)}_{i,{\rm HIS}} &= \frac{1}{2} (n^{1/2}_{i-1}n^{1/2}_{i+1}+n^{3/2}_{i-1}n^{3/2}_{i+1})
\begin{pmatrix} 1 & -1 \\ -1 & 1\end{pmatrix}_i \\
P^{(1)}_{i,{\rm HIS}} &= \frac{1}{2} (n^{1/2}_{i-1}n^{1/2}_{i+1}+n^{3/2}_{i-1}n^{3/2}_{i+1})
\begin{pmatrix} 1 & 1 \\ 1 & 1\end{pmatrix}_i  +
\frac{1}{2} (n^{1/2}_{i-1}n^{3/2}_{i+1}+n^{3/2}_{i-1}n^{1/2}_{i+1})
\begin{pmatrix} 1 & -1 \\ -1 & 1\end{pmatrix}_i \\
P^{(2)}_{i,{\rm HIS}} &= \frac{1}{2} (n^{1/2}_{i-1}n^{3/2}_{i+1}+n^{3/2}_{i-1}n^{1/2}_{i+1})
\begin{pmatrix} 1 & 1 \\ 1 & 1\end{pmatrix}_i
\end{align}


\subsubsection{The su(2)$_5$ spin-$1$ chain}

Using notation
$d_1= 1+2\cos(2\pi/7)$ and $d_2 = 2 \cos(\pi/7)$, the projectors are given by
\begin{multline}
P^{(1)}_i = n_{i-1}^{0} n_{i+1}^{1} + n_{i-1}^{1} n_{i+1}^{0} +
\frac{1}{d_1^4} n_{i-1}^{1}n_{i+1}^{1}
\begin{pmatrix}
d_1^3 & -d_1^{3/2} & -d_1^2 d_2^{3/2} \\
-d_1^{3/2} & 1 & \sqrt{d_1} d_2^{3/2} \\
-d_1^2 d_2^{3/2} & \sqrt{d_1} d_2^{3/2} & d_1 d_2^3
\end{pmatrix}_i
\\
+ \frac{d_2}{d_1^2} (n_{i-1}^{1}n_{i+1}^{2}+n_{i-1}^{2}n_{i+1}^{1})
\begin{pmatrix}
d_2 & -\sqrt{d_2} \\
-\sqrt{d_2} & 1
\end{pmatrix}_i
+ \frac{1}{d_1 d_2} n_{i-1}^{2}n_{i+1}^{2}
\begin{pmatrix}
d_2 & \sqrt{d_1 d_2} \\
\sqrt{d_1 d_2} & d_1
\end{pmatrix}_i
\end{multline}

\begin{equation}
P^{(2)}_i = n_{i-1}^{0} n_{i+1}^{2} + n_{i-1}^{2} n_{i+1}^{0} +
\frac{1}{d_1^4} n_{i-1}^{1}n_{i+1}^{1}
\begin{pmatrix}
d_1^2 d_2 & d_1^{3/2} d_2^2 & d_1 d_2^{3/2} \\
d_1^{3/2} d_2^2 & d_1 d_2^3 & \sqrt{d_1} d_2^{5/2} \\
d_1 d_2^{3/2} & \sqrt{d_1} d_2^{5/2} & d_2^2
\end{pmatrix}_i
+ \frac{d_2}{d_1^2} (n_{i-1}^{1}n_{i+1}^{2}+n_{i-1}^{2}n_{i+1}^{1})
\begin{pmatrix}
1 & \sqrt{d_2} \\
\sqrt{d_2} & d_2
\end{pmatrix}_i
\end{equation}

\subsubsection{The su(2)$_6$ spin-$1$ chain}
\label{spin1_chain_k=6}

In the following, we use the notation
$d_{1/2} = 2\cos(\pi/8)$,
$d_{1} = 1+2\cos(\pi/4)=1+\sqrt{2}$
and $d_{3/2} = 2 \sqrt{2} \cos(\pi/8)$.
In the integer sector (IS), the projectors onto the different channels can be written as follows,
\begin{multline}
P^{(1)}_{i,{\rm IS}} =
n^{0}_{i-1}n^{1}_{i+1} + n^{1}_{i-1}n^{0}_{i+1}  +
n^{2}_{i-1}n^{3}_{i+1} + n^{3}_{i-1}n^{2}_{i+1}  +
\frac{1}{2}(n^{1}_{i-1}n^{2}_{i+1} + n^{2}_{i-1}n^{1}_{i+1})
\begin{pmatrix} 1 & -1 \\ -1 & 1 \end{pmatrix}_{i} \\ +
\frac{1}{d_{1/2} d_{1} d_{3/2}} n^{1}_{i-1}n^{1}_{i+1}
\begin{pmatrix}
d_{1/2} d_{3/2} & -\sqrt{d_{1/2}d_{3/2}} & -d_{1} \sqrt{d_{1/2}d_{3/2}} \\
-\sqrt{d_{1/2}d_{3/2}} & 1 & d_{1} \\
-d_{1} \sqrt{d_{1/2}d_{3/2}} & d_{1} & d_{1}^2
\end{pmatrix}_{i} \\ +
\frac{1}{d_{1/2} d_{1} d_{3/2}} n^{2}_{i-1}n^{2}_{i+1}
\begin{pmatrix}
d_{1}^2 & d_{1} & -d_{1} \sqrt{d_{1/2}d_{3/2}} \\
d_{1} & 1 & -\sqrt{d_{1/2}d_{3/2}} \\
-d_{1} \sqrt{d_{1/2}d_{3/2}} & -\sqrt{d_{1/2}d_{3/2}} & d_{1/2} d_{3/2}
\end{pmatrix}_{i}
\end{multline}
\begin{multline}
P^{(2)}_{i,{\rm IS}} =
n^{0}_{i-1}n^{2}_{i+1} + n^{2}_{i-1}n^{0}_{i+1}  +
n^{1}_{i-1}n^{3}_{i+1} + n^{3}_{i-1}n^{1}_{i+1}  +
\frac{1}{2}(n^{1}_{i-1}n^{2}_{i+1} + n^{2}_{i-1}n^{1}_{i+1})
\begin{pmatrix} 1 & 1 \\ 1 & 1 \end{pmatrix}_{i} \\ +
\frac{1}{d_{1/2} d_{1} d_{3/2}} n^{1}_{i-1}n^{1}_{i+1}
\begin{pmatrix}
d_{1/2} d_{3/2} & d_{1} \sqrt{d_{1/2}d_{3/2}} & \sqrt{d_{1/2}d_{3/2}} \\
d_{1} \sqrt{d_{1/2}d_{3/2}} & d_{1}^2 & d_{1} \\
\sqrt{d_{1/2}d_{3/2}} & d_{1} & 1
\end{pmatrix}_{i} \\ +
\frac{1}{d_{1/2} d_{1} d_{3/2}} n^{2}_{i-1}n^{2}_{i+1}
\begin{pmatrix}
1 & d_{1} & \sqrt{d_{1/2}d_{3/2}} \\
d_{1} & d_{1}^2 & d_{1} \sqrt{d_{1/2}d_{3/2}} \\
\sqrt{d_{1/2}d_{3/2}} & d_{1} \sqrt{d_{1/2}d_{3/2}} & d_{1/2} d_{3/2}
\end{pmatrix}_{i}
\end{multline}
In the half-integer sector (HIS), the projectors onto the different channels can be written as follows,
\begin{multline}
P^{(1)}_{i,{\rm HIS}} =
\frac{1}{d_{1}^2}
n^{1/2}_{i-1}n^{1/2}_{i+1}
\begin{pmatrix}
d_{1/2}^2 & d_{1/2} \sqrt{d_1} \\ d_{1/2} \sqrt{d_1} & d_1
\end{pmatrix}_{i} + 
\frac{1}{\sqrt{d_{1/2} d_{1} d_{3/2}}}
(n^{1/2}_{i-1}n^{3/2}_{i+1} + n^{3/2}_{i-1}n^{1/2}_{i+1})
\begin{pmatrix}
1 & -\sqrt{d_{1}} \\ -\sqrt{d_{1}} & d_{1}
\end{pmatrix}_{i} \\ +
\frac{1}{2}
n^{3/2}_{i-1}n^{3/2}_{i+1}
\begin{pmatrix} 1 & 0 & -1 \\ 0 & 0 & 0 \\ -1 & 0 & 1 \end{pmatrix}_{i}
\frac{1}{\sqrt{d_{1/2} d_{1} d_{3/2}}}
(n^{3/2}_{i-1}n^{5/2}_{i+1} + n^{5/2}_{i-1}n^{3/2}_{i+1})
\begin{pmatrix}
d_{1} & -\sqrt{d_{1}} \\ -\sqrt{d_{1}} & 1
\end{pmatrix}_{i} +
\frac{1}{d_{1}^2}
n^{5/2}_{i-1}n^{5/2}_{i+1}
\begin{pmatrix}
d_1 & d_{1/2} \sqrt{d_1} \\ d_{1/2} \sqrt{d_1} & d_{1/2}^2
\end{pmatrix}_{i} 
\end{multline}
\begin{multline}
P^{(2)}_{i,{\rm HIS}} =
\frac{1}{\sqrt{d_{1/2} d_{1} d_{3/2}}}
(n^{1/2}_{i-1}n^{3/2}_{i+1} + n^{3/2}_{i-1}n^{1/2}_{i+1})
\begin{pmatrix}
d_{1} & \sqrt{d_{1}} \\ \sqrt{d_{1}} & 1
\end{pmatrix}_{i} +
\frac{1}{d_{1/2} d_{1}^2 d_{3/2} }n^{3/2}_{i-1}n^{3/2}_{i+1}
\begin{pmatrix}
d_{1}^2 & d_{1}^{3/2} d_{3/2} & d_{1}^2 \\  
d_{1}^{3/2} d_{3/2} & d_{1/2}^3 d_{3/2} & d_{1}^{3/2} d_{3/2} \\
d_{1}^2 & d_{1}^{3/2} d_{3/2} & d_{1}^2
\end{pmatrix}_{i} \\ +
\frac{1}{\sqrt{d_{1/2} d_{1} d_{3/2}}}
(n^{3/2}_{i-1}n^{5/2}_{i+1} + n^{5/2}_{i-1}n^{3/2}_{i+1})
\begin{pmatrix}
1 & \sqrt{d_{1}} \\ \sqrt{d_{1}} & d_{1}
\end{pmatrix}_{i}
\end{multline}

\subsubsection{The su(2)$_7$ spin-$1$ chain}
In the following, we use the notation
$d_1 = 1+2\cos(2\pi/9)$,
$d_2 = 1+2\cos(\pi/9)$
and $d_3 = 2 \cos(\pi/9)$.
The projector $P^{(1)}_i$ takes the form
\begin{multline}
P^{(1)}_i = n_{i-1}^{0} n_{i+1}^{1} + n_{i-1}^{1} n_{i+1}^{0}
+ n_{i-1}^{1}n_{i+1}^{1}
\begin{pmatrix}
\frac{1}{d_1} & -\frac{1}{\sqrt{d_1}d_2} & -\frac{d_3}{d_1\sqrt{d_2}} \\
-\frac{1}{\sqrt{d_1}d_2} & \frac{1}{d_2^2} & \frac{d_3}{\sqrt{d_1 d_2}}\\
-\frac{d_3}{d_1\sqrt{d_2}} & \frac{d_3}{\sqrt{d_1 d_2}} & \frac{d_3^2}{d_1 d_2}
\end{pmatrix}_i
+ (n_{i-1}^{1} n_{i+1}^{2} + n_{i-1}^{2} n_{i+1}^{1})
\begin{pmatrix}
\frac{d_3^2}{d_2^2} & -\frac{\sqrt{d_1}d_3^{3/2}}{d_2^2} \\ 
-\frac{\sqrt{d_1}d_3^{3/2}}{d_2^2} & \frac{d_1d_3}{d_2^2} 
\end{pmatrix}_i
\\
 + n_{i-1}^{2}n_{i+1}^{2}
\begin{pmatrix}
\frac{d_1^2 d_3}{d_2^3} & \frac{\sqrt{d_1}d_3}{d_2^{7/2}} & -\frac{\sqrt{d_1}d_3^{3/2}}{d_2^2} \\
\frac{\sqrt{d_1}d_3}{d_2^{7/2}} & \frac{d_3}{d_1d_2^{4}} & -\frac{d_3^{3/2}}{d_1d_2^{5/2}} \\
-\frac{\sqrt{d_1}d_3^{3/2}}{d_2^2} & -\frac{d_3^{3/2}}{d_1d_2^{5/2}} & \frac{d_3^2}{d_1d_2}
\end{pmatrix}_i
+ (n_{i-1}^{2} n_{i+1}^{3} + n_{i-1}^{3} n_{i+1}^{2})
\begin{pmatrix}
\frac{d_3}{d_1} & -\frac{d_3}{d_1\sqrt{d_2}} \\ 
-\frac{d_3}{d_1\sqrt{d_2}} & \frac{d_3}{d_1d_2} 
\end{pmatrix}_i
+ n_{i-1}^{3} n_{i+1}^{3}
\begin{pmatrix}
\frac{1}{d_1} & \frac{\sqrt{d_2}}{d_1\sqrt{d_3}} \\
\frac{\sqrt{d_2}}{d_1\sqrt{d_3}} & \frac{d_2}{d_1d_3}
\end{pmatrix}_i
\end{multline}
The projector $P^{(2)}_i$ is given by
\begin{multline}
P^{(2)}_i = n_{i-1}^{0} n_{i+1}^{2} + n_{i-1}^{2} n_{i+1}^{0}
 + n_{i-1}^{1}n_{i+1}^{1}
\begin{pmatrix}
\frac{d_2}{d_1^2} & \frac{d_3}{d_1^{3/2}} & \frac{d_3}{d_1^2\sqrt{d_2}} \\
\frac{d_3}{d_1^{3/2}} & \frac{d_3^2}{d_1d_2} & \frac{d_3^2}{d_1^{3/2}d_2^{3/2}} \\
\frac{d_3}{d_1^2\sqrt{d_2}} & \frac{d_3^2}{d_1^{3/2}d_2^{3/2}} & \frac{d_3^2}{d_1^2d_2^2}
 \end{pmatrix}_i
+ (n_{i-1}^{1}n_{i+1}^{2}+n_{i-1}^{2}n_{i+1}^{1})
\begin{pmatrix}
\frac{d_1d_3}{d_2^2} & \frac{\sqrt{d_1}d_3^{3/2}}{d_2^2}\\
\frac{\sqrt{d_1}d_3^{3/2}}{d_2^2} & \frac{d_3^2}{d_2^2}
\end{pmatrix}_i
\\
+n_{i-1}^{1} n_{i+1}^{3} + n_{i-1}^{3} n_{i+1}^{1}
+ n_{i-1}^{2}n_{i+1}^{2}
\begin{pmatrix}
\frac{d_3^2}{d_2^3} & \frac{d_3^3}{\sqrt{d_1}d_2^{5/2}} & \frac{d_3^{3/2}}{\sqrt{d_1}d_2^2} \\
\frac{d_3^3}{\sqrt{d_1}d_2^{5/2}} & \frac{d_3^4}{d_1d_2^2} & \frac{d_3^{5/2}}{d_1d_2^{3/2}} \\
\frac{d_3^{3/2}}{\sqrt{d_1}d_2^2} & \frac{d_3^{5/2}}{d_1d_2^{3/2}} & \frac{d_3}{d_1d_2}
\end{pmatrix}_i
+ (n_{i-1}^{2}n_{i+1}^{3}+n_{i-1}^{3}n_{i+1}^{2})
\begin{pmatrix}
\frac{d_3}{d_1d_2} & \frac{d_3}{d_1\sqrt{d_2}} \\
\frac{d_3}{d_1\sqrt{d_2}} & \frac{d_3}{d_1}
\end{pmatrix}_i
\end{multline}

\end{widetext}

\section{Exact form of the AKLT states}
\label{app:AKLT-states}

In this section, we present the explicit form of the zero energy ground states
of the periodic anyonic spin-1 chains for $k$ odd at the anyonic equivalent of the AKLT point. 
In the main text, we
discussed the case $k=5$.

At the AKLT point, the Hamiltonian contains only the projector  onto the
anyon spin-$2$ channel, i.e., the fusion of neighboring spin-$1$ anyons into a
spin-$2$ anyon is penalized. First, we note that the fusion  of anyons of types $1$ and $(k-1)/2$ (the
latter being the largest integer`spin' for an anyon in the su(2)$_k$ theory) 
results in
$1\times (k-1)/2 = (k-3)/2 + (k-1)/2$. 
In addition, we have that
$1\times (k-3)/2 = (k-5)/2 + (k-3)/2 + (k-1)/2$. Thus, 
a local basis for which $x_{i-1} = x_{i+1} = (k-1)/2$ 
implies that
$x_{i} = (k-3)/2$ or $x_{i} = (k-1)/2$.
 It follows that
after the local basis transformation, $\tilde{x}_i$ can only take two possible values,
namely $\tilde{x}_i = 0,1$  (consider 
$2 \times (k-1)/2 = (k-5)/2 + (k-3)/2$, and let $\tilde{x}_i=2$, $x_{i-1}=(k-1)/2$,
then $x_{i+1}$ could only take values $(k-5)/2$ and $(k-3)/2$ but not $(k-1)/2$).
This, in turn, means that a
 choice of
local variables $x_{i-1}$ and $x_{i+1}$ does not give rise to non-zero contributions at the AKLT point
as fusion of neighboring spin-$1$ anyons in the chain cannot result in $\tilde{x}_i=2$.
 We thus obtain a zero-energy ground state of the form
$\ket{v_{0}} = \ket{(k-1)/2,(k-1)/2,\ldots,(k-1)/2}$.

To construct the other ground states, we make use of the topological symmetry
operators $Y_l$. These operators mutually commute, and they commute with the
Hamiltonian. The state $v_{0}$ is not an eigenstate of the operators $Y_l$ (with $l>0$),
and hence alternative zero-energy ground states of the Hamiltonian are given by
$\ket{v_l} = Y_{l} \ket{v_{0}}$.
These ground states $\ket{v_l}$ ($l>0$) can be obtained explicitly.
The local basis states  take values 
$x_i = p-l$ or $x_i= p-l+1$, where $p=(k-1)/2$.
The states $\ket{v_{l}}$ are a sum over all possible labelings of the fusion tree 
with these two values of $x_i$.
We introduce the following notation: $\# l$ denotes the
number of local basis states for which$x_i=l$, and $\#(l,m)$ denotes the number
of local basis states for which  $x_i = l$ and $x_{i+1} = m$, where we use periodic
boundary conditions, $x_{L} = x_{0}$.
For $l>0$, we  obtain 
\begin{equation}
\ket{v_{l}} = \sum_{x_i \in \{p-l,p-l+1 \}} f_l (\{ x_i \}) \ket{x_0,x_1,\ldots,x_{L-1}} \ .
\end{equation}
The coefficients $f_l (\{ x_i \})$  ($0<l<p$) are given by
\begin{equation*}
\begin{split}
f_l (\{ x_i \}) =
\left( \frac{d_{l+1}}{d_{l}d_1}\right)^{L/2} (-1)^{\#p-l+1}
\left(\sqrt{\frac{d_{l-1}}{d_{l+1}}}\right)^{\#(p-l+1,p-l+1)}\\\times
\left(\sqrt{d_{p-l}d_{p-l+1}}\frac{d_{p}}{d_{l+1}}\right)^{\#(p-l+1,p-l)} \ .
\end{split}
\end{equation*}
For $l=p$, this results in
\begin{equation*}
f_p (\{ x_i \}) =
\left( \frac{d_{p}}{d_{p-1}}\right)^{L/2} (-\frac{d_{p-1}}{d_{p}\sqrt{d_1}})^{\#1} \ .
\end{equation*}

We denoted the ground states by $\ket{v_{l}}$ for the following reason. 
In section~\ref{sec:top-sym},
we explained that the operators $Y_l$ can be thought of as fusing an anyon with `spin' $l$
into the chain, effectively changing the `overall fusion channel', or flux thought the chain.
If we take a state
$\ket{v_{j_2}}$, and act on it with the operator $Y_{j_1}$, we find that
$Y_{j_1} \ket{v_{j_2}} = \sum_{j_3 \in j_1\times j_2} \ket{v_{j_3}}$, where the sum
is over those $j_3$ which appear in the fusion $j_1 \times j_2$.
Thus, the ground states
of the AKLT anyonic spin chain form a `representation' of the fusion algebra of su(2)$_k$.
This implies that  eigenstates of the topological operators
$Y_l$ can be constructed because  the modular $S$-matrix diagonalizes the fusion rules.
In particular orthogonal (not normalized) ground states at the AKLT point are
written as
$\ket{\psi_{\rm AKLT,i}} = \sum_{j=0}^{(k-1)/2}  S_{i,j} \ket{v_{j}}$, where $S_{i,j}$ is
the modular $S$ matrix for su(2)$_k$, and the sum is over integer values.


\section{Conformal field theories of interest}

In this appendix, we summarize the most important aspects of the conformal field theories relevant to this paper.
In the following,  `primary fields' refers
to Virasoro primary fields. Detailed discussions  of conformal field theories can
be found in  Ref.~\onlinecite{byb} and Ref~\onlinecite{bpz84}.

\subsection{Virasoro minimal models}
\label{app:minmod}

The unitary minimal models\cite{bpz84}, which can also be described in terms of the 
coset $\frac{su(2)_1\times su(2)_{k-1}}{su(2)_{k}}$, have
a central charge $c=1-\frac{6}{(k+1)(k+2)}$ ($k\geq 2$).
The primary fields are labeled by integers $r$ and $s$, where
$1 \leq r \leq k$ and $1\leq s \leq k+1$. Their conformal dimensions
are given by
\begin{equation}
h_{r,s}=\frac{(r(k+2)-(k+1)s)^2-1}{4(k+1)(k+2)}\ .
\label{h_minmodel}
\end{equation}
Typically, the minimal models are labeled by a
parameter $m=k+1$.

 \begin{table}[t]
	 \begin{center}
	 \begin{tabular}{c||c|c|c}  
	 $p$, $p'$ & $(A,D)$ & $(r,s)$ & mult. $2$ \\ \hline \hline
	 $p'=2(2n+1)$ & $(D_{p'/2+1},A_{p-1})$    &$r$ odd &  $r=p'-r$ \\ \hline 
	 $p=2(2n+1)$ &  $(A_{p'-1},D_{p/2+1})$    &$s$ odd &  $s=p-s$ \\ \hline
	   \end{tabular}
	   \caption{{\it Modular invariants of the Virasoro minimal models.}
For a given  pair of indices, $(p,p')=(m+1,m)$ and $n$ integer,  only the fields with indices $(r,s)$ as specified
	    in the third column appear ($1\le r<p'$, $1\le s<p$). Some fields have multiplicity two, as indicated in
		column four.
  }	   \label{table_modular}
	   \end{center}
	   \end{table}

Apart from the so-called diagonal models, there exist  modular invariants that
give rise to conformal field theories with a different field content\cite{ciz87a,ciz87b}.
More information on these modular invariants can be found in table~\ref{table_modular}.

\subsection{$\mathcal{N}=1$ superconformal minimal models}
\label{app:supercft}

The $\mathcal{N}=1$ superconformal minimal models\cite{fqs84} are described
by the coset
$$
\frac{su(2)_2\times su(2)_{k-2}}{su(2)_{k}} \ ,
$$
and have central charge $c=\frac{3}{2}-\frac{12}{k(k+2)}$.
The primary fields have conformal dimension
\begin{equation}
h_{(r,s)} = \frac{(r(k+2)-sk)^2-4}{8k(k+2)}+\frac{1}{32}(1-(-1)^{r-s}) \ ,
\label{superCFT_scalingdim}
\end{equation}
where $1\leq r \leq k-1$ and $1\leq s \leq k+1$. The fields
with $r+s$ even, i.e. the fields in the Neveu-Schwarz sector, 
have a super partner, whose conformal dimensions are given by 
\begin{equation}
h'_{(r,s)} = h_{(r,s)} + 1/2 + \delta_{r+s,2} \qquad \text{for $r+s$ even}\ .
\end{equation}

\subsection{$S_3$ minimal models}
\label{app:s3_models}

The class of $S_3$ symmetric minimal models\cite{fz87,zf87}
are described by the coset theory
$$
\frac{su(2)_4 \times su(2)_{k-4}}{su(2)_{k}} \ ,
$$
and have central charge
\begin{equation}
c=2-\frac{24}{(k-2)(k+2)} \ .
\end{equation}
There are two main sets of primary fields. The first set has conformal weights
\begin{equation}
h_{(r,s)} = \frac{(r(k+2)-s (k-2))^2-16}{16 (k-2)(k+2)} 
+\frac{1-\cos^4(\pi(r-s)/4)}{12} \ .
\end{equation}
The second set has scaling dimensions
\begin{multline}
h'_{(r,s)} = h_{(r,s)} +\\
\frac{1+\sin^2(\pi(r-s)/4)}{3} + \delta_{r,1}\delta_{s,1}+\delta_{r,1}\delta_{s,2}
+2 \delta_{r,2}\delta_{s,1}\ ,
\end{multline}
where for both sets
$1\leq r \leq k-3$ and $1\leq s \leq k+1$. There are additional (Virasoro)
primary fields, with scaling dimensions differing by integers from the scaling dimensions
listed above. These additional primary fields are not relevant to this work.

\subsection{The $Z_k$ parafermion CFT.}
\label{app:zk-pf}

The $Z_k$ parafermions\cite{zf85}  can be described in terms of the coset
$$
\frac{su(2)_k}{u(1)_{2k}} \ ,
$$
where u(1)$_{2k}$ denotes the $c=1$ boson, compactified on a circle
of radius $R=\sqrt{2k}$. The central charge is given by
$c=\frac{2(k-1)}{k+2}$, and the conformal dimensions of the primary fields
are given by
\begin{equation}
h_{(l,m)} = \frac{l(l+2)}{4(k+2)} - \frac{m^2}{4k} \ .
\end{equation}
Here, the indices run over values
$l=0,1,\ldots,k$ and $m=-l+2,-l+4,\ldots,l$.

\subsection{The $\mathbb{Z}_2$ orbifold theories}
\label{App:orbifolds}

We will briefly discuss the $\mathbb{Z}_2$ orbifold of the compactified boson at squared radius
$R^2=2p$.
For a detailed account, we refer to \cite{dijkgraaf}.
The number of primary fields is given $p+7$, where $p=1,2,\ldots$. For $p=1$,
the CFT is Abelian, and it is equivalent to a the compactified boson theory with
eight primary fields. In general, the following fields are present.

\begin{itemize}
\item
The identity field $\id$, with scaling dimension $h_\id=0$ and quantum dimension $d_\id = 1$.
\item
The field $\Theta$, with dimension $h_\Theta = 1$ and quantum dimension $d_\Theta = 1$.
\item
Two `degenerate' fields $\Phi^{1}$ and $\Phi^{2}$, with scaling dimension
$h_\Phi = \frac{p}{4}$ and quantum dimension $d_\Phi = 1$.
\item
The twist fields $\sigma^{1}$, $\sigma^{2}$ and $\tau^{1}$, $\tau^{2}$, with
scaling dimensions $h_\sigma = \frac{1}{16}$ and $h_\tau = \frac{9}{16}$ and
quantum dimensions $d_\sigma=d_\tau = \sqrt{p}$.
\item
The fields $\phi_\lambda$, with $\lambda = 1,2,\ldots,p-1$, with scaling dimensions
$h_\lambda = \frac{\lambda^2}{4p}$ and quantum dimensions $d_\lambda = 2$.
\end{itemize}

\subsubsection{The $S$-matrix}
To verify that the assignment of the topological symmetry sectors of states of the
critical su(2)$_4$ spin-1 anyonic chains are compatible with the fusion rules of the
orbifold CFTs describing the critical behavior, we need the fusion rules of the orbifold
CFTs. We will not give these fusions rules explicitly here, but specify the
modular $S$-matrix. The fusion rules can be obtained from the modular $S$-matrix
by means of the Verlinde formula\cite{Verlinde88}.

The modular $S$-matrix can be written in a compact way as follows. In the basis
$(\id,\Theta,\Phi^{i},\sigma^{i},\tau^{i},\phi_\lambda)$ for the rows and
$(\id,\Theta,\Phi^{j},\sigma^{j},\tau^{j},\phi_\mu)$ for the columns,
where $i,j=1,2$ and $\lambda,\mu = 1,2,\ldots,p-1$, the modular $S$-matrix is given by
\begin{equation}
S = \frac{1}{\sqrt{8p}}
\begin{pmatrix}
1 & 1 & 1 & \sqrt{p} & \sqrt{p} & 2 \\
1 & 1 & 1 & -\sqrt{p} & -\sqrt{p} & 2 \\ 
1 & 1 & (-1)^p & b_{i,j} & b_{i,j} & (-1)^\mu 2 \\
\sqrt{p} & -\sqrt{p} & b_{i,j} & a_{i,j} & -a_{i,j} & 0\\
\sqrt{p} & -\sqrt{p} & b_{i,j} & -a_{i,j} & a_{i,j} & 0\\
2 & 2 & (-1)^\lambda 2 & 0 & 0 & c_{\lambda,\mu}
\end{pmatrix} \ .
\end{equation}
Here, the matrices $a,b,c$ have the elements
\begin{eqnarray}
 a_{i,j} & = & \sqrt{p/2} (1+(2\delta_{i,j}-1)) e^{-\pi i p/2} \,, \nonumber \\ 
 b_{i,j} & = & (-1)^{p+\delta_{i,j}} \sqrt{p} e^{\pi i p/2} \,,  \nonumber \\
 c_{\lambda,\mu} & = & 4\cos(\pi \lambda\mu/p) \,. \nonumber
\end{eqnarray}
We note that we used a simplified notation in the above definition: for the matrix elements that do not
depend $i$ or $j$, the particular element does not depend on $i$ and $j$.
\vspace{1 cm}


\end{document}